\documentclass[aps,preprint,floatfix,nofootinbib,showpacs]{revtex4-1}
\pdfoutput=1
\usepackage{dcolumn}
\usepackage{bm}
\usepackage{graphicx}
\usepackage{amssymb,amsmath}
\usepackage{multirow}
\usepackage{units}
\usepackage{color,url}
\usepackage{slashed}
\usepackage{tabu}
\usepackage{array}
\usepackage[colorlinks=true,urlcolor=blue,anchorcolor=blue
,citecolor=blue,filecolor=blue,linkcolor=blue,menucolor=blue
,linktocpage=true,pdfproducer=medialab,pdfa=true]{hyperref}
\usepackage{colordvi}
\usepackage{subcaption}
\def\ga{\mathrel{\raise.3ex\hbox{$>$\kern-.75em\lower1ex\hbox{$\sim$}}}}
\def\la{\mathrel{\raise.3ex\hbox{$<$\kern-.75em\lower1ex\hbox{$\sim$}}}}
\def\beqa{\begin{eqnarray}}
\def\eeqa{\end{eqnarray}}
\newcommand{\ov}{\overline}

\begin{document}
\preprint{KIAS-P21001}
\title{
  Exploring properties of long-lived particles in \\ inelastic dark matter models at Belle II }
\def\slash#1{#1\!\!/}

\renewcommand{\thefootnote}{\arabic{footnote}}
\renewcommand{\thefootnote}{\arabic{footnote}}

\author{
  Dong Woo Kang$^1$, P. Ko$^1$, Chih-Ting Lu$^1$ }
\affiliation{
 $^1$ School of Physics, KIAS, Seoul 130-722, Republic of Korea
}
\date{\today}

\begin{abstract}

The inelastic dark matter model is one kind of popular models for the light dark matter (DM) 
below $O(1)$ GeV.  If the mass splitting between DM  excited and ground states is small enough, 
the co-annihilation becomes the dominant channel for thermal relic density and the DM excited 
state can be long-lived at the collider scale. We study scalar and fermion inelastic dark matter 
models for $ {\cal O}(1) $ GeV DM at Belle II with $ U(1)_D $ dark gauge symmetry broken into its 
$Z_2$ subgroup. 
We focus on dilepton displaced vertex signatures from decays of the DM excited state. 
With the help of precise displaced vertex detection ability at Belle II, we can explore the DM spin, 
mass and mass splitting between DM excited and ground states. Especially, we show scalar and 
fermion DM candidates can be discriminated and the mass and mass splitting of DM sector can be 
determined within the percentage of deviation for some benchmark points.
Furthermore, the allowed parameter space to explain the excess of muon 
$(g-2)_\mu$ is also studied and it can be covered in our displaced vertex analysis during the 
early stage of Belle II experiment.   
\end{abstract}

\maketitle

\section{Introduction}

The nature of dark matter (DM) in our Universe is still a great mysterious issue. As we know, DM 
plays a major role in the structure formation~\cite{Frenk:2012ph}, and its abundance is about 
5.5 times larger than the ordinary matter in the present Universe~\cite{Aghanim:2018eyx}.
However, even all robust evidences to support the existence of DM until now are only connected 
with gravitational interactions.  Still it is believed that there are non-negligible couplings between  Standard Model (SM) particles and DM in addition to gravitational interactions. 
Especially, the Weakly Interacting Massive Particle (WIMP) DM candidate  with the freeze-out 
mechanism has been overwhelming in both theoretical and experimental communities during 
the past decades as shown in Ref.~\cite{Plehn:2017fdg} and  references therein. 
This kind of WIMP DM particles can be searched with direct, indirect detections and also at colliders experiments. The null signal result from direct detection provides severe bounds on the cross section 
of DM and nuclear scattering above a few GeV~\cite{Aprile:2018dbl,Ren:2018gyx,Aprile:2019dbj}. 
For instance, the spin independent (spin dependent) DM and nuclear scattering cross section is 
restricted to $ \sigma_{SI}\lesssim 10^{-46}~{\rm cm}^2 $~\cite{Aprile:2018dbl} ($ \sigma_{SD}
\lesssim 2\times 10^{-41}~{\rm cm}^2 $~\cite{Aprile:2019dbj}) for $M_{DM}=100$ GeV.
Nevertheless, lighter DM candidates are still less constrained 
because the restrictions of fine energy threshold are required from these direct detection experiments. Therefore, those dark sector models with MeV to GeV DM candidates become more and more popular for phenomenological studies and new experimental searches~\cite{Knapen:2017xzo,Lin:2019uvt}.

For the GeV scale DM searches, the high-intensity machines such as the BESIII~\cite{Ablikim:2009aa}, BaBar~\cite{TheBABAR:2013jta}, Belle II~\cite{Adachi:2018qme,Kou:2018nap} and also the fixed 
target experiments~\cite{Bjorken:2009mm} can be more powerful than the high-energy machines 
such as the Tevatron and LHC~\cite{Boveia:2018yeb}. 
On the other hand, DM is not the lonely particle in the dark sector for most of DM portal
models~\cite{Pospelov:2007mp,ArkaniHamed:2008qn,Pospelov:2008zw,Baek:2013qwa}. 
In order to connect the SM sector with the dark sector, a mediator is required. 
Among these mediator candidates, the dark photon via the kinetic mixing portal is an attractive  type~\cite{Holdom:1985ag} for the dark sector with Abelian gauge symmetry. 
In order to be consistent with the relic density bound, it's natural for both DM and dark photon 
in a similar energy scale. 
Therefore, the searches for them at high-intensity machines are in full swing and relevant constraints can be set up as shown in Ref.~\cite{Izaguirre:2015zva,Duerr:2019dmv}. 
Even though the original purpose for these high-intensity machines is to study the properties of 
$ J/\psi $ and $ B $ mesons, we can also apply them to explore some models with light DM candidates via the mono-photon signature~\cite{Zhang:2019wnz,Lees:2017lec,Izaguirre:2015zva,Duerr:2019dmv}. 
Furthermore, as pointed out in Ref.~\cite{Izaguirre:2015zva,Mohlabeng:2019vrz,Duerr:2019dmv}, 
we can not only study the mono-photon signature, but also the displaced DM signature at B-factories 
for the inelastic DM models. Especially, searches for displaced DM signature can cover some 
parameter space which the invisible DM signature cannot reach at high-intensity machines.
Hence, we will focus on exploring the properties of long-lived particles in dark sectors with 
$ U(1)_D $ gauge symmetry at Belle II in this work. 

The inelastic (or excited) DM models with extra $U(1)_D$ gauge symmetry~\cite{TuckerSmith:2001hy,Baek:2014kna,Ko:2019wxq,Baek:2020owl} is one of the most popular dark sector models with light DM candidates\footnote{
In Ref. \cite{TuckerSmith:2001hy}, the $U(1)_D$ symmetry is explicitly and softly 
broken by a dim--2 operator for scalar DM, and there is no dark Higgs boson there.
On the other hand, in the models considered in this paper in that there is dark Higgs boson which 
plays important roles in DM phenomenology.}.
There are at least two states in the dark sectors and there is an inelastic transition between them 
via the new $U(1)_D$ gauge boson. If the mass splitting between these two states is small enough, 
the co-annihilation channel could be the dominant one of DM relic density in early Universe~\cite{Griest:1990kh}.   It is one of the unique features of this sort of models. 
The co-annihilation production for light DM via thermal freeze out is still consistent with the Cosmic Microwave Background (CMB) constraint for the amount of parameter space \cite{Baek:2020owl,Slatyer:2015jla}.
On the other hand, the constraint from DM and nuclear inelastic scattering is much weaker than the elastic one in the direct detection experiments\footnote{
The bounds from direct detection experiments are weak for $\sim O(1)$ GeV DM.
Hence we can safely ignore possible contributions to the DM-nucleon elastic scattering 
from the SM Higgs boson and dark Higgs boson in the direct detection experiments.}. 
It makes more allowed parameter space can be explored in the inelastic DM models. 
Besides, there are also other rich phenomenons in this model. For example, it is possible to explain the muon $g-2$ anomaly~\cite{Mohlabeng:2019vrz} and XENON1T excess~\cite{Harigaya:2020ckz,Bell:2020bes,Lee:2020wmh,Baryakhtar:2020rwy,Bramante:2020zos,An:2020tcg,Baek:2020owl,Borah:2020smw}.  Last but not least, the excited DM state can naturally become long-lived and leave displaced vertex inside detectors after it has been produced such that these novel signatures 
can be searched for at colliders.

The fundamental properties of DM particle include its mass, charge, and spin. Since we assume DM is electrically neutral or millicharged, only its mass and spin are left to be determined. Actually, the mass range for WIMP DM can cover from few MeV to a few hundred TeV and its spin can be $ 0, 1/2, 1, 3/2 $ 
or even higher. Most of the time the precise determination of DM spin and mass are challenging tasks. 
However, there are already some previous studies about DM spin and mass determinations. 
For the DM spin determination, it has been studied for direct detection experiments in Ref.~\cite{Catena:2017wzu} and collider experiments in Ref.~\cite{Cheng:2010yy,Melia:2011cu,Edelhauser:2012xb,Christensen:2013sea,Choi:2018sqc,Abdallah:2019tpo}. For the DM mass determination, it has been studied for direct detection in Ref.~\cite{Shan:2009ym,Kavanagh:2013wba} and collider experiments in Ref.~\cite{Cheng:2011ya,HarlandLang:2011ih, HarlandLang:2012gn,Kobach:2013tga,Harland-Lang:2013wxa,Christensen:2014yya,Konar:2015hea,Ko:2016xwd,Xiang:2016jni,Kamon:2017yfx,Dutta:2017sod,Kang:2019ukr,Kim:2019prx,Banerjee:2019ktv,Bae:2020dwf}.
Especially, using the displaced vertex information for DM mass determination starts from Ref.~\cite{Kang:2019ukr}. Thanks to the precise track resolution in the inner detector of ATLAS 
and CMS experiments, the DM mass reconstruction can be studied in some BSM models of the 
specific cascade processes involving long-lived particles. 
Furthermore, the resolutions of displaced vertex, charged lepton and photon momentum can be even improved at detectors of Belle II experiments~\cite{Adachi:2018qme}.
Therefore, our goal in this paper is to explore the spin and mass properties of inelastic DM models 
with dilepton displaced vertex signatures at Belle II. As we will see in Sec.~\ref{Sec:Simulation}, 
the scalar and fermion inelastic DM models can be well discriminated
and the DM mass and mass splitting between DM excited and ground states can be determined 
within the percentage of deviation for those benchmark points.

The organization of this paper is as follows. We first review scalar and fermion inelastic DM models 
with $U(1)_D$ gauge symmetry in Sec.~\ref{Sec:Model}. 
Both analytical representations and numerical results for cross sections of relevant signal processes 
are displayed in Sec.~\ref{Sec:Xsec}. We also point out how to distinguish scalar and fermion inelastic 
DM models via the size and kinematic distribution of their cross sections in this section.
Detailed simulations and methods to determine DM mass, and mass splitting between DM excited 
and ground states are shown in Sec.~\ref{Sec:Simulation}. Finally, we conclude our studies in Sec.~\ref{Sec:Conclusion}. Some supplemental formulae for Sec.~\ref{Sec:Xsec} can be found in the Appendix~\ref{Sec:LAB_rep} and~\ref{Sec:2to4_rep}.

\section{Inelastic dark matter models}\label{Sec:Model}

The scalar and fermion inelastic (or excited) DM models with $U(1)_D$ gauge symmetry are reviewed 
in this section. After the spontaneous symmetry breaking (SSB) of this $U(1)_D$ gauge symmetry, we expect 
the accidentally residual $ Z_2 $ symmetry, $\phi_1\rightarrow -\phi_1$ (scalar) or $\chi_1\rightarrow -\chi_1$ (fermion), 
can be left such that $ \phi_1 $ or $ \chi_1 $ are stable and become DM candidates in our Universe. 

\subsection{The scalar model \cite{Baek:2014kna,Baek:2020owl}}

We consider a dark sector with two singlet complex scalars $\Phi$ and 
$ \phi = (\phi_2 +i\phi_1)/\sqrt{2} $ as the dark Higgs and dark matter sectors, respectively. 
Both $\Phi$ and $\phi$ are charged under $U(1)_D$, but neutral of the SM gauge symmetry.
The $U(1)_D$ charges for them are assigned as $ Q_D(\Phi)= +2 $ and $ Q_D(\phi)= +1 $. 
Besides, the SM-like Higgs doublet and other SM particles do not carry $U(1)_D$ charges. 

The scalar part of the renormalizable and gauge invariant Lagrangian density is
\begin{equation}
{\cal L}_{scalar} = 
| D_{\mu}H |^2 + | D_{\mu}\Phi |^2 + | D_{\mu}\phi |^2 - V(H,\Phi,\phi),
\label{eq:Ls}
\end{equation}
with
\begin{equation}
D_{\mu}H = (\partial_\mu +i\frac{g}{2}\sigma_a W^a_{\mu}+i\frac{g'}{2}B_{\mu})H,
\label{eq:kin1}
\end{equation}
\begin{equation}
D_{\mu}\Phi = (\partial_\mu +ig_D Q_D(\Phi) X_{\mu})\Phi,
\label{eq:kin2}
\end{equation}
\begin{equation}
D_{\mu}\phi = (\partial_\mu +ig_D Q_D(\phi) X_{\mu})\phi,
\end{equation}
where $ W^a_{\mu} $, $ B_{\mu} $, and $ X_{\mu} $ are the gauge potentials of the $SU(2)_L$, $U(1)_Y$ and $U(1)_D$ with gauge couplings $g$, $g'$ and $g_D$, respectively. 
The $ \sigma_a $ is the Pauli matrix and $ a $ runs from 1 to 3.
The scalar potential in Eq.(\ref{eq:Ls}) is given by
\begin{align}
V(H,\Phi,\phi) = & 
-\mu^2_H H^{\dagger}H +\lambda_H (H^{\dagger}H)^2 -\mu^2_\Phi \Phi^{\ast}\Phi +\lambda_\Phi (\Phi^{\ast}\Phi)^2
\nonumber  \\ &
-\mu^2_\phi \phi^{\ast}\phi +\lambda_\phi (\phi^{\ast}\phi)^2
+(\mu_{\Phi\phi}\Phi^{\ast}\phi^2 +H.c.)
\nonumber  \\ &
+\lambda_{H\Phi}(H^{\dagger}H)(\Phi^{\ast}\Phi) +\lambda_{H\phi}(H^{\dagger}H)(\phi^{\ast}\phi) +\lambda_{\Phi\phi}(\Phi^{\ast}\Phi)(\phi^{\ast}\phi),
\label{eq:Vs}
\end{align}
where all parameters are assumed to be real for simplicity. 

We then expand $H,\Phi$ fields around the vacuum with the unitary gauge,
\begin{equation}
H(x) = \frac{1}{\sqrt 2} 
\left(
\begin{tabular}{c}
0
\\
$v + h(x)$
\end{tabular}
\right)
\;\;\; , \;\;\; 
\Phi (x) = \frac{1}{\sqrt 2} \left( v_D + h_D (x) \right),
\label{eq:expand}
\end{equation}
and highlight some important relations in the below :
\begin{itemize}
\item The four-points and off-diagonal interactions of the DM sector and new gauge boson can be 
obtained from the $ | D_{\mu}\phi |^2 $ term:
\begin{equation}
| D_{\mu}\phi |^2 = (\partial_{\mu}\phi_2)^2 +(\partial_{\mu}\phi_1)^2 + g^2_D X_{\mu}X^{\mu}(\phi^2_2 +\phi^2_1) +g_D X_{\mu}(\phi_2\partial^{\mu}\phi_1 -\phi_1\partial^{\mu}\phi_2)
\end{equation} 
\item The DM sector mass splitting and dark Higgs, DM sector trilinear interactions are derived from 
$ \mu_{\Phi\phi}\Phi^{\ast}\phi^2 +H.c. $ terms:
\begin{equation}
\mu_{\Phi\phi}\Phi^{\ast}\phi^2 +H.c. = \mu_{\Phi\phi}(v_D + h_D (x))(\phi^2_2-\phi^2_1).
\end{equation}
\item Finally, the $ \phi_1 $, $ \phi_2 $ masses and their mass splitting can be represented as
\begin{equation}
M_{\phi_{1,2}} = \sqrt{\frac{1}{2}(-\mu^2_{\phi}+\lambda_{H\phi}v^2 +\lambda_{\Phi\phi}v^2_D)\mp \mu_{\Phi\phi}v_D},
\end{equation}
and
\begin{equation}
\Delta_{\phi}\equiv M_{\phi_2} -M_{\phi_1} = \frac{2\mu_{\Phi\phi}v_D}{M_{\phi_1} + M_{\phi_2}}.
\end{equation}
\end{itemize}
Notice the interaction states $ (h, h_D) $ would be rotated to the mass states $ (h_1, h_2) $ 
via the mixing angle $ \theta $. The $ \sin\theta $ is chosen to be small enough in our analysis such that it is consistent with the SM-like Higgs boson data at the LHC\footnote{The long-lived dark Higgs phenomenology study in the inelastic DM model at Belle II can be found in Ref.~\cite{Duerr:2020muu}.}.

Finally, since all SM fermions don't carry $U(1)_D$ charges, the only way for the new $ X_\mu $ 
boson and SM fermions to interact is via the kinetic mixing between $ B_{\mu\nu} $ and 
$ X_{\mu\nu} $. The Lagrangian density of this part can be represented as
\begin{equation}
{\cal L}_{X,gauge} = -\frac{1}{4}  X_{\mu\nu}X^{\mu\nu} 
-\frac{\sin\epsilon}{2} B_{\mu\nu}X^{\mu\nu},
\label{eq:Zps}
\end{equation}
where $ \epsilon $ is the kinetic mixing parameter between these two $U(1)$s.
If we apply the linear order approximation in $ \epsilon $, the extra interaction terms for SM fermions and $ Z^{\prime} $ boson can be written as
\begin{equation}
{\cal L}_{Z^{\prime}f\overline{f}} = -\epsilon e c_W \sum_{f} x_f \ov{f} \slashed{Z}^{\prime} f,
\label{eq:Zpffbar}
\end{equation}
where $ c_W $ is the weak mixing angle and $ x_l = -1 $, $ x_{\nu} = 0 $, $ x_q = \frac{2}{3} $ or $ -\frac{1}{3} $ depending on the electrical charge of quark.
The $ Z^{\prime} $ boson mass can be approximated as
\begin{equation}
m_{Z^{\prime}} \simeq g_D Q_D(\Phi) v_D.
\label{eq:Zpmass}
\end{equation}
Notice the correction from the kinetic mixing term is second order in $ \epsilon $ which can be safely neglected here.


\subsection{The fermion model \cite{Ko:2019wxq,Baek:2020owl}} 

Here we consider a dark sector with the singlet complex scalar $\Phi$ and Dirac fermion $\chi$ 
as the dark Higgs and dark matter sectors, respectively. 
Both $\Phi$ and $\chi$ are charged under $U(1)_D$, but neutral of the SM gauge symmetry.
The $U(1)_D$ charges for them are assigned as $ Q_D(\Phi)= +2 $ and $ Q_D(\chi)= +1 $. Again, the SM-like Higgs doublet and other SM particles do not carry $U(1)_D$ charges.

The scalar part of the renormalizable and gauge invariant Lagrangian density is
\begin{equation}
{\cal L}_{scalar} = 
| D_{\mu}H |^2 + | D_{\mu}\Phi |^2 - V(H,\Phi),
\label{eq:Lf}
\end{equation}
where $ D_{\mu}H $ and $ D_{\mu}\Phi $ are the same in Eq.(\ref{eq:kin1}) and~(\ref{eq:kin2}).
The scalar potential in Eq.(\ref{eq:Lf}) is given by
\begin{align}
V(H,\Phi) = & 
-\mu^2_H H^{\dagger}H +\lambda_H (H^{\dagger}H)^2 -\mu^2_\Phi \Phi^{\ast}\Phi +\lambda_\Phi (\Phi^{\ast}\Phi)^2 
\nonumber  \\ &
+\lambda_{H\Phi}(H^{\dagger}H)(\Phi^{\ast}\Phi),
\label{eq:Vs}
\end{align}
where all parameters are assumed to be real for simplicity.

The Lagrangian density of dark matter sector part is 
\begin{equation}
{\cal L}_{\chi} = 
\ov{\chi}(i{\rlap{\,/}\partial}+g_D{\rlap{\,/}X}- M_{\chi})\chi -(\frac{f}{2}\ov{\chi^c}\chi\Phi^{\ast}+H.c.),
\label{eq:LDMf}
\end{equation}
where $f$ is assumed to be a real parameter. 
In order to decompose the Dirac fermion $\chi$ into a pair of two independent Majorana fermions, 
$\chi_1$ and $\chi_2$, we set 
\begin{equation}
\chi_{1,2}(x)=\frac{1}{\sqrt{2}}(\chi(x)\mp\chi^{c}(x)).
\end{equation}
After expanding $H,\Phi$ fields around the vacuum with the unitary gauge as shown in Eq.(\ref{eq:expand}),
the Eq.(\ref{eq:LDMf}) can be written as 
\begin{align}
{\cal L}_{\chi} = & 
\frac{1}{2}\ov{\chi_2}(i{\rlap{\,/}\partial}-M_{\chi_2})\chi_2 +\frac{1}{2}\ov{\chi_1}(i{\rlap{\,/}\partial}
-M_{\chi_1})\chi_1  
\nonumber  \\ &
-i\frac{g_D}{2}(\ov{\chi_2}{\rlap{\,/}X}\chi_1 -\ov{\chi_1}{\rlap{\,/}X}\chi_2) -\frac{f}{2}h_D (\ov{\chi_2}\chi_2 -\ov{\chi_1}\chi_1),
\end{align}
where $ \chi_1 $, $ \chi_2 $ masses and their mass splitting can be represented as
\begin{equation}
M_{\chi_{1,2}} = M_{\chi}\mp fv_D,
\end{equation}
and
\begin{equation}
\Delta_{\chi}\equiv (M_{\chi_2} - M_{\chi_1}) = 2fv_D.
\end{equation}

Finally, the $ Z^{\prime} $ boson mass and its interactions with SM fermions are the same with Eq.(\ref{eq:Zpmass}) and~(\ref{eq:Zpffbar}).

\subsection{The target parameter space and decay width of $ \phi_2 (\chi_2) $}

In this work, we focus on the scenario $ m_{Z^{\prime}} > M_{\phi_1} + M_{\phi_2} $ or $ m_{Z^{\prime}} > M_{\chi_1} + M_{\chi_2} $ such that the $ Z^{\prime}\rightarrow\phi_1\phi_2 (\chi_1\chi_2) $ decay mode is kinematically allowed and becomes the dominant one. 
On the other hand, we are interested in the co-annihilation dominant channel for DM relic density 
in early Universe, so we restrict ourselves to the compressed mass spectrum with 
$ \Delta_{\phi,\chi} < 0.5 M_{\phi_1,\chi_1} $ in inelastic DM models~\cite{Duerr:2019dmv}. 
Finally, we concentrate on $ M_{\phi_1,\chi_1} > 100 $ MeV such that the parameter space is free from the Big Bang Nucleosynthesis (BBN) constraint~\cite{Boehm:2012gr,Depta:2019lbe} 
\footnote{Notice that the DM mass can be extended to lower regions  if the light dark Higgs is included as shown in Ref.~\cite{Baek:2020owl}.}. 

The $ \phi_1 (\chi_1) $ is the only DM candidate in our analysis. If the mass splitting $ \Delta_{\phi,\chi} $ is not ignorable compared with the $ M_{\phi_1 (\chi_1)} $, $ \phi_2 (\chi_2) $ is not stable and will decay to $ \phi_1 (\chi_1) $ and a SM fermion pair via the off-shell $ Z^{\prime} $. The full analytical formulas for the total width of $ \phi_2 (\chi_2) $ three-body decay can be found in Appendix B of Ref.~\cite{Giudice:2017zke}. In the mass range of our interest, there are three kind of $ \phi_2 (\chi_2) $ decay modes \footnote{Since the light $ Z^{\prime} $ is isosinglet, it can mix with $ \omega $ meson 
and decay to three pions, $Z^{\prime}\rightarrow \pi^+ \pi^- \pi^0$, if kinematically allowed. However, according to Fig.1 in Ref.~\cite{Ilten:2018crw}, the contributions from $Z^{\prime}\rightarrow \pi^+ \pi^- \pi^0$ mode are important only in the adjacent regions of $ m_{Z^{\prime}}\approx m_{\omega} = 0.782 $ GeV. Hence, we don't specifically study this regions in our analysis.}, $ \phi_2 (\chi_2)\rightarrow\phi_1 (\chi_1) e^{+}e^{-}$, $\phi_1 (\chi_1)\mu^{+}\mu^{-}$ and $\phi_1 (\chi_1)\pi^{+}\pi^{-} $.  In our numerical calculations, the total decay width of $\phi_2 (\chi_2)$ is automatically calculated in \textbf{MadGraph5 aMC@NLO}~\cite{Alwall:2014hca}.  
For the partial decay width of $ \phi_2 (\chi_2)\rightarrow\phi_1 (\chi_1)\pi^{+}\pi^{-} $, we rescale the partial decay width of $ \phi_2 (\chi_2)\rightarrow\phi_1 (\chi_1)\mu^{+}\mu^{-} $ with the measured $ R(s) $ values in Ref.~\cite{Zyla:2020zbs}.

\section{The relevant cross sections}\label{Sec:Xsec}

In this section, we show both analytical representations and numerical results for cross sections of $ e^+ e^-\rightarrow \phi_1\phi_2 (\chi_1\chi_2) $ and $ e^+ e^-\rightarrow \phi_1\phi_2 (\chi_1\chi_2)\gamma $ processes. Especially, the method to distinguish spin-0 and spin-1/2 DM candidates in the inelastic DM models are also discussed here. 

\subsection{The analytical representations}

The differential cross section for $ e^+ e^-\rightarrow \phi_1\phi_2 $ via the s-channel $ Z^{\prime} $ 
can be represented as
\begin{equation}
\frac{d\sigma (e^+ e^-\rightarrow \phi_1\phi_2)}{dt} = \frac{\epsilon^2 e^2 g^2_D}{8\pi s^2} 
\frac{\left[ (M^2_{\phi_1}-t)(M^2_{\phi_2}-t)-st\right]}{\left[ (s-m^2_{Z^{\prime}})^2 +m^2_{Z^{\prime}}\Gamma^2_{Z^{\prime}}\right]},
\label{eq:22scalar0}
\end{equation}
where $s$, $t$ are Mandelstam variables and $ \Gamma_{Z^{\prime}} $ is the total width of $ Z^{\prime} $ boson.
In order to study the differential angular distributions of cross sections, we transfer from $ d\sigma /dt $ to $ d\sigma /d\cos\theta $, where $ \theta $ is the polar angle of $\phi_2 $ and it is defined as the direction of $ \phi_2 $ relative to the positron beam direction. 

The $ d\sigma /d\cos\theta $ in the centre-of-mass (CM) frame can be simply written as
\begin{equation}
\left.\frac{d\sigma (e^+ e^-\rightarrow \phi_1\phi_2)}{d\cos\theta}\right|_{CM}=\frac{1}{64\pi}\frac{\epsilon^2 e^2 g^2_D E^2_{CM}}{[(E^2_{CM}-m^2_{Z'})^2 +m^2_{Z'}\Gamma^2_{Z'}]}\xi^{3/2}(1-\cos^2\theta),
\label{eq:22scalar1}
\end{equation}
and
\begin{equation}
\xi = \sqrt{1-\frac{2(M^2_{\phi_2}+M^2_{\phi_1})}{E^2_{CM}}+\frac{(M^2_{\phi_2}-M^2_{\phi_1})^2}{E^4_{CM}}},
\label{eq:xi}
\end{equation}
where $ E_{CM} $ is the centre-of-mass energy. 
We can find Eq.(\ref{eq:22scalar1}) is always proportional to $ 1-\cos^2\theta $. Consequently, it indicates if the s-channel $Z'$ is on-shell produced, it is longitudinally polarized (helicity $=0$).

However, the formula for $ d\sigma /d\cos\theta $ in the Belle II laboratory (LAB) frame is more tedious, so we show it in Eq.(\ref{eq:22scalarLAB}) of the Appendix~\ref{Sec:LAB_rep}. 
As we will see the numerical results, the differential angular distributions of cross sections in the LAB frame are just shifted to the initial electron beam direction for the Belle II machine compared with the ones in the CM frame. 

The differential cross section for $ e^+ e^-\rightarrow \chi_1\chi_2 $ via s-channel $ Z^{\prime} $ can be represented as
\begin{equation}
\frac{d\sigma (e^+ e^-\rightarrow \chi_1\chi_2)}{dt} = \frac{\epsilon^2 e^2 g^2_D}{8\pi s^2} 
\frac{\left[ s^2 +2(t-M^2_{\chi_1})(t-M^2_{\chi_2})+s(2t-(M_{\chi_2}-M_{\chi_1})^2)\right]}{ \left[ (s-m^2_{Z^{\prime}})^2 +m^2_{Z^{\prime}}\Gamma^2_{Z^{\prime}}\right]},
\label{eq:22fermion0}
\end{equation}
and $ d\sigma /d\cos\theta $ in the CM frame can be written as
\begin{align}
\left.\frac{d\sigma (e^+ e^-\rightarrow \chi_1\chi_2)}{d\cos\theta}\right|_{CM}= & \frac{1}{32\pi}\frac{\epsilon^2 e^2 g^2_D E^2_{CM}}{[(E^2_{CM}-m^2_{Z'})^2 +m^2_{Z'}\Gamma^2_{Z'}]}\times\nonumber \\ & 
\left[(1-\frac{(M^2_{\chi_2}-M^2_{\chi_1})^2}{E^4_{CM}}+\frac{4M_{\chi_1}M_{\chi_2}}{E^2_{CM}})\xi +\xi^{3/2}\cos^2\theta\right],
\label{eq:22fermion1}
\end{align}
where $ \xi $ can be found in Eq.(\ref{eq:xi}) with $ M_{\phi_{1,2}}\rightarrow M_{\chi_{1,2}} $. 
For the $ M_{\chi_{1,2}}\rightarrow 0 $ limit, Eq.(\ref{eq:22fermion1}) is proportional to 
$ 1+\cos^2\theta $, which is well known. It shows if the s-channel $Z'$ is on-shell produced, it is transversely polarized (helicity $=\pm$). 
However, the mass effects of $ M_{\chi_{1,2}} $ spoil parts of this property.
Similarly, we show the formula for $ d\sigma /d\cos\theta $ in the LAB frame in Eq.(\ref{eq:22fermionLAB}) 
of the Appendix~\ref{Sec:LAB_rep}. Again, there is a skewing behavior for the differential angular distributions of cross sections in the LAB frame compared with the ones in the CM frame. 

The initial state radiation (ISR) photon is an useful trigger for signals with missing energy or 
soft objects at lepton colliders.  Especially, for the process with on-shell $Z^{\prime}$ production, 
the ISR photon is used not only for background rejection, but also for $Z^{\prime}$ invariant mass reconstruction.
To include the ISR photon in   
$ e^+ e^-\rightarrow \phi_1\phi_2 (\chi_1\chi_2) $ process, the differential cross section can be written as~\cite{Birkedal:2004xn}
\begin{equation}
\frac{d\sigma (e^+ e^-\rightarrow \phi_1\phi_2 (\chi_1\chi_2)\gamma)}{dzd\cos\theta^{\prime}}\simeq {\cal P}(z,\cos\theta^{\prime})\widehat{\sigma}(e^+ e^-\rightarrow \phi_1\phi_2 (\chi_1\chi_2)),
\label{eq:factorization}
\end{equation}
and the splitting kernal $ {\cal P}(z,\cos\theta^{\prime}) $ is
\begin{equation}
{\cal P}(z,\cos\theta^{\prime}) = \frac{\alpha}{\pi}\frac{1+(1-z)^2}{z}\frac{1}{\sin^2\theta^{\prime}},
\end{equation}
where $ \alpha $ is the fine structure constant, $ z = \frac{E_{\gamma}}{E_{CM}/2} $ is the energy 
fraction of ISR photon from the initial electron/positron and $ \theta^{\prime} $ is the polar angle of 
ISR photon and it is defined as the direction of $ \gamma $ to the positron beam direction. 
The differential form of $ \widehat{\sigma}(e^+ e^-\rightarrow \phi_1\phi_2 (\chi_1\chi_2)) $ can be found in Eq.(\ref{eq:22scalar0}) and~(\ref{eq:22fermion0}).
In the CM frame, if we assign the four momentum of the initial electron/positron and ISR photon as $ p_{e^{\pm}} = \left(E_{CM}/2,0,0,\pm E_{CM}/2 \right) $ and $ p_{ISR} = \left(E_{\gamma}, p_{x,\gamma}, p_{y,\gamma}, p_{z,\gamma}\right) $, the four momentum of electron/positron after the radiation is $ p_{e^{\prime\pm}} = \left(E_{CM}-E_{\gamma}, -p_{x,\gamma}, -p_{y,\gamma}, \pm E_{CM}/2-p_{z,\gamma}\right) $. Let's consider two limit scenarios to intuitively catch up the behavior of $ d\widehat{\sigma}(e^+ e^-\rightarrow \phi_1\phi_2 (\chi_1\chi_2))/d\cos\theta $ distributons.
First, if the ISR photon is soft ($ z\approx 0 $), $ p_{e^{\prime\pm}}\approx p_{e^{\pm}} $ and $ d\widehat{\sigma}(e^+ e^-\rightarrow \phi_1\phi_2 (\chi_1\chi_2))/d\cos\theta $ distributions are close to Eq.(\ref{eq:22scalar1}) and~(\ref{eq:22fermion1}). On the contrary, if the ISR photon takes almost all energy from the initial electron/positron ($ z\rightarrow 1 $), the $ \phi_2 (\chi_2) $ is obvious in the forward or backward direction. 
Since $ d\widehat{\sigma}(e^+ e^-\rightarrow \phi_1\phi_2 (\chi_1\chi_2))/d\cos\theta $ distributions 
are highly dependent on the $ z $ parameter of the ISR photon, the $ e^+ e^-\rightarrow \phi_1\phi_2 (\chi_1\chi_2)\gamma $ processes are not ideal to distinguish scalar and fermion inelastic DM models.

Finally, we focus on the parameter space $ m_{Z^{\prime}} > M_{\phi_2 (\chi_2)} + M_{\phi_1 (\chi_1)} $ in this work, so only the three-body decay of $ \phi_2 (\chi_2) $ via the off-shell $ Z^{\prime} $ is possible. The full analytical representations of $ e^+ e^-\rightarrow \phi_1\phi_2 (\chi_1\chi_2)\rightarrow \phi_1\phi_1 (\chi_1\chi_1) f\overline{f} $ are shown in the Appendix~\ref{Sec:2to4_rep}.

\subsection{The numerical results}

\begin{figure}
\centering
\includegraphics[width=3.0in]{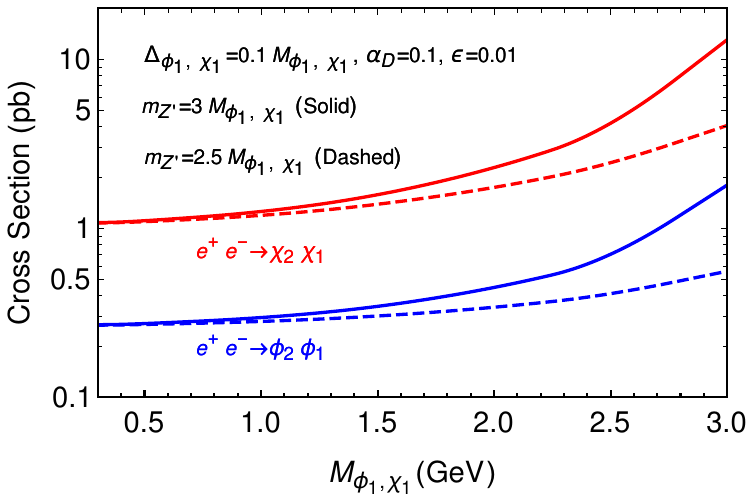}
\includegraphics[width=3.0in]{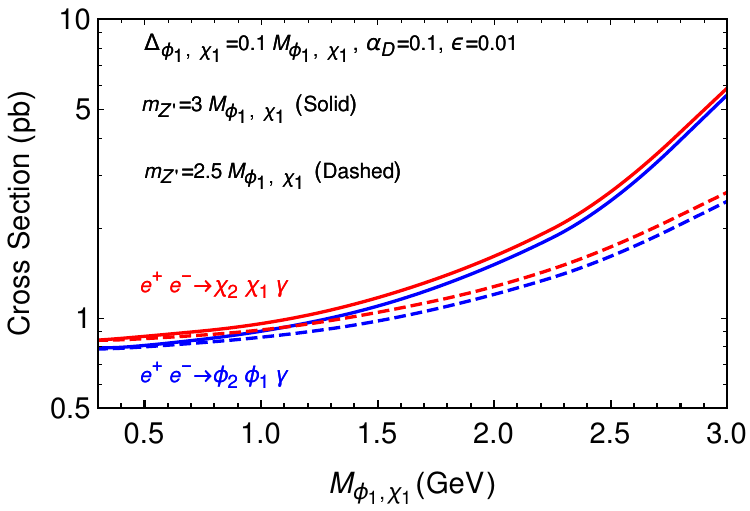}
\caption{
The relations between $ M_{\phi_1,\chi_1} $ (GeV) and $ \sigma $ (pb) for $ e^+ e^- \rightarrow \phi_2 \phi_1 $ and $ e^+ e^- \rightarrow \chi_2 \chi_1 $ processes (left panel) and $ e^+ e^- \rightarrow \phi_2 \phi_1\gamma $ and $ e^+ e^- \rightarrow \chi_2 \chi_1\gamma $ processes (right panel).
}\label{fig:cross_section}
\end{figure}

We first generate both scalar and fermion inelastic DM UFO model files from \textbf{FeynRules}~\cite{Alloul:2013bka}, and then calculate cross sections of $ e^+ e^- \rightarrow \phi_2\phi_1 (\chi_2\chi_1) $ and $ e^+ e^- \rightarrow \phi_2\phi_1 (\chi_2\chi_1)\gamma $ processes via \textbf{MadGraph5 aMC@NLO}\footnote{The $ Z^{\prime} $ boson total decay width is automatically calculated in MadGraph5 aMC@NLO.}. The beam energies are set to be $ E(e^+)=4.0$ GeV and $ E(e^-)=7.0 $ GeV which are consistent with the  Belle II experiment. In order to display the relations of cross sections with $ M_{\phi_1,\chi_1} $ and $ m_{Z'} $, we study $  m_{Z'}=2.5M_{\phi_1,\chi_1} $ and $ m_{Z'}=3M_{\phi_1,\chi_1} $ with 
fixed $ \alpha_D \equiv g^2_D/4\pi = 0.1 $, $ \epsilon = 0.01 $ and $ \Delta_{\phi ,\chi} = 0.1 M_{\phi_1,\chi_1} $. 
The relation between $M_{\phi_1,\chi_1}$ (GeV) and $\sigma$ (pb) for $ e^+ e^- \rightarrow \phi_2 \phi_1 (\chi_2 \chi_1) $ processes are shown in the left panel of Fig.~\ref{fig:cross_section}. 
Since all of these cross sections are proportional to $ \epsilon^2\alpha_D $, it is straightforward to rescale cross sections with different values of $ \epsilon $ and $ \alpha_D $. On the other hand, the influence from the changes of $ \Delta_{\phi,\chi} $ to cross sections are mild for $ \Delta_{\phi,\chi} < 0.5 M_{\phi_1,\chi_1} $.  
The scalar and fermion pair production cross sections can be scaled by $ \beta^{3/2} $ and 
$ \beta^{1/2} $ respectively, where $ \beta $ is the velocity of the final state particle in the CM frame. 
Because of this extra $ \beta $ factor for the scalar case, cross sections for 
$ e^+ e^- \rightarrow \phi_2 \phi_1 $ are suppressed compared with 
$ e^+ e^- \rightarrow \chi_2 \chi_1 $.

On the other hand, the relation between $ M_{\phi_1,\chi_1} $ (GeV) and $ \sigma $ (pb) for 
$ e^+ e^- \rightarrow \phi_2 \phi_1 (\chi_2 \chi_1)\gamma $ processes are shown in the right panel 
of Fig.~\ref{fig:cross_section} with the same parameter settings.
Here, the basic cuts $ E(\gamma) > 0.1 $ GeV and $ \eta < 2.203 $ are applied for the ISR photon. 
We find the main contributions of $ e^+ e^-\rightarrow\phi_2\phi_1 (\chi_2\chi_1)\gamma $ come 
from $ e^+ e^-\rightarrow Z^{\prime}\gamma\rightarrow\phi_2\phi_1 (\chi_2 \chi_1)\gamma $ 
processes. It explains why scalar and fermion pair cross sections in this process are very close 
to each other in the right panel of Fig.~\ref{fig:cross_section}.

\begin{figure}
\centering
\includegraphics[width=6.0in]{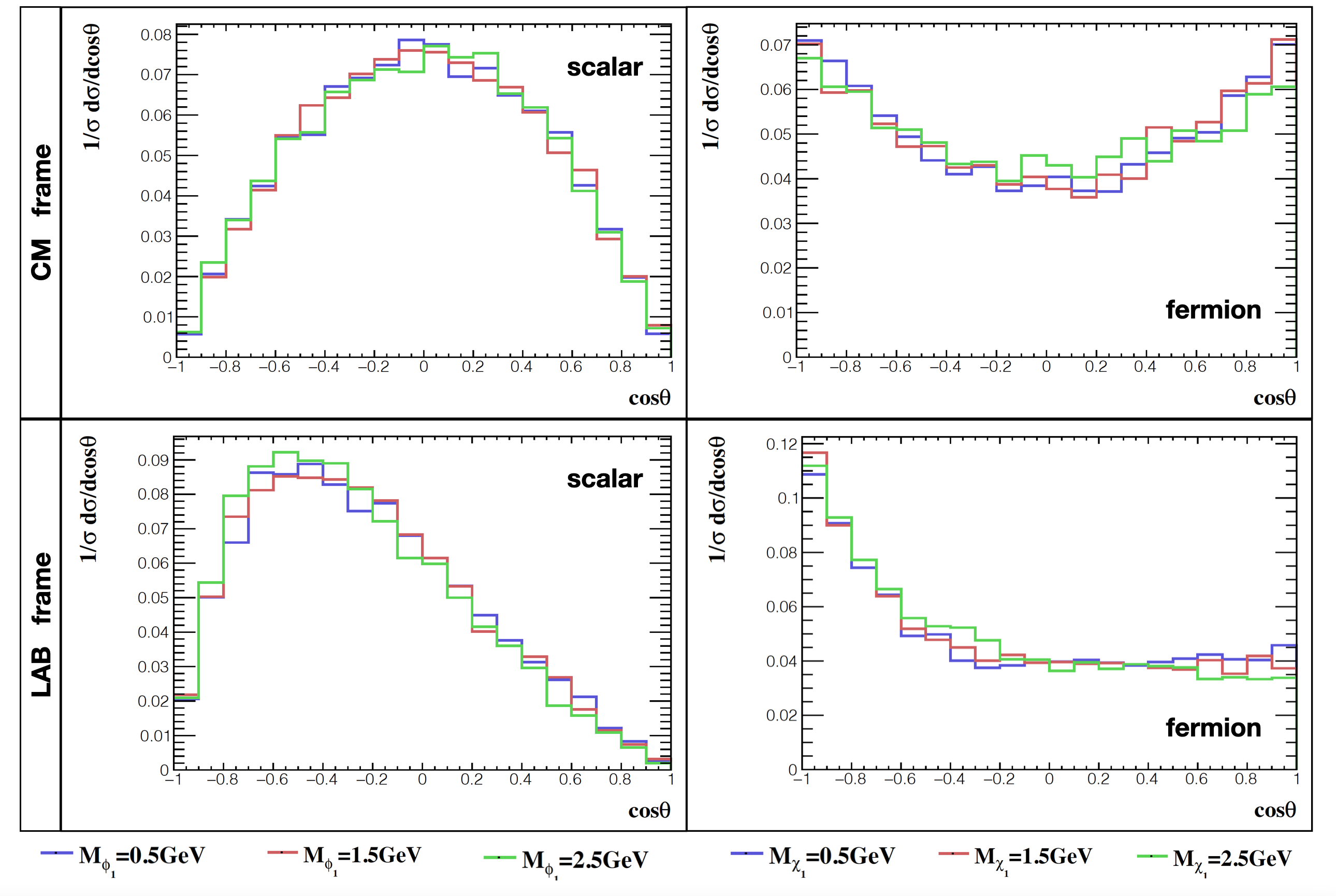}
\caption{
The $ (1/\sigma)(d\sigma /d\cos\theta) $ distributions for $ e^+ e^- \rightarrow \phi_2 \phi_1 $ in CM frame (top left panel) and Belle II LAB frame (bottom left panel) and $ e^+ e^- \rightarrow \chi_2 \chi_1 $ in CM frame (top right panel) and Belle II LAB frame (bottom right panel). 
}\label{fig:theta}
\end{figure}

We then turn to the study of $ (1/\sigma)(d\sigma /d\cos\theta) $ distributions for $ e^+ e^- \rightarrow \phi_2 \phi_1 $ and $ e^+ e^- \rightarrow \chi_2 \chi_1 $ processes with fixed $ \alpha_D = 0.1 $, $ \epsilon = 0.01 $, $ \Delta_{\phi ,\chi} = 0.1 M_{\phi_1,\chi_1} $, and $ m_{Z'}=3M_{\phi_1,\chi_1} $. The results in the CM and Belle II LAB frame are shown in Fig.~\ref{fig:theta}. 
It is clear that the $ (1/\sigma)(d\sigma /d\cos\theta) $ distributions for $ e^+ e^- \rightarrow \phi_2 \phi_1 $ in the CM frame is proportional to $ 1-\cos^2\theta $ and $ e^+ e^- \rightarrow \chi_2 \chi_1 $ behaves close to $ 1+\cos^2\theta $ with some distortions from $ M_{\chi_{1,2}} $ effects.
However, since the measurement of time-of-flight for the long-lived particle is poor at the Belle II machine, we will lose this information and cannot make the Lorentz transformation for the four-vector 
of displaced vertex from the Belle II LAB frame to the CM frame. Hence, the distributions in the CM 
frame  is only a reference for the comparison with the ones in the Belle II LAB frame. Hopefully, the real 
situation for the distributions in the Belle II LAB frame is just shifted to the electron beam direction 
for the Belle II machine compared with the ones in the CM frame. It is still clear to see the different 
distributions between scalar and fermion inelastic DM models.

\begin{figure}
\centering
\includegraphics[width=6.0in]{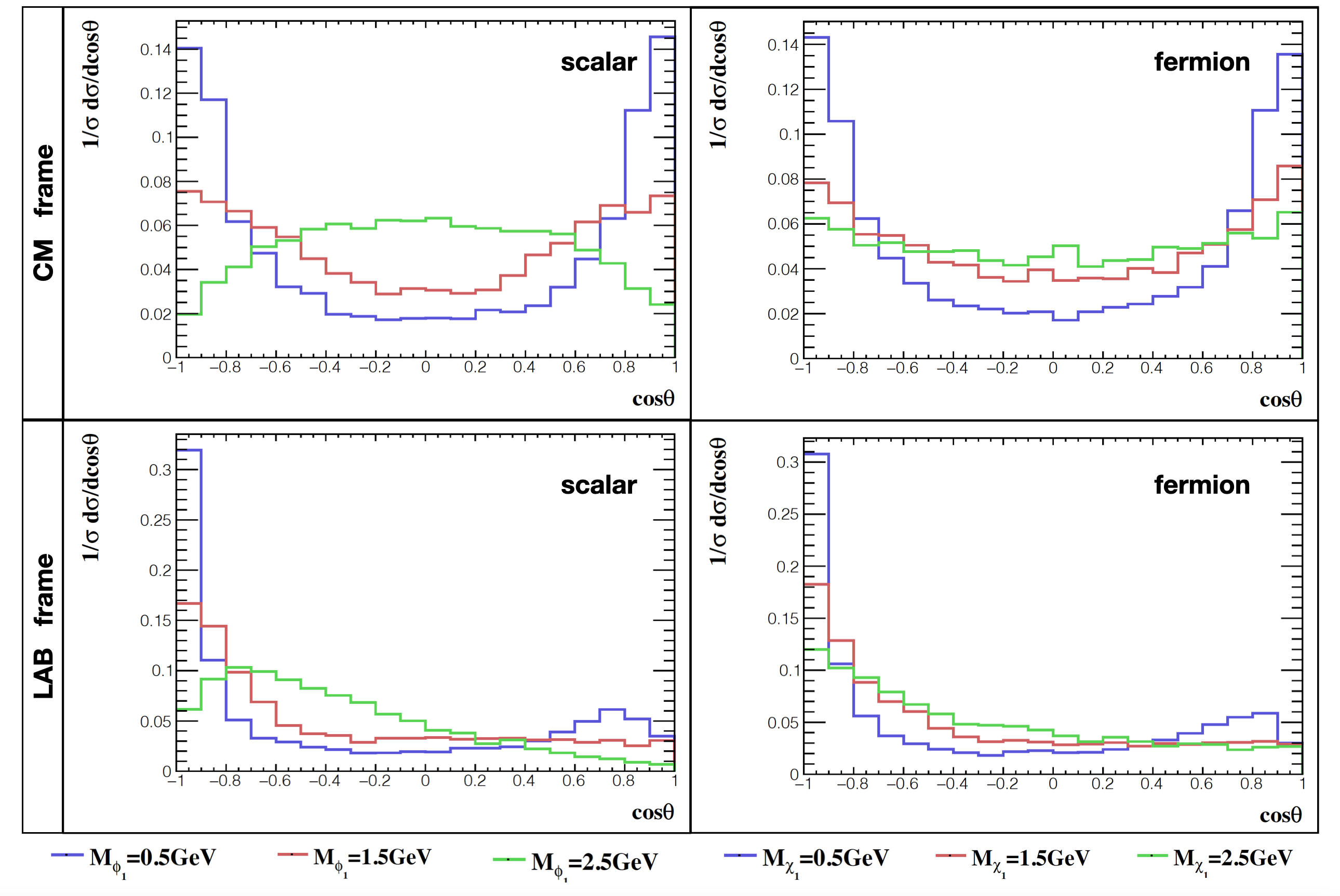}
\caption{
The $ (1/\sigma)(d\sigma /d\cos\theta) $ distributions for $ e^+ e^-\rightarrow\phi_2\phi_1\gamma $ in CM frame (top left panel) and Belle II LAB frame (bottom left panel) and $ e^+ e^-\rightarrow\chi_2 \chi_1\gamma $ in CM frame (top right panel) and Belle II LAB frame (bottom right panel). 
}\label{fig:theta_a}
\end{figure}

Finally, we fix the same parameter settings as Fig.~\ref{fig:theta} for the $ (1/\sigma)(d\sigma /d\cos\theta) $ distributions of $ e^+ e^-\rightarrow\phi_2\phi_1 (\chi_2\chi_1)\gamma $ processes in the CM and LAB frames in Fig.~\ref{fig:theta_a}. 
Again, the distributions in the CM frame is only for the comparison and it is clear to see the skewing behavior for the differential angular distributions of cross sections in the LAB frame compared with the ones in the CM frame. 
As discussed in the previous subsection, because these distributions are highly dependent on the $ z $ parameter of the ISR photon, the $ e^+ e^-\rightarrow \phi_1\phi_2 (\chi_1\chi_2)\gamma $ processes are not ideal for the DM spin discrimination. 
Only when the $ Z^{\prime} $ is on-shell produced and heavy, the ISR photon becomes soft 
($z$ is small) such that differences of $ (1/\sigma)(d\sigma /d\cos\theta) $ distributions from scalar 
and fermion inelastic DM models show up.   

In summary, the size for cross sections of $ e^+ e^- \rightarrow \phi_2\phi_1 (\chi_2\chi_1) $ and their $ (1/\sigma)(d\sigma /d\cos\theta) $ distributions in the Belle II LAB frame can help us to statistically distinguish fermion inelastic DM model from the scalar one even for the same parameter settings.

\section{Search for long-lived particles in inelastic dark matter models at Belle II}\label{Sec:Simulation}

In this section, we study how to search for long-lived particles in inelastic DM models at Belle II experiment. We first briefly overview the Belle II experiment and signal signatures, and then 
show some interesting kinematic distributions based on four signal benchmark points. 
We further set up event selections for dilepton displaced signature and relevant results are discussed. 
Finally, we solve kinematic equations of 
$ e^+ e^-\rightarrow\chi_1\chi_2\rightarrow\chi_1\chi_1 l^{+}l^{-} $ and 
$ e^+ e^-\rightarrow Z^{\prime}\gamma\rightarrow\chi_1\chi_2\gamma\rightarrow
\chi_1\chi_1 l^{+}l^{-}\gamma $ processes to determine both $M_{\chi_1}$ and $M_{\chi_2}$. 
 
\subsection{The Belle II experiment and signal signatures}\label{Sec:detector}
The SuperKEKB accelerator of Belle II experiment is a circular asymmetric $ e^{+}e^{-} $ collider 
with the nominal collision energy of $ \sqrt{s}=10.58 $ GeV. The beam parameters are $E(e^{+})=4$ 
GeV and $E(e^{-})=7$ GeV.
The planned full integrated luminosity for the final dataset is $ 50~{\rm ab}^{-1} $. 
In this work, the following Belle II sub-detectors are relevant : the tracking system including vertex detectors (VXD) and central drift chamber (CDC), the electromagnetic calorimeter (ECAL), and muon system.

We consider the following two kinds of processes in inelastic DM models\footnote{Because the major contribuitons of $ e^+ e^-\rightarrow\phi_1\phi_2(\chi_1\chi_2)\gamma $ come from the on-shell $ Z^{\prime} $ production, hence, we only focus on the process $ e^+ e^-\rightarrow Z^{\prime}\gamma\rightarrow\phi_1\phi_2(\chi_1\chi_2)\gamma $ in the following analysis unless noted otherwise.}:
\begin{align}
& e^+ e^-\rightarrow\phi_1\phi_2 (\chi_1\chi_2)\rightarrow\phi_1\phi_1 (\chi_1\chi_1) l^{+}l^{-} \ ,\label{eq:woISR} \\ &
e^+ e^-\rightarrow Z^{\prime}\gamma\rightarrow\phi_1\phi_2 (\chi_1\chi_2) \gamma\rightarrow\phi_1\phi_1 (\chi_1\chi_1) l^{+}l^{-}\gamma \ .
\label{eq:wISR}
\end{align}
And there are five possible signatures can be searched for at the Belle II experiment :
\begin{itemize}
\item Mono-$\gamma$ : If the excited DM state decays outside the detector or the decay products are too soft to be detected in Eq.(\ref{eq:wISR}).  
\item Mono-$\gamma$ with prompt lepton pair: The excited DM state promptly decays and the visible products can be successfully detected and defined in Eq.(\ref{eq:wISR}).
\item Mono-$\gamma$ with displaced lepton pair : The excited DM state is long-lived and decays inside the detector leaving the displaced vertex in Eq.(\ref{eq:wISR}).
\item Only prompt lepton pair : The same as the second one but without ISR photon in Eq.(\ref{eq:woISR}).
\item Only displaced lepton pair : The same as the third one but without ISR photon in Eq.(\ref{eq:woISR}).  
\end{itemize}
The Mono-$\gamma$ signature is well-studied as shown in Ref.~\cite{Zhang:2019wnz,Lees:2017lec,Duerr:2019dmv}.
Since we focus on searching for long-lived particles in this paper, only signatures with displaced vertices are studied. 
On the other hand, the signature with displaced charged pion pairs is also possible from the long-lived excited DM state decay. However, for simplicity, here we only study signatures with displaced electron and muon pairs. 

In order to make our results more realistic, we follow Ref.~\cite{Adachi:2018qme} to involve the detector resolution effects with Gaussian smearing at Belle II experiment. 
The tracking resolution of electron/muon momentum in the CDC is given by
\begin{equation}
\sigma_{p^l_t}/p^l_t = 0.0011 p^l_t~[\rm{GeV}]\oplus 0.0025/\beta,
\end{equation}
where $ p^l_t $ is the transverse momentum of electron/muon track and $ \beta $ is its velocity in the natural unit. We conservatively apply $ \sigma_{p_l}/p_l = 0.005 $ in our event analysis. On the other hand, the muon efficiency in the muon system is approximated to $ 0.98 $.
The photon momentum resolution in the ECAL is approximated to $ \sigma_{E_{\gamma}}/E_{\gamma} = 2\% $ where $ E_{\gamma} $ is the energy of photon. 
Finally, we use the resolution of $ \sigma_{r_{DV}} = 26 \mu$m for the displaced vertex vector of lepton pair in our analysis. 

\subsection{Benchmark points and kinematic distributions}

In this study, we use the following four benchmark points (BPs) to display our analysis :
\begin{itemize}
\item (I) $ M_{\phi_1,\chi_1} = 0.3 $ GeV, $ \Delta_{\phi_1,\chi_1} = 0.4 M_{\phi_1,\chi_1} $, $ m_{Z'} = 3 M_{\phi_1,\chi_1} $ and $ \epsilon = 2\times 10^{-2} $
\item (II) $ M_{\phi_1,\chi_1} = 3.0 $ GeV, $ \Delta_{\phi_1,\chi_1} = 0.1 M_{\phi_1,\chi_1} $, $ m_{Z'} = 3 M_{\phi_1,\chi_1} $ and $ \epsilon = 2\times 10^{-3} $
\item (III) $ M_{\phi_1,\chi_1} = 1.0 $ GeV, $ \Delta_{\phi_1,\chi_1} = 0.4 M_{\phi_1,\chi_1} $, $ m_{Z'} = 2.5 M_{\phi_1,\chi_1} $ and $ \epsilon = 10^{-3} $
\item (IV) $ M_{\phi_1,\chi_1} = 2.0 $ GeV, $ \Delta_{\phi_1,\chi_1} = 0.2 M_{\phi_1,\chi_1} $, $ m_{Z'} = 2.5 M_{\phi_1,\chi_1} $ and $ \epsilon = 10^{-3} $
\end{itemize}
with fixed $ \alpha_D = 0.1 $\footnote{Notice $ Z'\rightarrow\phi_1\phi_2 $ or $ \chi_1\chi_2 $ is the dominant decay channel for these four BPs since $ Z' $ can only couple to SM fermion pairs via the kinetic mixing. We have checked the branching ratio for $ Z' $ to SM particles is less than $10^{-3}$ for these four BPs. Therefore, we can safely ignore relevant constraints from visible dark photon searches.}. The first BP is inspired from the allowed parameter space to explain the muon anomalous magnetic moment excess~\cite{Mohlabeng:2019vrz}. The other three BPs are the ones which cannot be covered from Mono-$\gamma$ searches from BaBar and Belle II~\cite{Duerr:2019dmv}. In addition, they can also be searched for in the future long-lived particle experiments, like CODEX-b~\cite{Gligorov:2017nwh}, SeaQuest~\cite{Berlin:2018pwi}, FASER~\cite{Feng:2017uoz}, MATHUSLA~\cite{Chou:2016lxi}, SHiP~\cite{Alekhin:2015byh}, ANUBIS~\cite{Bauer:2019vqk} and AL3X~\cite{Gligorov:2018vkc}.

\begin{figure}
\includegraphics[width=3.0in]{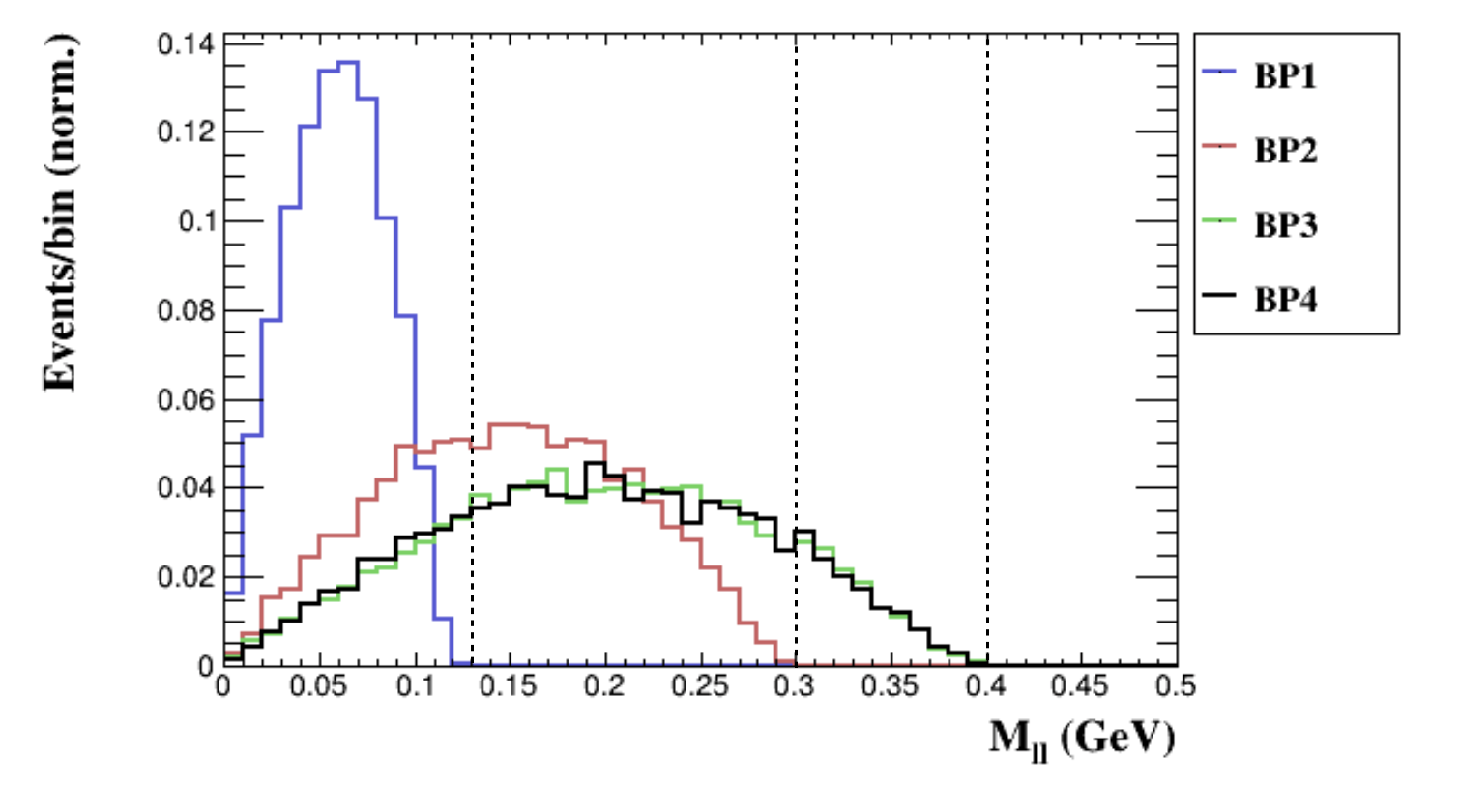}
\includegraphics[width=3.0in]{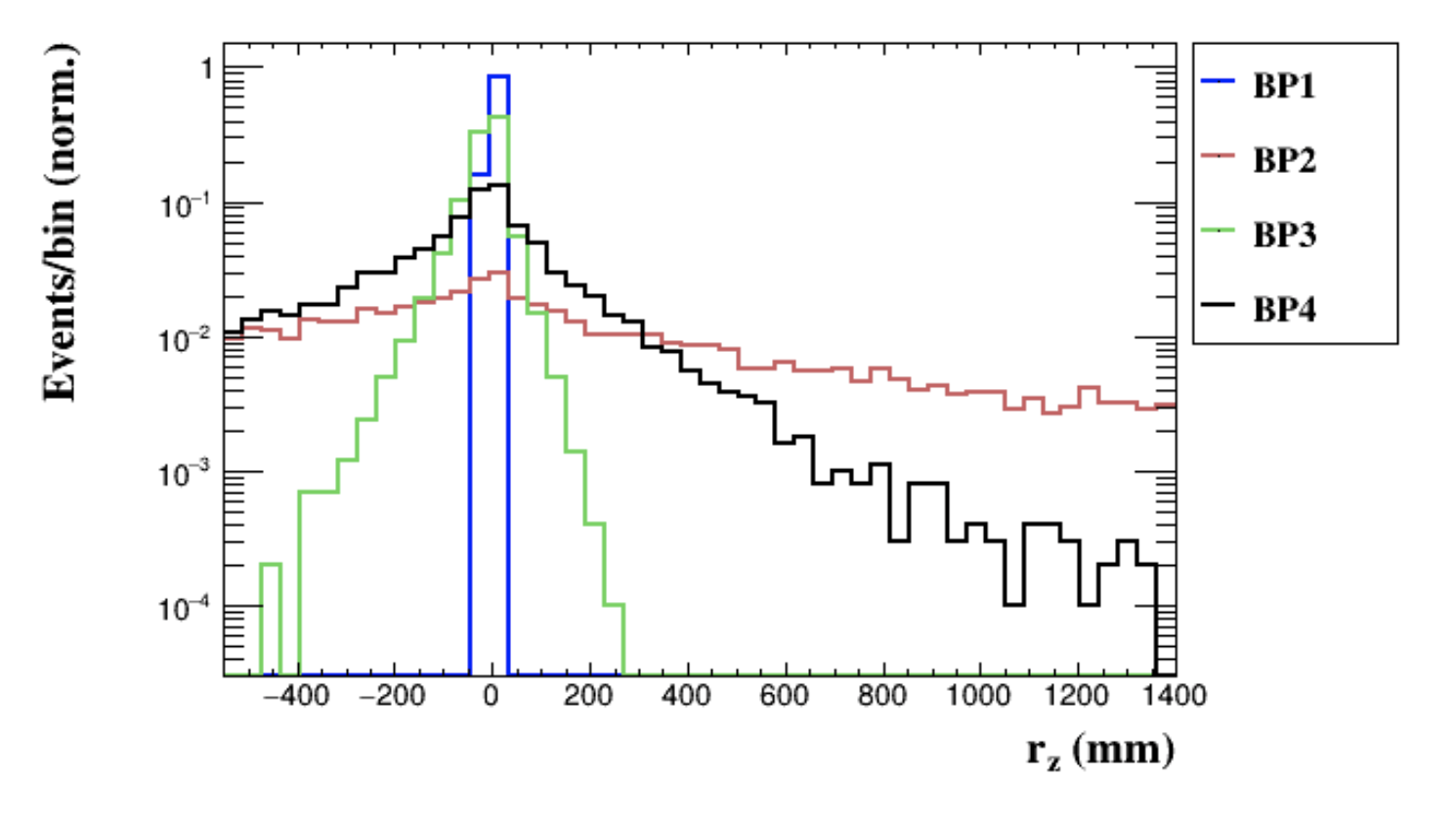}
\includegraphics[width=3.0in]{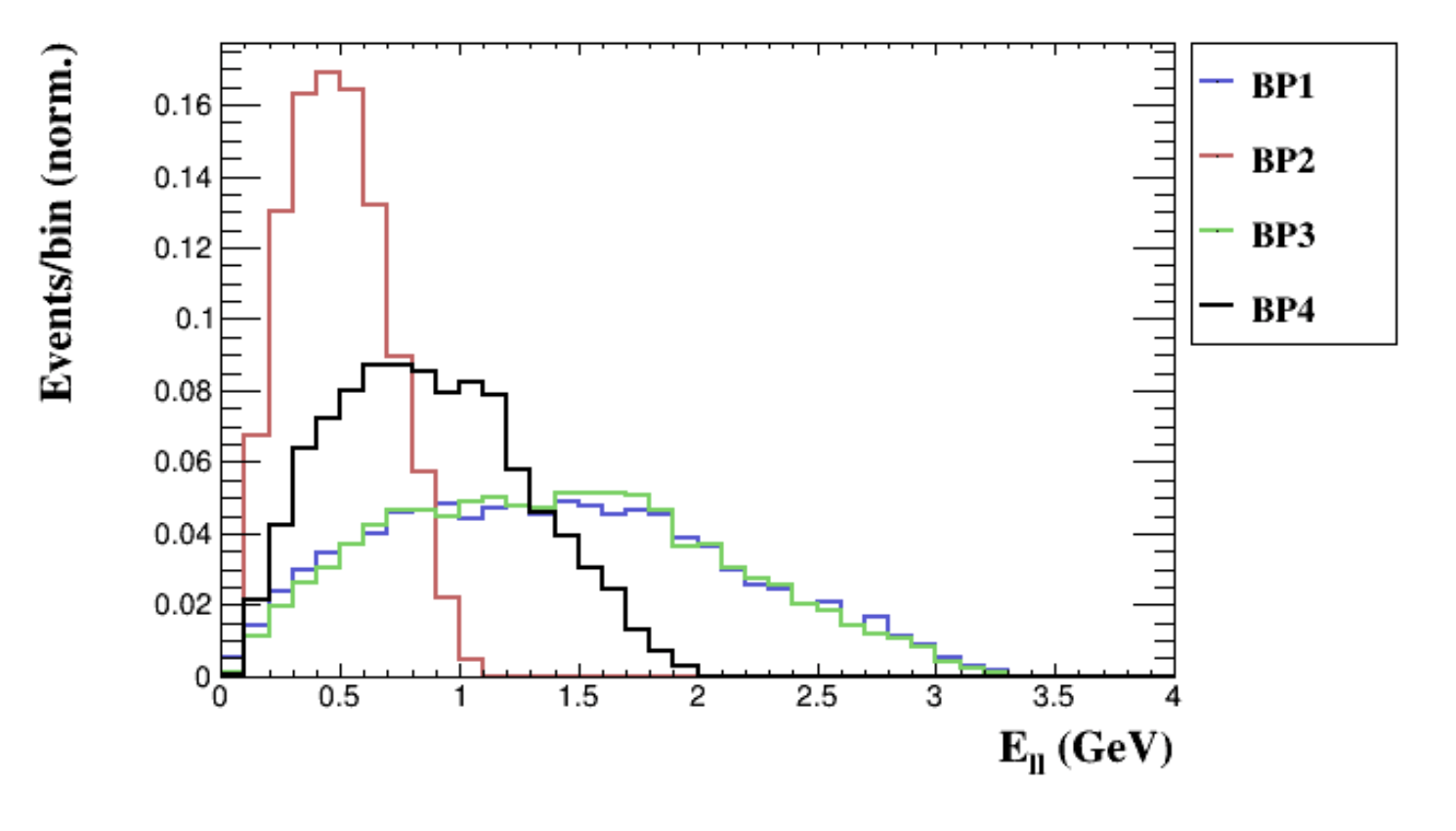}
\includegraphics[width=3.0in]{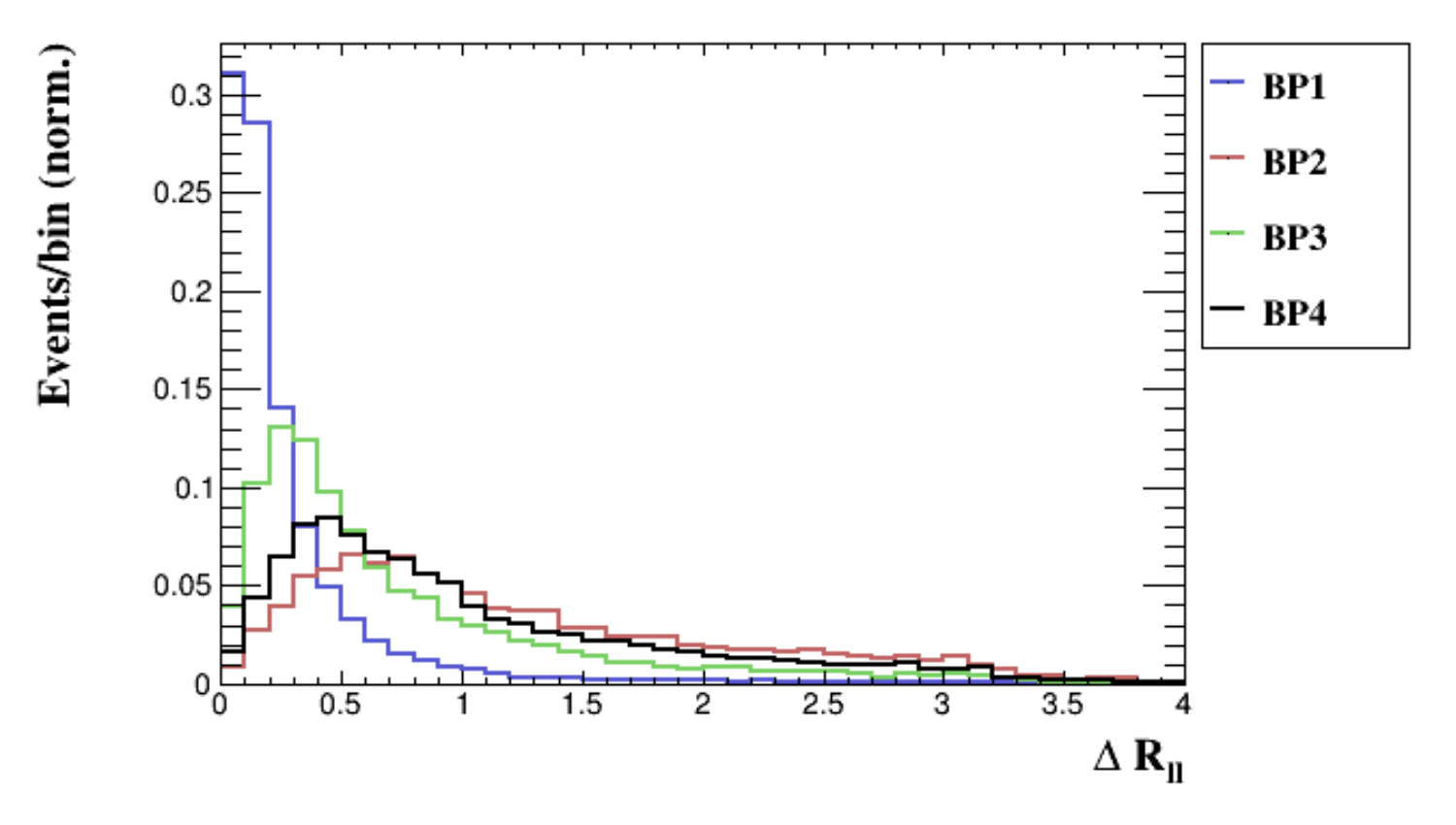}
\caption{
Various kinematic distributions for $ e^+ e^-\rightarrow\chi_1\chi_2\rightarrow\chi_1\chi_1 l^{+}l^{-} $ based on the four BPs in the main text. Top left panel: lepton pair invariant mass, $ M_{ll} $ (GeV), top 
right panel: projection of the decay vertex distance on $z$ axis, $ r_z $ (mm), bottom left panel: lepton pair energy, $ E_{ll} $ (GeV), bottom right panel: angular distance between lepton pair, $ \Delta R_{ll} $.  
}\label{fig:kinematic1}
\end{figure}

\begin{figure}
\includegraphics[width=3.0in]{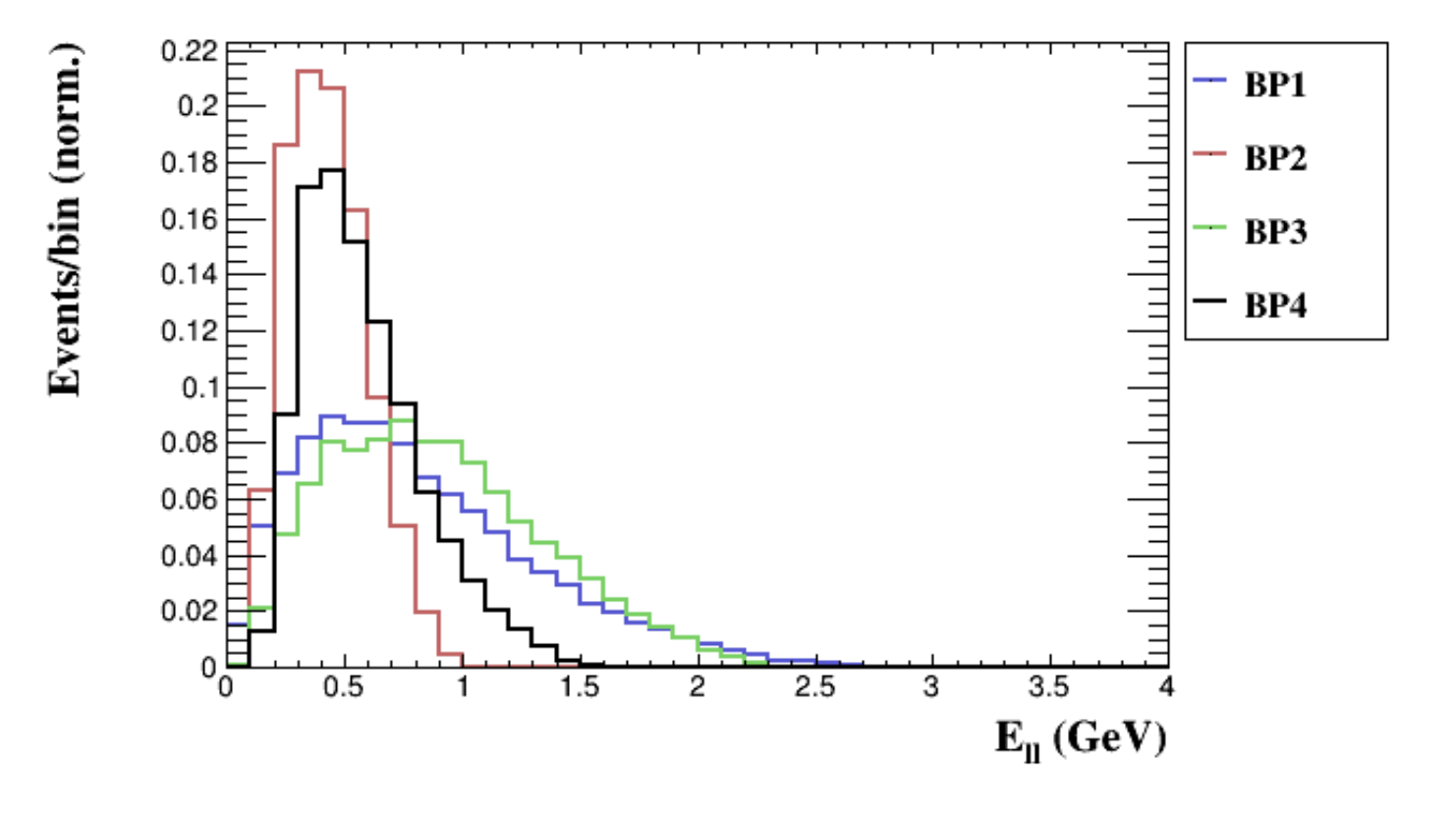}
\includegraphics[width=3.0in]{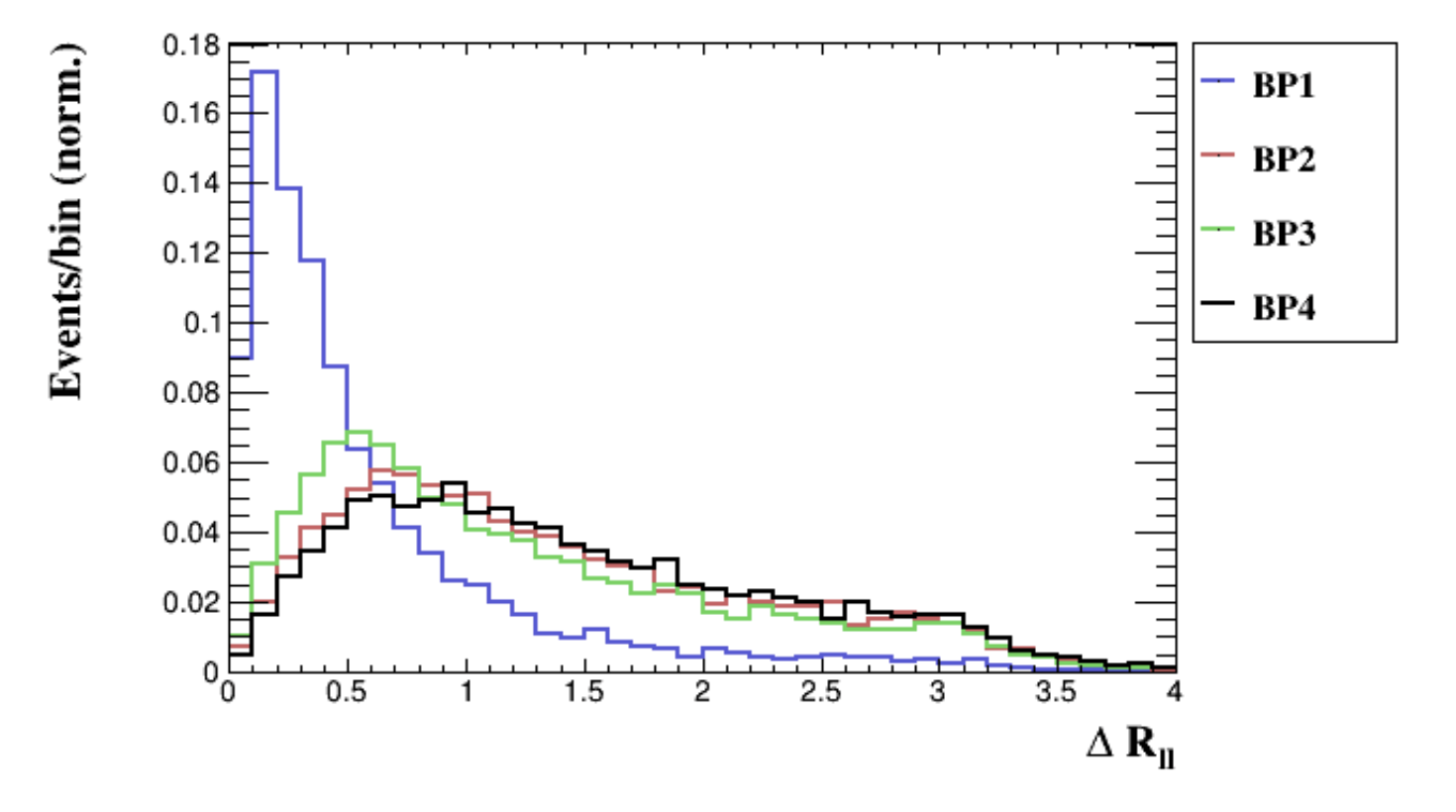}
\includegraphics[width=3.0in]{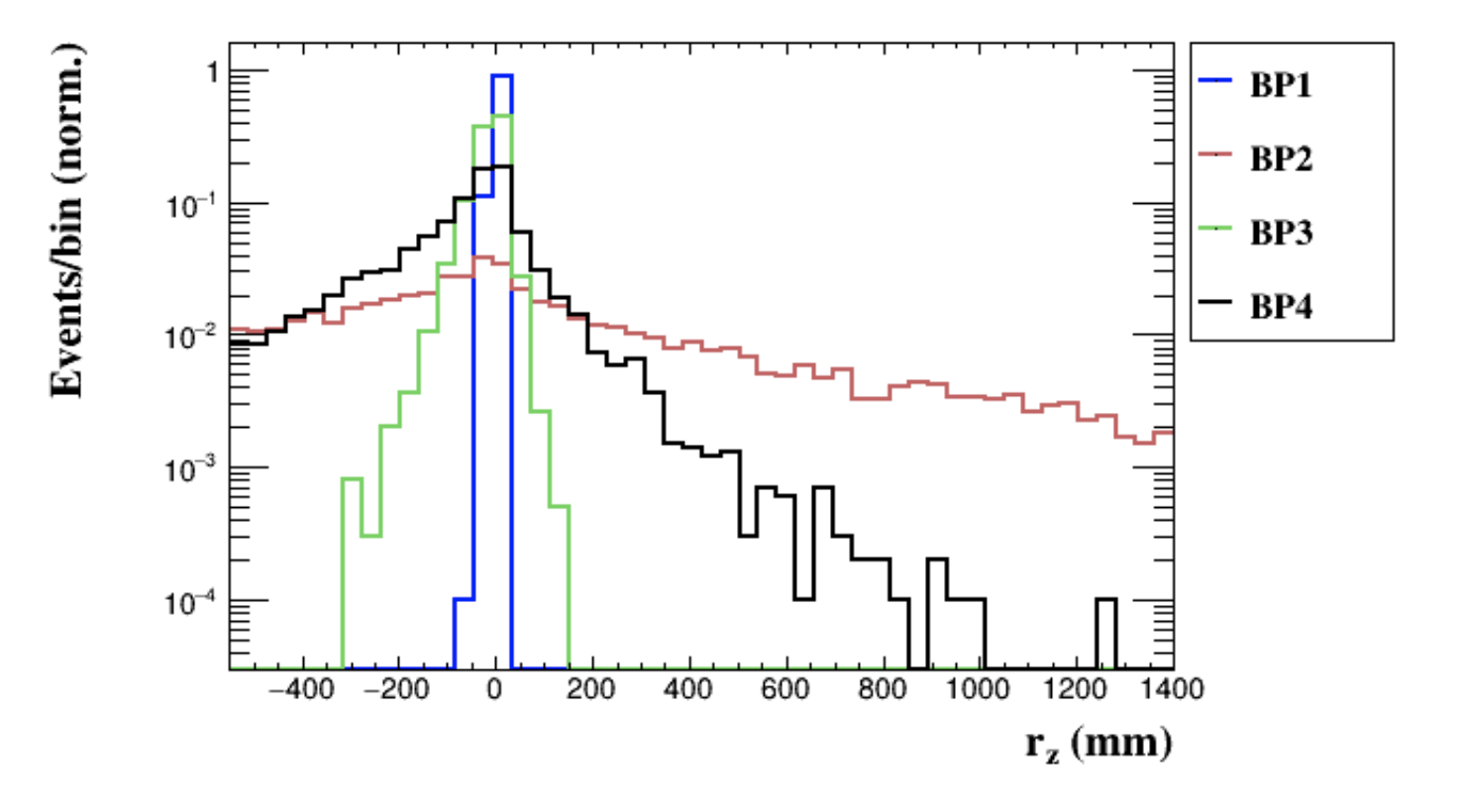}
\includegraphics[width=3.0in]{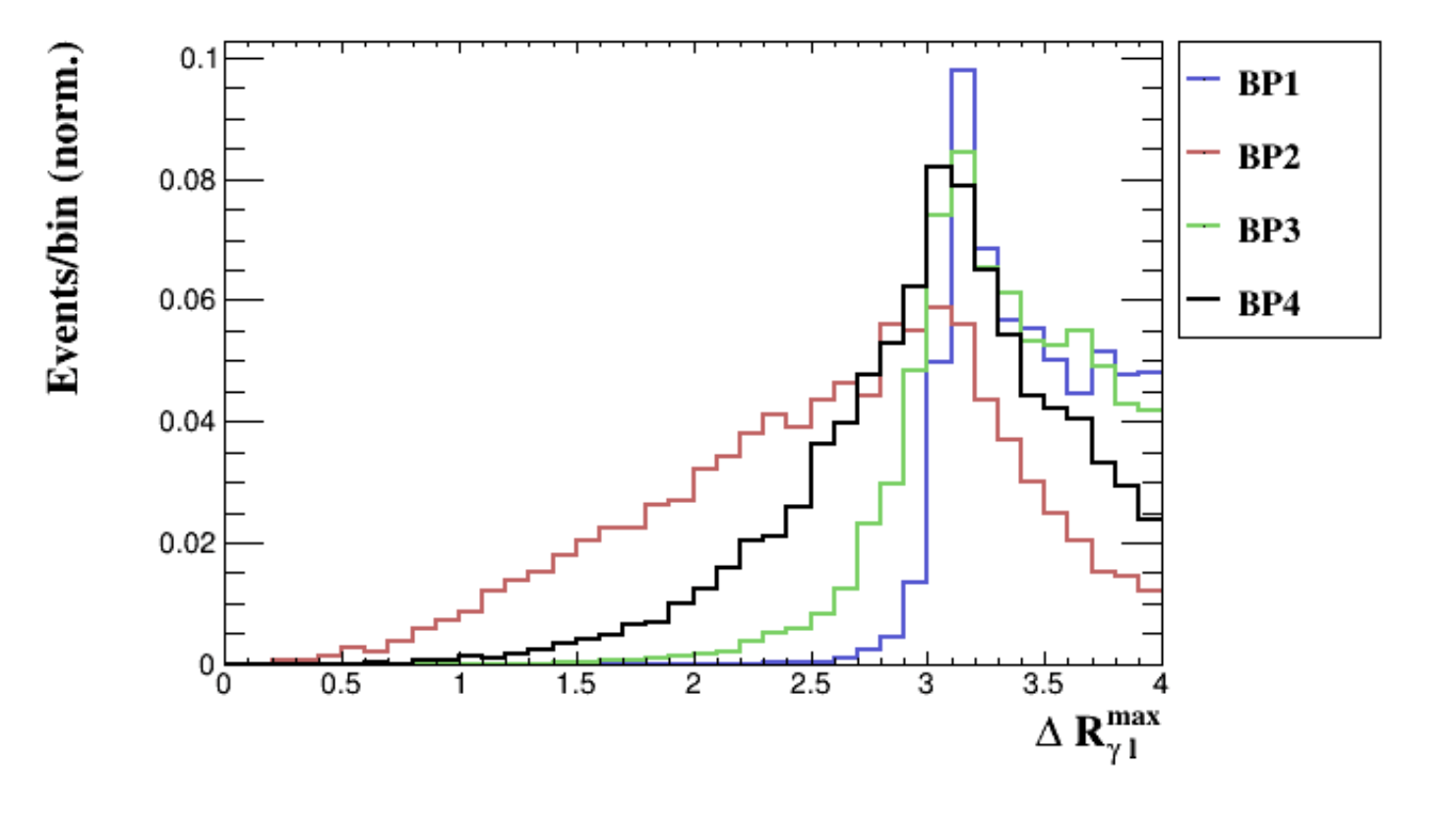}
\includegraphics[width=3.0in]{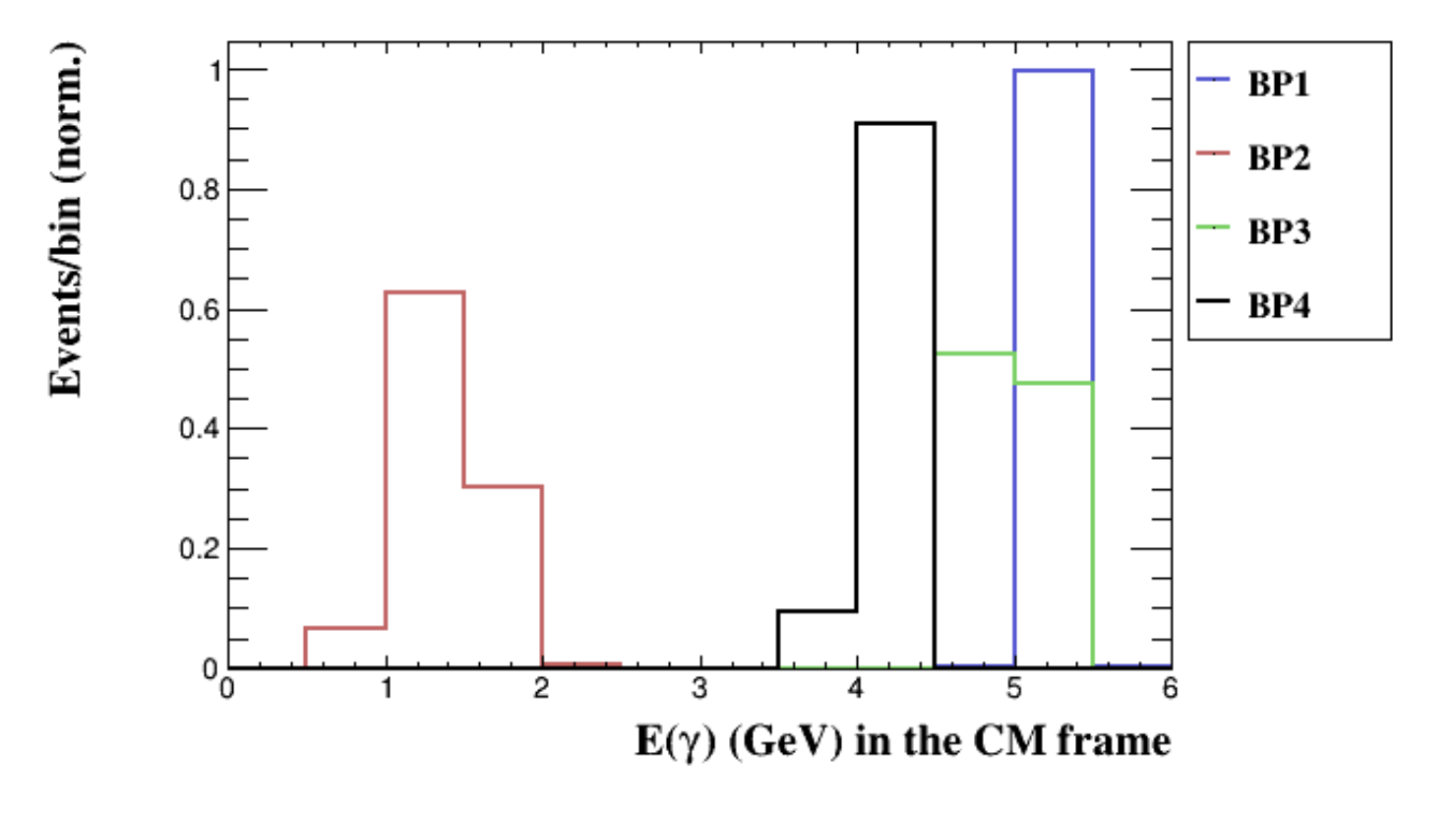}
\includegraphics[width=3.0in]{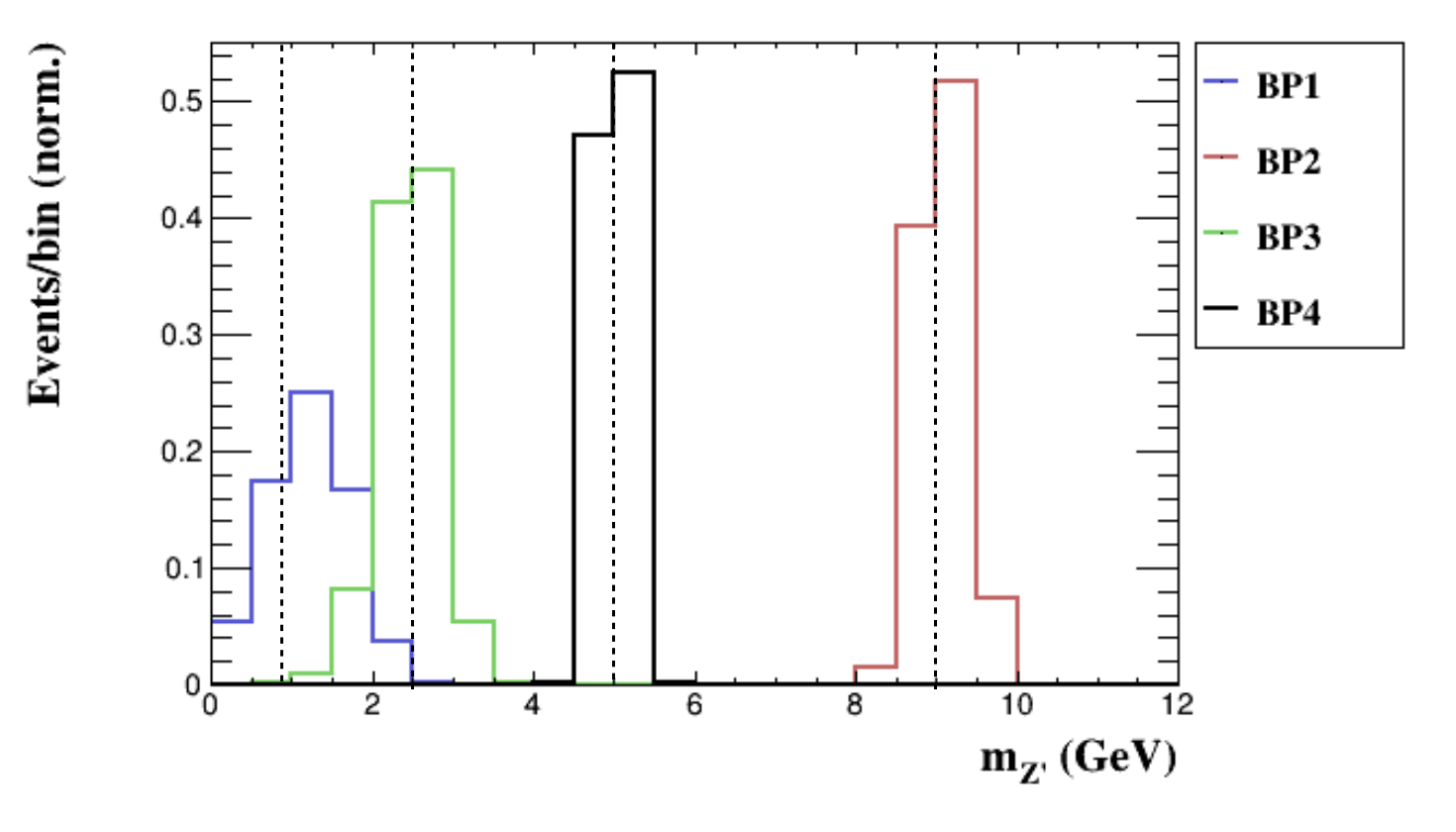}
\caption{
Various kinematic distributions for $ e^+ e^-\rightarrow Z^{\prime}\gamma\rightarrow\chi_1\chi_2\gamma\rightarrow\chi_1\chi_1 l^{+}l^{-}\gamma $ based on the four BPs in the main text. Top left panel: lepton pair energy, $ E_{ll} $ (GeV), top right panel: angular distance between lepton pair, $ \Delta R_{ll} $, middle left panel: projection of the decay vertex distance on $z$ axis, $ r_z $ (mm), middle right panel: the maximum angular distance between ISR photon and one of the lepton in lepton pair, $ \Delta R^{max}_{\gamma l} $, bottom left panel: photon energy in the CM frame,  $ E(\gamma) $ (GeV), bottom right panel : reconstructed $ Z^{\prime} $ invariant mass, $ m_{Z^{\prime}} $ (GeV). 
}\label{fig:kinematic2}
\end{figure}

We show some interesting kinematic distributions based on the above four BPs for $ e^+ e^-\rightarrow\chi_1\chi_2\rightarrow\chi_1\chi_1 l^{+}l^{-} $  and $ e^+ e^-\rightarrow Z^{\prime}\gamma\rightarrow\chi_1\chi_2\gamma\rightarrow\chi_1\chi_1 l^{+}l^{-}\gamma $ in 
Fig.~\ref{fig:kinematic1} and Fig.~\ref{fig:kinematic2}, respectively. 
First, the lepton pair invariant mass is smaller or equal to the mass splitting $ \Delta_{\chi} $ (dashed lines), $ 2 m_l \leq M_{ll}\leq\Delta_{\chi} $. Once we have enough signal events, the threshold value of $ M_{ll} $ can help us to roughly determine the mass splitting $ \Delta_{\chi} $ in inelastic DM models.
Second, the distance of displaced vertex of $ \chi_2 $ is not only dependent on $ M_{\chi_2} $, 
$ \Delta_{\chi} $, $ \epsilon $ and $ \alpha_D $, but also the boost of $ \chi_2 $ in the LAB frame. 
Here we show projection of the decay vertex distance on $z$ axis, $ r_z $. It is clear to see that 
the $ \chi_2 $ decay length in BP1 (BP2) is the shortest (longest) one on average.  
The distributions for projection of the decay vertex distance on the transverse plane, $ R_{xy} $, are similar. 
On the other hand, $ E_{ll} $ and $ \Delta R_{ll} $ are energy and angular distance for the lepton pair. 
We can find the boost of $ \chi_2 $ is smaller for the process involving ISR photon by comparing $ r_z $, $ E_{ll} $ and $ \Delta R_{ll} $ distributions in Fig.~\ref{fig:kinematic1} and Fig.~\ref{fig:kinematic2}. 
The larger boost on $ \chi_2 $ causes the longer $ r_z $, larger $ E_{ll} $, and smaller $ \Delta R_{ll} $ distributions. 
Besides, we will see larger $ E_{ll} $, and smaller $ r_z $ distributions of BP1 suffer from more severe smearing effects from detector resolutions such that it is more challenge to reconstruct DM mass and mass splitting between DM excited and ground states in the real experiment.   
Third, if the $ Z^{\prime} $ is light enough, the ISR photon and $ Z^{\prime} $ are almost back-to-back produced such that the lepton from $ Z^{\prime} $ decay can have large angular distance with the ISR photon. Here we show the maximum angular distance between ISR photon and one of the lepton in lepton pair, $ \Delta R^{max}_{\gamma l} $. 
Fourth, we boost the $ E(\gamma) $ distribution of ISR photon from the Belle II LAB frame to the CM frame and it is clear to see its mono-energetic behavior. This feature can largely reduce SM 
backgrounds which are continuous in the photon energy spectrum. 
Finally, the $ Z^{\prime} $ invariant mass can be reconstructed from Eq.(\ref{eq:Zp4momenta}) with measurable observables. 
Because of the detector resolution, it is clear to find smearing effects from the true $ Z^{\prime} $ 
mass (dashed lines), especially for the case of light $ Z^{\prime} $. We will see this phenomenon 
critically affects abilities to determine DM mass and the mass splitting between DM excited and 
ground states for the process in Eq.(\ref{eq:wISR}). 
Notice those kinematic distributions for the scalar inelastic DM model are similar to Fig.~\ref{fig:kinematic1} and Fig.~\ref{fig:kinematic2}, so we do not show them again.

\subsection{Event selections and results}
\label{Sec:selections}

We closely follow Ref.~\cite{Duerr:2019dmv} to set up event selections for the displaced signature at Belle II. We only conservatively consider the following two background-free regions after event selections in our analysis :
\begin{itemize}
\item low $ R_{xy} $ region ($100\%$ detection efficiency) : $0.2 < R_{xy}\leq 0.9 $ cm (electron), $0.2 < R_{xy}\leq 17.0 $ cm (muon). 
\item high $ R_{xy} $ region ($30\%$ detection efficiency) : $17.0 < R_{xy}\leq 60.0 $ cm (electron, muon).
\end{itemize}
According to the arguments in Ref.~\cite{Duerr:2019dmv}, backgrounds from photon conversion can be further reduced to a negligible level by requiring the lepton pair invariant mass $ M_{ll}\geq 0.03 $ GeV, and their opening angle larger than $ 0.1 $ rad. Notice backgrounds from the photon conversion to muon pair in the region $0.9 < R_{xy}\leq 17.0 $ cm are negligible such that the analysis for muon in low $ R_{xy} $ region can be extended.

\begin{table}[bt]
\centering
\caption{Event selections for the displaced vertex analysis at Belle II.} 
\label{Tab:selection}
 \begin{tabular}{| l | l |}
  \hline
  Objects & Selections \\
  \hline \hline
   \multirow{2}{8em}{displaced vertex} & (i) $\unit[-55]{cm} \leq z \leq \unit[140]{cm}$\\ 
                                       & (ii) $ 17^\circ\leq \theta^{\text{DV}}_\text{LAB} \leq 150^\circ$ \\
   \hline
   \multirow{3}{8em}{electrons} &  (i) both $E(e^+)$ and $E(e^-) > \unit[0.1]{GeV}$\\
             &  (ii) opening angle of pair $ \theta_{ee} > 0.1$ \,rad\\
             &  (iii) invariant mass of pair $m_{ee}> \unit[0.03]{GeV}$\\
             \hline
   \multirow{4}{8em}{muons}     &  (i) both $p_\text{T}(\mu^+)$ and $p_\text{T}(\mu^-) > \unit[0.05]{GeV}$\\ 
             &  (ii) opening angle of pair $ \theta_{\mu\mu} > 0.1$ \,rad\\
             &  (iii) invariant mass of pair $m_{\mu\mu} > \unit[0.03]{GeV}$\\
             &  (iv) veto $ \unit[0.48]{GeV}\leq m_{\mu\mu}\leq \unit[0.52]{GeV}$ \\
             \hline
   \multirow{2}{8em}{photons} & (i) $E^{\gamma}_\text{LAB} > \unit[0.5]{GeV}$ \\ 
                              & (ii) $ 17^\circ\leq \theta^{\gamma}_\text{LAB} \leq 150^\circ$  \\
   \hline
 \end{tabular}
\end{table}

We summarize all event selections in our analysis in Table~\ref{Tab:selection}. For the muon pair in the final state, we veto the invariant mass region, $ 0.48\leq m_{\mu^{+}\mu^{-}}\leq 0.52 $ GeV, to reject backgrounds from $ K^0_S $ decay. More discussions on the trigger issue can be found in Ref.~\cite{Duerr:2019dmv}. Here, event selections in Table~\ref{Tab:selection} are done offline and we simply assume these events are already triggered and stored. 

\begin{table}[bt]
\centering
\caption{The analysis results of four BPs for the process in Eq.(\ref{eq:woISR}). The first column is the type of inelastic DM models. The second column is labeled by BP which $e$ and $\mu$ denote the type of lepton pair in the final state. The third column is production cross sections for process in Eq.(\ref{eq:woISR}). The fourth and fifth columns are efficiencies of low $ R_{xy} $ and high $ R_{xy} $ regions defined in the main text, respectively. Final column is the number of signal events with an integrated luminosity of $ 1 ab^{-1} $.} 
\label{Tab:Eff1}
 \begin{tabular}{|l|l|l|l|l|l|}
  \hline
Type & BP & $ \sigma $ (fb) & Eff.(low $ R_{xy} $) & Eff.(high $ R_{xy} $) & $ N_{event} $ \\
  \hline \hline
   \multirow{7}{7em}{scalar} & BP1$e$ & $948.14$ & $16.98$   & $0\%$    & $1.61\times 10^5$ \\ 
                             & BP2$e$ & $58.39$  & $0.15\%$  & $2.48\%$ & $1.54\times 10^3$ \\
                             & BP2$\mu$ & $6.15$ & $0.21\%$ & $ 3.33\%$ & $217.71$ \\                             
                             & BP3$e$ & $1.86$   & $10.06\%$ & $0.70\%$ & $200.09$ \\
                             & BP3$\mu$ & $0.61$ & $11.25\%$ & $0.74\%$ & $73.14$ \\                             
                             & BP4$e$ & $2.23$   & $1.56\%$  & $9.34\%$ & $243.26$ \\   
                             & BP4$\mu$ & $0.74$ & $1.72\%$ & $10.78\%$ & $92.50$ \\                                                                            
  \hline \hline
   \multirow{7}{7em}{fermion} & BP1$e$ & $3856.00$ & $14.26\%$ & $0\%$   & $5.50\times 10^5$ \\ 
                              & BP2$e$ & $422.80$  & $0.17\%$  & $2.35\%$ & $1.07\times 10^4$ \\
                              & BP2$\mu$ & $44.63$ & $0.22\%$ & $2.97\%$ & $1.42\times 10^3$ \\
                              & BP3$e$ & $7.99$    & $10.20\%$ & $0.42\%$ & $848.54$ \\
                              & BP3$\mu$ & $2.69$ & $11.20\%$ & $ 0.46\% $ & $313.65$ \\
                              & BP4$e$ & $11.71$   & $1.57\%$  & $7.82\%$ & $1.10\times 10^3$ \\                                                      
                              & BP4$\mu$ & $3.88$ & $1.69\%$ & $8.75\%$ & $405.07$ \\                                                      
   \hline
 \end{tabular}
\end{table}

\begin{table}[bt]
\centering
\caption{The same as Table~\ref{Tab:Eff1}, but for process in Eq.(\ref{eq:wISR}).} 
\label{Tab:Eff2}
 \begin{tabular}{|l|l|l|l|l|l|}
  \hline
Type & BP & $ \sigma $ (fb) & Eff.(low $ R_{xy} $) & Eff.(high $ R_{xy} $) & $ N_{event} $ \\
  \hline \hline
   \multirow{7}{7em}{scalar} & BP1$e$ & $2472.70$ & $6.70\%$ & $0\%$    & $1.66\times 10^5$ \\ 
                             & BP2$e$ & $159.85$  & $0.16\%$ & $2.27\%$ & $3.88\times 10^3$ \\
                             & BP2$\mu$ & $16.85$ & $0.20\%$ & $2.87\%$ & $517.30$ \\
                             & BP3$e$ & $5.13$    & $7.64\%$ & $0.02\%$ & $392.96$ \\
                             & BP3$\mu$ & $1.69$ & $8.83\%$ & $0.03\%$ & $149.73$ \\
                             & BP4$e$ & $7.14$    & $1.86\%$ & $3.29\%$ & $367.71$ \\      
                             & BP4$\mu$ & $2.35$ & $2.02\%$ & $2.87\%$ & $114.92$ \\                          
  \hline \hline
   \multirow{7}{7em}{fermion} & BP1$e$ & $2503.60$ & $6.14\%$  & $0\%$   & $1.54\times 10^5$ \\ 
                              & BP2$e$ & $167.10$  & $0.16\%$  & $2.16\%$ & $3.87\times 10^3$ \\
                              & BP2$\mu$ & $17.66$ & $0.18\%$ & $2.67\%$ & $503.31$ \\
                              & BP3$e$ & $5.05$    & $7.77\%$  & $0.02\%$ & $393.40$ \\
                              & BP3$\mu$ & $1.70$ & $8.89\%$ & $0.02\%$ & $151.47$ \\
                              & BP4$e$ & $7.14$    & $1.95\%$  & $3.14\%$ & $363.43$ \\                                                      
                              & BP4$\mu$ & $2.37$ & $2.05\%$ & $3.44\%$ & $130.11$ \\
   \hline
 \end{tabular}
\end{table}

The results for four BPs are shown in Table~\ref{Tab:Eff1} and~\ref{Tab:Eff2} for processes in Eq.(\ref{eq:woISR}) and~(\ref{eq:wISR}), respectively. 
Eff.(low $ R_{xy} $) and Eff.(high $ R_{xy} $) are efficiencies of low $ R_{xy} $ and high $ R_{xy} $ regions after involving the event selections. Here we use an integrated luminosity of $ 1~{\rm ab}^{-1} $ to calculate number of signal events ($N_{event}$). 
We can find most events in BP1 and BP3 (BP2 and BP4) are located in low (high) $ R_{xy} $ regions. The discrimination of scalar and fermion inelastic DM models with $N_{event}$ in Table~\ref{Tab:Eff1} and~\ref{Tab:Eff2} is shown in Appendix~\ref{Sec:pol_ang}.

\begin{figure}
\centering
\includegraphics[width=3.0in]{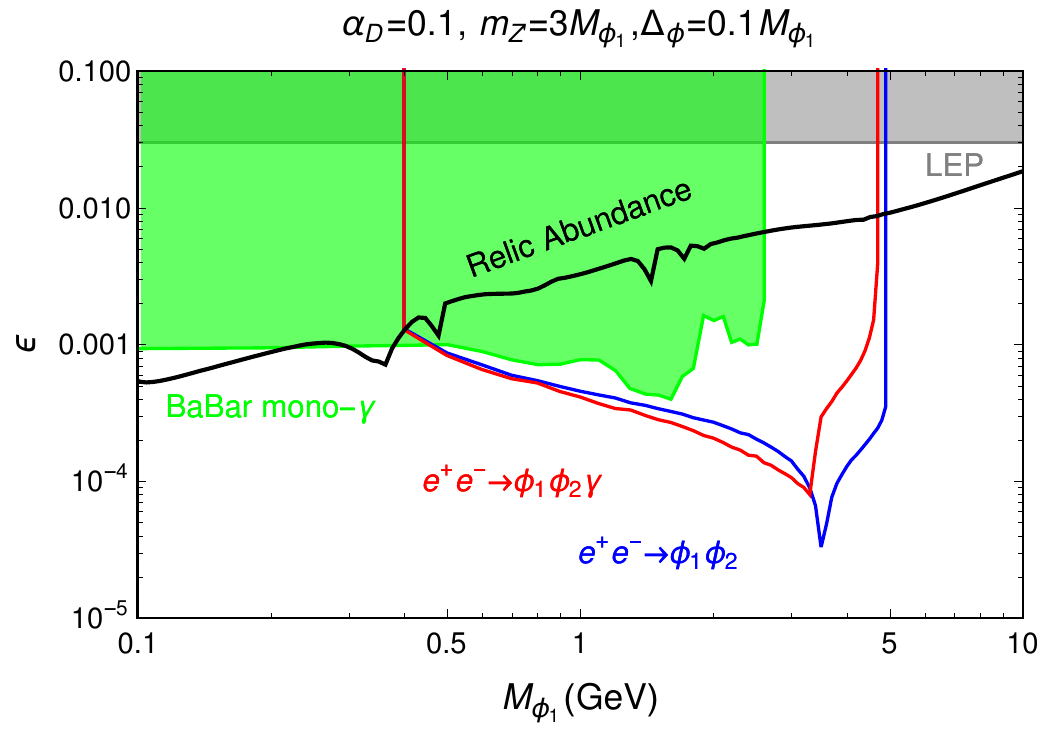}
\includegraphics[width=3.0in]{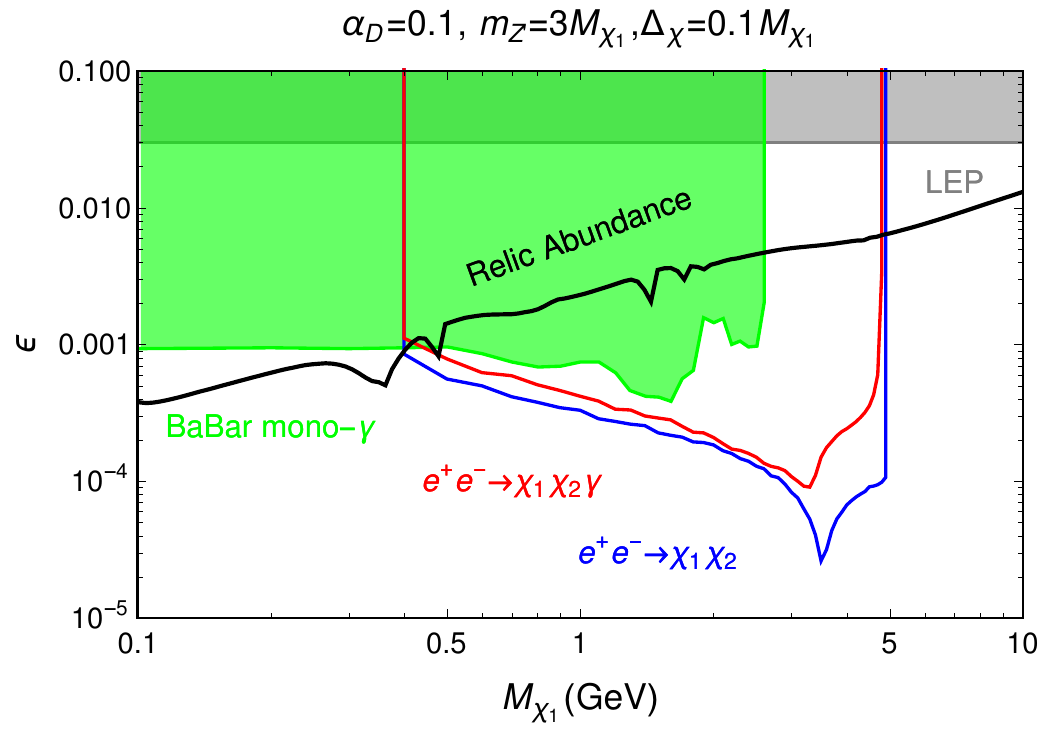}
\caption{
The future bounds from $ e^{+}e^{-}\rightarrow\phi_1\phi_2 (\chi_1\chi_2) $ and $ e^{+}e^{-}\rightarrow\phi_1\phi_2 (\chi_1\chi_2)\gamma $ processes for event selections in Table~\ref{Tab:selection} with the integrated luminosity of $ 50~{\rm ab}^{-1} $. Here parameters $ \alpha_D = 0.1 $, $ m_{Z^{\prime}} = 3 M_{\phi_1,\chi_1} $ and $ \Delta_{\phi,\chi} = 0.1 M_{\phi_1,\chi_1} $ are fixed and $ 90\% $ C.L. contours which correspond to an upper limit of 2.3 events with the assumption of background-free are applied. The model-independent LEP bound~\cite{Hook:2010tw}, BaBar mono-$\gamma$ 
bound~\cite{Lees:2017lec} and correct relic abundance lines
are also shown. 
}\label{fig:summary1}
\end{figure}

Finally, we use the most optimistic value of  
$ 50~{\rm ab}^{-1} $ for event selections in Table~\ref{Tab:selection} at Belle II for scalar and fermion inelastic DM models and predict the future bounds from $ e^{+}e^{-}\rightarrow\phi_1\phi_2 (\chi_1\chi_2) $ and $ e^{+}e^{-}\rightarrow\phi_1\phi_2 (\chi_1\chi_2)\gamma $ processes in Fig.~\ref{fig:summary1}.
Notice the off-shell $Z^{\prime}$ production in the second process is also considered.
Here we fix the parameters, $ \alpha_D = 0.1 $, $ m_{Z^{\prime}} = 3 M_{\phi_1,\chi_1} $ and $ \Delta_{\phi,\chi} = 0.1 M_{\phi_1,\chi_1} $, and apply $ 90\% $ C.L. contours which correspond to an upper limit of 2.3 events with the assumption of background-free. Notice constraints from 
model-independent LEP bound~\cite{Hook:2010tw} and BaBar mono-$\gamma$ bound~\cite{Lees:2017lec} are added for the comparison. We closely follow Ref.~\cite{Duerr:2019dmv} for the recasting of BaBar mono-$\gamma$ constraints in inelastic DM models. 
The correct relic abundance lines are also shown in Fig.~\ref{fig:summary1}. Here the dominant contribution comes from $\phi_1\phi_2 (\chi_1\chi_2)$ coannihilation 
process~\cite{Izaguirre:2015zva}. In our scenario, $\phi_1\phi_1 (\chi_1\chi_1)$ annihilation processes with dark Higgs boson or SM-like Higgs boson are either ignorable or kinematically forbidden.
For $M_{\phi_1,\chi_1} < 0.4 $ GeV, even dilepton displaced vertex is located in our target regions, the lepton pair in the final state is too soft and collinear to pass the event selections in Table~\ref{Tab:selection}. On the contrary, for $M_{\phi_1,\chi_1} \gtrsim 4.9 $ GeV, the lepton pair in the final state becomes more energetic and well-separated, but dilepton displaced vertex is too short to reach our target regions and production cross sections are also highly suppressed. 
We find our results in the second process for fermion inelastic DM model are consistent with the ones in Ref.~\cite{Duerr:2019dmv} and we further include detector resolution effects in our analysis. Besides, we believe our results for the first process in both scalar and fermion inelastic DM models are first shown in the literature. Most importantly, the future bounds from the first process can be even stronger than the usual ones in the second process.

\begin{figure}
\centering
\includegraphics[width=3.0in]{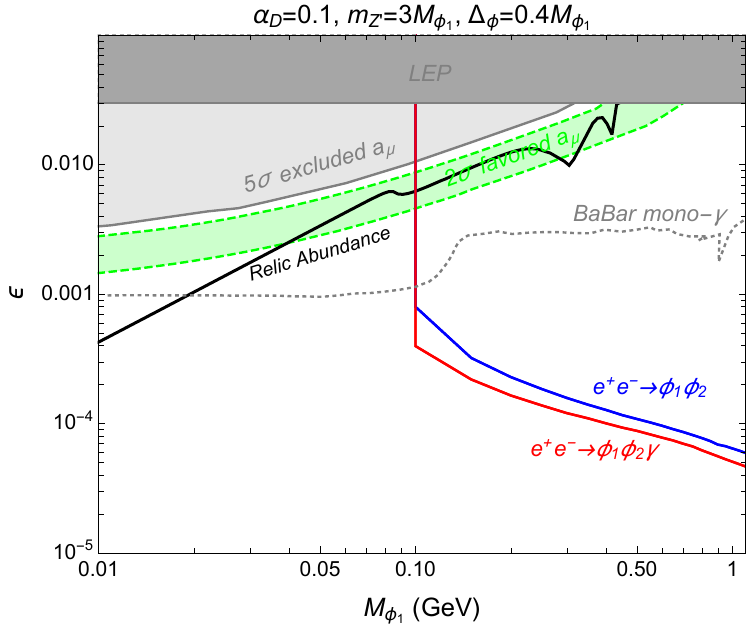}
\includegraphics[width=3.0in]{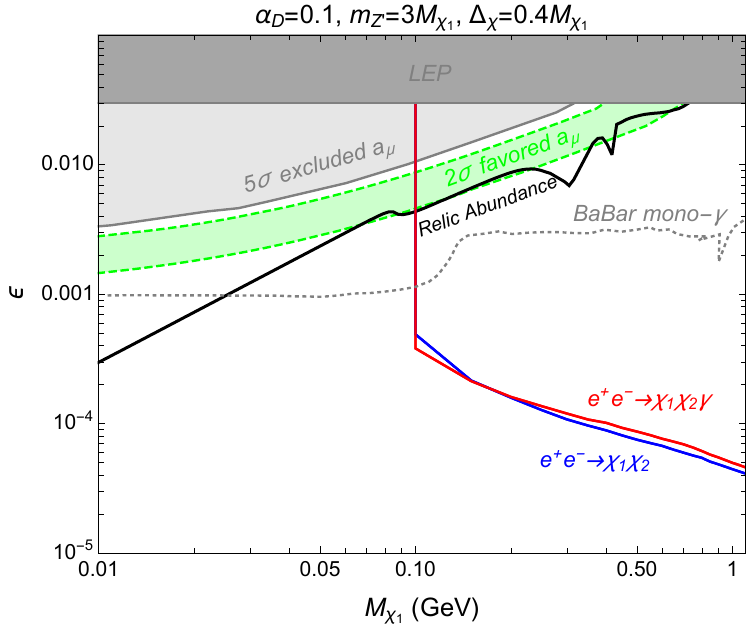}
\caption{
The same as Fig.~\ref{fig:summary1} but for $m_{Z^{\prime}} = 3 M_{\phi_1,\chi_1} $ and $ \Delta_{\phi,\chi} = 0.1 M_{\phi_1,\chi_1}$. The green shaded region bounded by the green dashed lines is the $2\sigma$ allowed region for the $(g-2)_\mu$ excess and the lighter gray region excluded by the $(g-2)_\mu$ at $5\sigma$ C.L.
}\label{fig:summary2}
\end{figure}

Moreover, inspired by BP1 which can explain the muon $(g-2)_\mu$~\cite{Mohlabeng:2019vrz}, 
we perform the same analysis in Fig.~\ref{fig:summary2} for $ \alpha_D = 0.1 $, $ m_{Z^{\prime}} = 3 M_{\phi_1,\chi_1} $ and $ \Delta_{\phi,\chi} = 0.4 M_{\phi_1,\chi_1} $ fixed. We show the 2$\sigma$ allowed and the 5$\sigma$ excluded regions for the muon $(g-2)_\mu$ as well as the model-independent LEP 
bound~\cite{Hook:2010tw}, BaBar mono-$\gamma$ 
bound~\cite{Lees:2017lec} and also correct relic abundance lines.
Once the mass of DM is increasing, the mass splitting between DM ground and excited states is also enhanced, hence, the final state lepton pair becomes more energetic. Therefore,
one can find the BaBar mono-$\gamma$ bound is slightly weakened when the displaced channel is open. 
Since the recasting in Ref~\cite{Mohlabeng:2019vrz} does not include the requirement on the angle $\theta_{LAB}^{DV}$ in Table~\ref{Tab:selection}, the results are weaker than the ones in Ref.~\cite{Duerr:2019dmv}. That is the reason why this parameter space is still allowed in Ref~\cite{Mohlabeng:2019vrz}. In order to make a more precise recasting, we follow Ref.~\cite{Duerr:2019dmv} for the BaBar mono-$\gamma$ bound and it is shown in dotted gray line in Fig.~\ref{fig:summary2} which can already cover the $(g-2)_\mu$ allowed window in the parameter space. Even the recasting BaBar mono-$\gamma$ bound can close this region, we find our result can give much stronger bound, $\epsilon \sim \mathcal{O}(10^{-5}-10^{-4}) $, especially in the parameter space where the mono-$\gamma$ searches are suppressed. 
Therefore we can explicitly cover this area using displaced lepton search at the Belle II in a very first stage of the run. 

In summary, we found the future bounds from the process without ISR photon can be stronger than the one with ISR photon in Fig.~\ref{fig:summary1}. Instead, in the lower $M_{\phi_1}$ region, the bound for process with ISR photon in Fig.~\ref{fig:summary1} and Fig.~\ref{fig:summary2} get stronger because its cross sections are larger than the ones without ISR photon as shown in Fig.~\ref{fig:cross_section}.

\subsection{Determination of the DM mass and mass splitting of DM sector}\label{Sec:kin_reco}

In general, we cannot uniquely determine the DM mass at colliders because of lacking enough constraints for invisible particles in the final state. We can take a mono-$\gamma$ search for DM 
pair production at B-factories as an example. If the DM is denoted by $ \chi $, the process for 
mono-$\gamma$ search  at B-factories is $ e^{+}e^{-}\rightarrow\chi\chi\gamma $ 
(or $ e^{+}e^{-}\rightarrow\chi\overline{\chi}\gamma $). There are eight unknown values from four-momentum of two DMs in the final state. 
Unfortunately, only five constraints in this process : four from the four-momentum conservation and one from the same mass for the DM pair. We still need three extra conditions to uniquely determine the DM mass for each event.

Now we turn our attention to the case of inelastic DM models. For the process in Eq.(\ref{eq:woISR}) or~(\ref{eq:wISR}), if the $ \phi_2 (\chi_2) $ is long-lived and leave the displaced vertex at the Belle II detectors, we will have two extra constraints. 
Again, there are still eight unknown values from four-momentum of two $\phi_1's (\chi_1's)$ in the final state. However, because of the charge neutrality of the $ \phi_2 (\chi_2) $, a three-momentum vector 
of $ \phi_2 (\chi_2) $ is proportional to the direction of displaced vertex (DV)~\cite{Kang:2019ukr,Bae:2020dwf} 
\begin{equation}
\overrightarrow{p}_{\phi_2 (\chi_2)} = |\overrightarrow{p}_{\phi_2 (\chi_2)}| \widehat{r}_{DV},
\end{equation}
where $ \overrightarrow{p}_{\phi_2 (\chi_2)} $ is the three-momentum vector of $ \phi_2 (\chi_2) $ and $ \widehat{r}_{DV} $ is the unit vector of displaced vertex from $ \phi_2 (\chi_2) $. Therefore, we have two more constraints, and there are seven constraints for this kind of processes in total. 
We still need one more condition to uniquely determine the DM mass. We first show some key kinematic equations for processes in Eq.(\ref{eq:woISR}) and~(\ref{eq:wISR}) and solve them event-by-event from our Monte Carlo samples and then involve detector resolution effects\footnote{Here we use the fermion inelastic DM model as an example. The same method can be applied to the scalar inelastic DM model, so we will not repeat it thereafter.}. 
More detalied  derivations of these kinematic equations can be found in Appendix~\ref{Sec:kin_eq}.  

\begin{figure}
\centering
\includegraphics[width=3.0in]{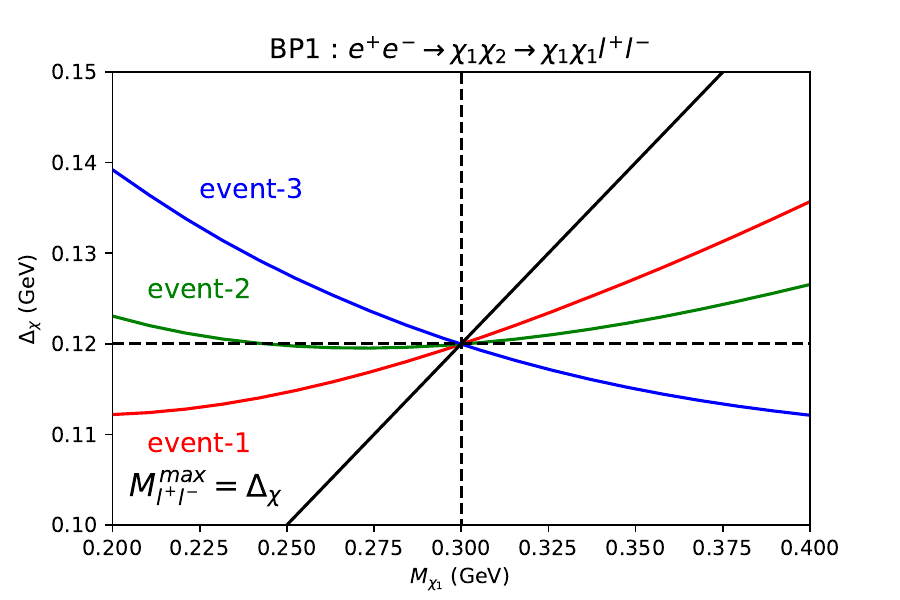}
\includegraphics[width=3.0in]{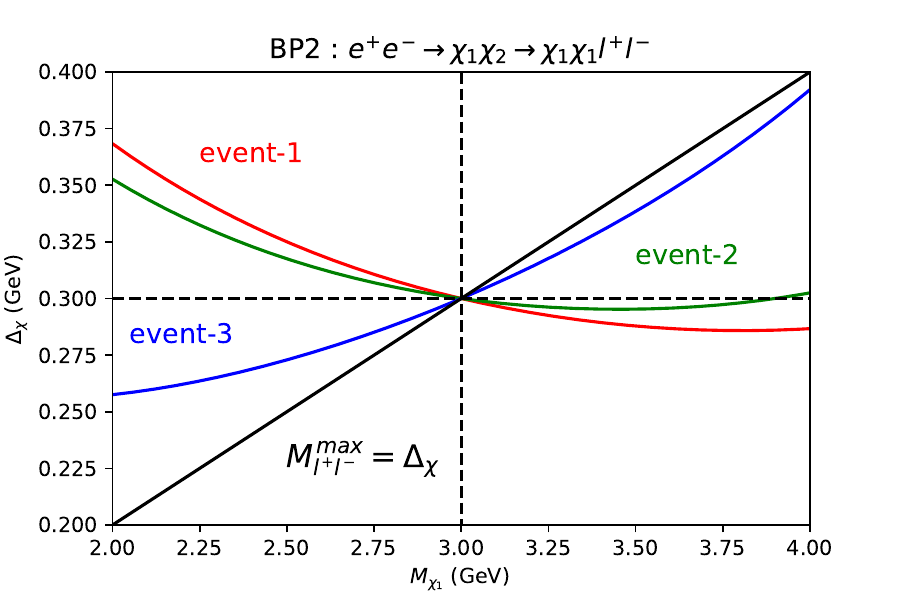}
\includegraphics[width=3.0in]{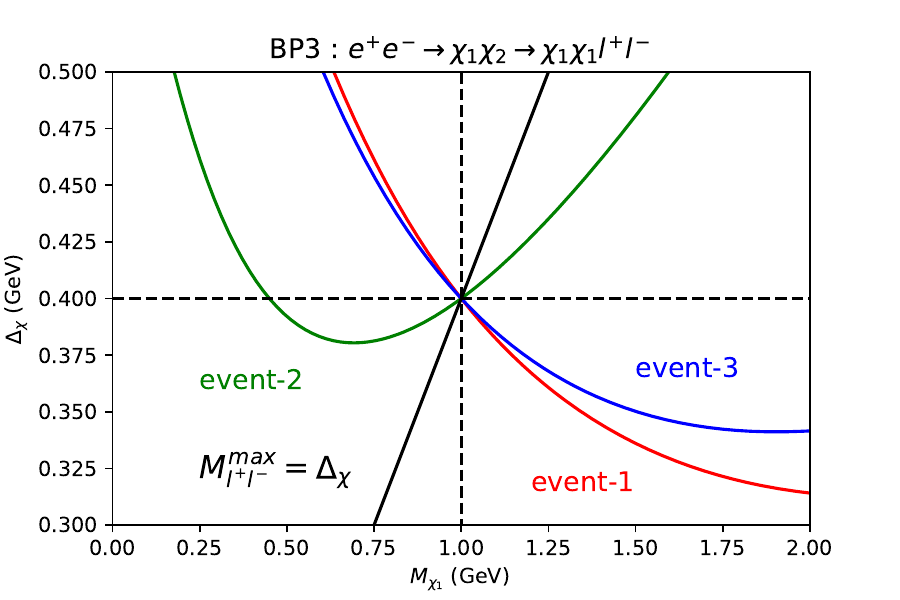}
\includegraphics[width=3.0in]{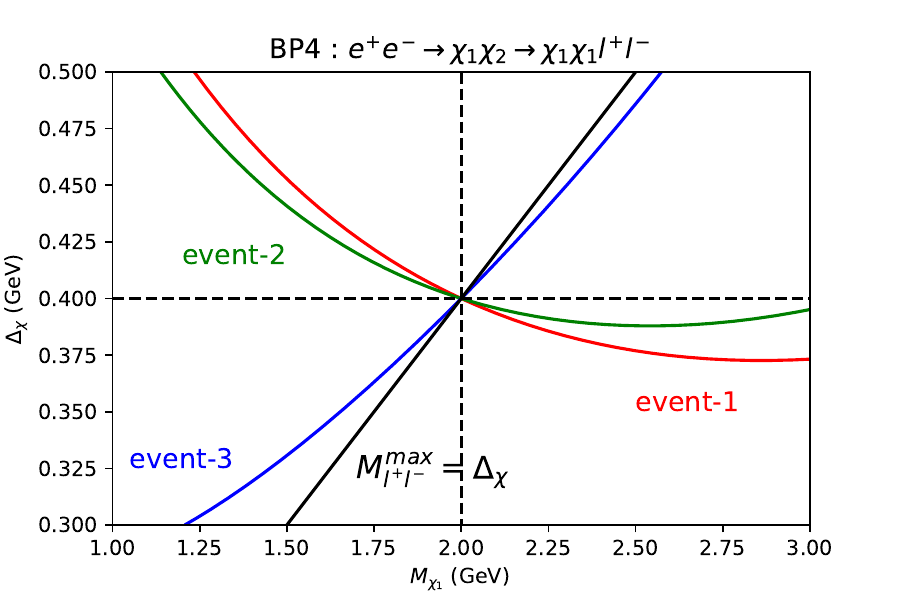}
\caption{
The solutions for Eq.(\ref{eq:Kinematic2}) with $ E_{\chi_2} $ in Eq.(\ref{eq:Echi2}) of $ e^+ e^-\rightarrow\chi_1\chi_2\rightarrow\chi_1\chi_1 l^{+}l^{-} $ process on the $ (M_{\chi_1}, \Delta_{\chi}) $ plane for four BPs. Here we display three arbitrary Monte Carlo events for each BP with red, green and blue lines. On the other hand, the black line shows the kinematic endpoint 
measurement of $ M^{max}_{l^{+}l^{-}} $. 
}\label{fig:KFP1}
\end{figure}

For $ e^+ e^-\rightarrow\chi_1\chi_2\rightarrow\chi_1\chi_1 l^{+}l^{-} $, we first solve the energy of $ \chi_2 $ for the subprocess $ e^+ e^-\rightarrow\chi_1\chi_2 $ as 
\begin{align}
E_{\chi_2} = & \frac{1}{2\left[ \sin^2\theta (E^2_{+}+E^2_{-})+2(1+\cos^2\theta)E_{+}E_{-}\right]} \left[ (E_{+}+E_{-})(4E_{+}E_{-}+M^2_{\chi_2}-M^2_{\chi_1})  \right. 
\nonumber  \\ &
\pm |(E_{-}-E_{+})\cos\theta | \{ (M^2_{\chi_1}-4E_{+}E_{-})^2 
-2[2\sin^2\theta (E^2_{+}+E^2_{-})+4\cos^2\theta E_{+}E_{-}
\nonumber  \\   &
\left. +M^2_{\chi_1}] M^2_{\chi_2}+M^4_{\chi_2} \}^{1/2} \right],
\label{eq:Echi2}
\end{align} 
where $ E_{-} $, $ E_{+} $, and $ \theta $ are $ e^{-} $, $ e^{+} $ beam energies 
and the polar angle of DV. 

We then turn to the subprocess    $ \chi_2\rightarrow\chi_1 l^{+}l^{-} $. 
With the help of energy and momentum conservation, we can receive the following equation 
for inputs of $ E_{V^{\prime}} $, $ \overrightarrow{p_{V^{\prime}}} $ and $ \widehat{r}_{DV} $,
\begin{equation}
M^2_{\chi_2}-M^2_{\chi_1}-2E_{\chi_2}E_{V^{\prime}}+E^2_{V^{\prime}}-|\overrightarrow{p_{V^{\prime}}}|^2 +2\sqrt{E_{\chi_2}^2 -M^2_{\chi_2}}(\widehat{r}_{DV}\cdot\overrightarrow{p_{V^{\prime}}})=0,
\label{eq:Kinematic2}
\end{equation}
where $ E_{\chi_2} $ is shown in Eq.(\ref{eq:Echi2}).
In Fig.~\ref{fig:KFP1}, we display three arbitrary Monte Carlo events on the 
$ (M_{\chi_1}, \Delta_{\chi}) $ plane for four BPs according to Eq.(\ref{eq:Kinematic2}).
On the other hand, the black line shows the kinematic endpoint measurement 
$ M^{max}_{l^{+}l^{-}} $. We can find that the lines from all events and $ M^{max}_{l^{+}l^{-}} $ cross to the same point which is the true $ (M_{\chi_1},\Delta_{\chi}) $ in our four BPs. This is a simple application of the Kinematic Focus Point Method proposed in Ref.~\cite{Kim:2019prx}.

\begin{figure}
\centering
\includegraphics[width=3.0in]{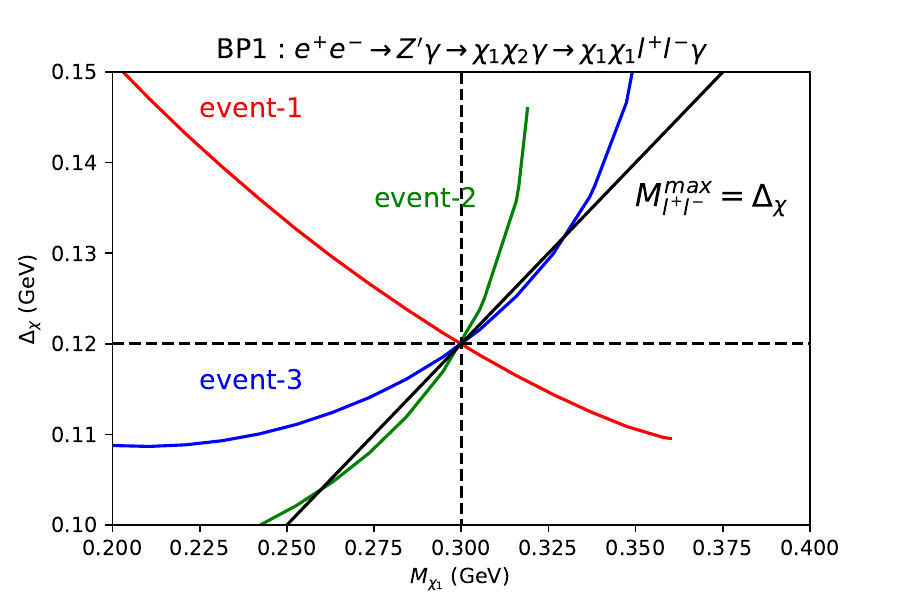}
\includegraphics[width=3.0in]{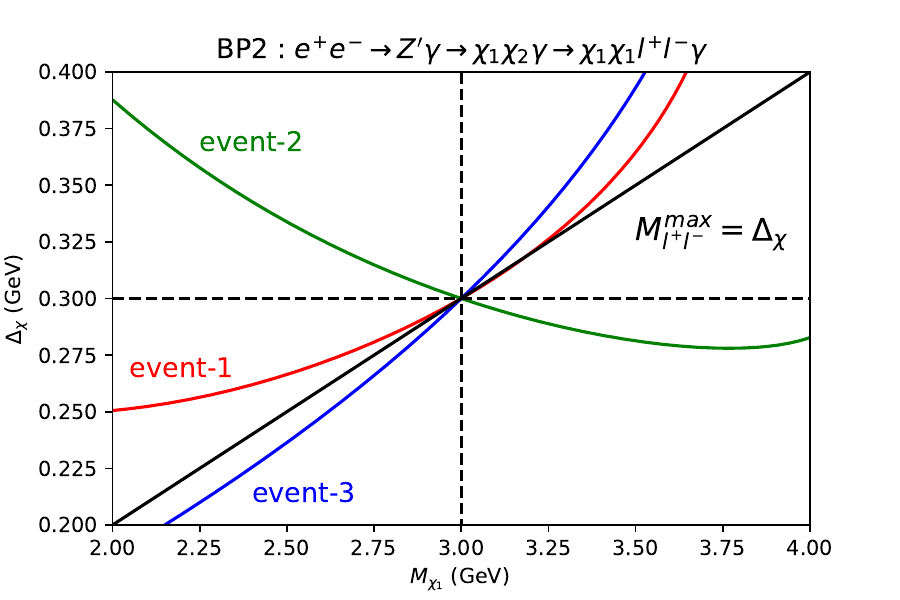}
\includegraphics[width=3.0in]{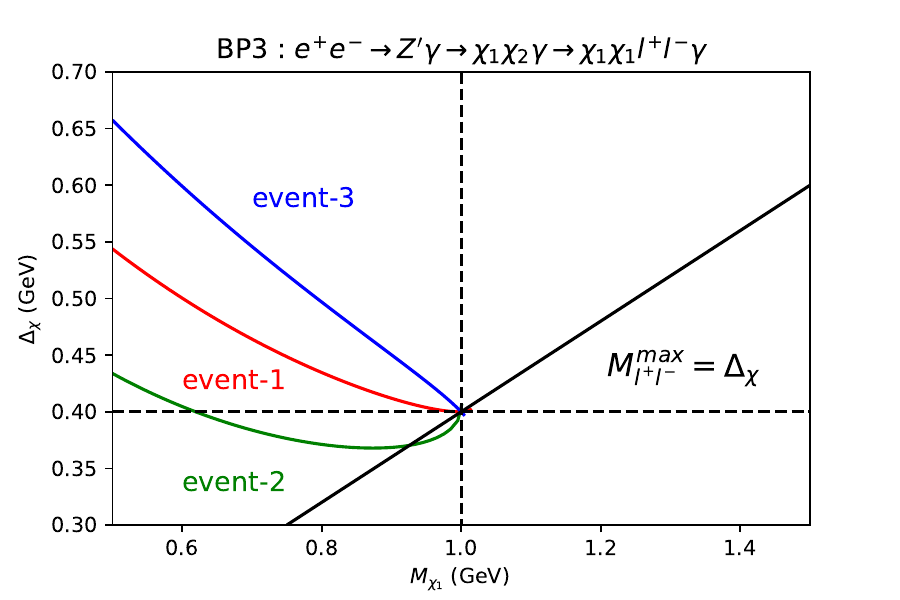}
\includegraphics[width=3.0in]{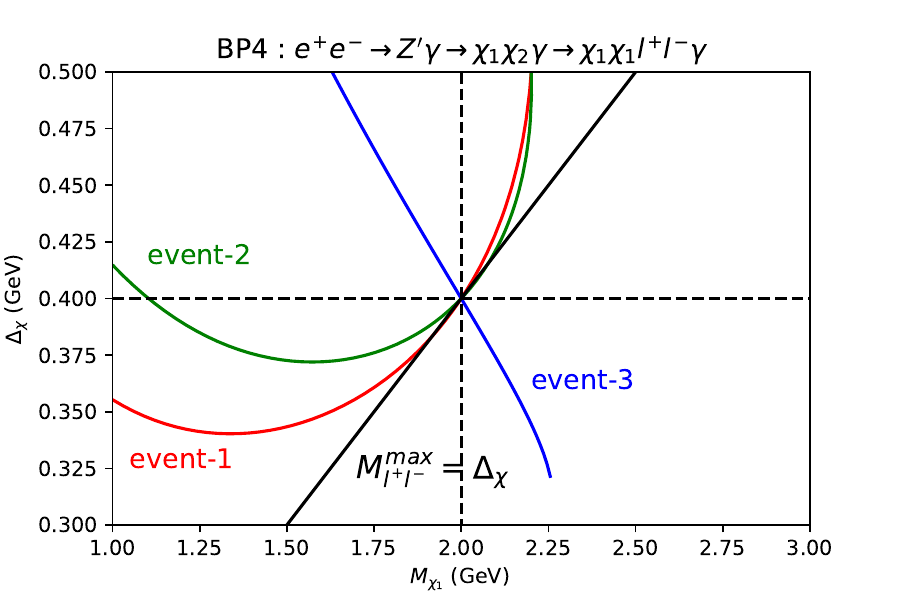}
\caption{
The same as Fig.~\ref{fig:KFP1}, but the solutions for Eq.(\ref{eq:Kinematic2}) with $ E_{\chi_2} $ in Eq.(\ref{eq:Echi2_ISR}) of $ e^+ e^-\rightarrow Z^{\prime}\gamma\rightarrow\chi_1\chi_2\gamma\rightarrow\chi_1\chi_1 l^{+}l^{-}\gamma $ process.  
}\label{fig:KFP2}
\end{figure}

Similarly, for $ e^+ e^-\rightarrow Z^{\prime}\gamma\rightarrow\chi_1\chi_2\gamma\rightarrow\chi_1\chi_1 l^{+}l^{-}\gamma $, we can also solve the energy of $ \chi_2 $ for the subprocess $ e^+ e^-\rightarrow Z^{\prime}\gamma\rightarrow\chi_1\chi_2\gamma $ as 
\begin{align}
E_{\chi_2} = & \frac{1}{2\left[ E^2_{Z^{\prime}}-(\widehat{r}_{DV}\cdot\overrightarrow{p_{Z^{\prime}}})^2\right]} [E_{Z^{\prime}}(m^2_{Z^{\prime}}+M^2_{\chi_2}-M^2_{\chi_1}) 
\nonumber  \\ &
\pm | \widehat{r}_{DV}\cdot\overrightarrow{p_{Z^{\prime}}}| \sqrt{(m^2_{Z^{\prime}}+M^2_{\chi_2}-M^2_{\chi_1})^2 -4M^2_{\chi_2}\left[ E^2_{Z^{\prime}}-(\widehat{r}_{DV}\cdot\overrightarrow{p_{Z^{\prime}}})^2\right]} ].
\label{eq:Echi2_ISR}
\end{align} 
Finally, the kinematic equation for $ \chi_2\rightarrow\chi_1 f\overline{f} $ is the same as Eq.(\ref{eq:Kinematic2}) with $ E_{\chi_2} $ in Eq.(\ref{eq:Echi2_ISR}).
Again, we show three arbitrary Monte Carlo events on the $ (M_{\chi_1}, \Delta_{\chi}) $ plane for four BPs according to Eq.(\ref{eq:Kinematic2}) in Fig.~\ref{fig:KFP2}.   All events and 
$ M^{max}_{l^{+}l^{-}} $ cross to the true $ (M_{\chi_1}, \Delta_{\chi}) $ in our four BPs.

\begin{figure}
\centering
\includegraphics[width=3.0in]{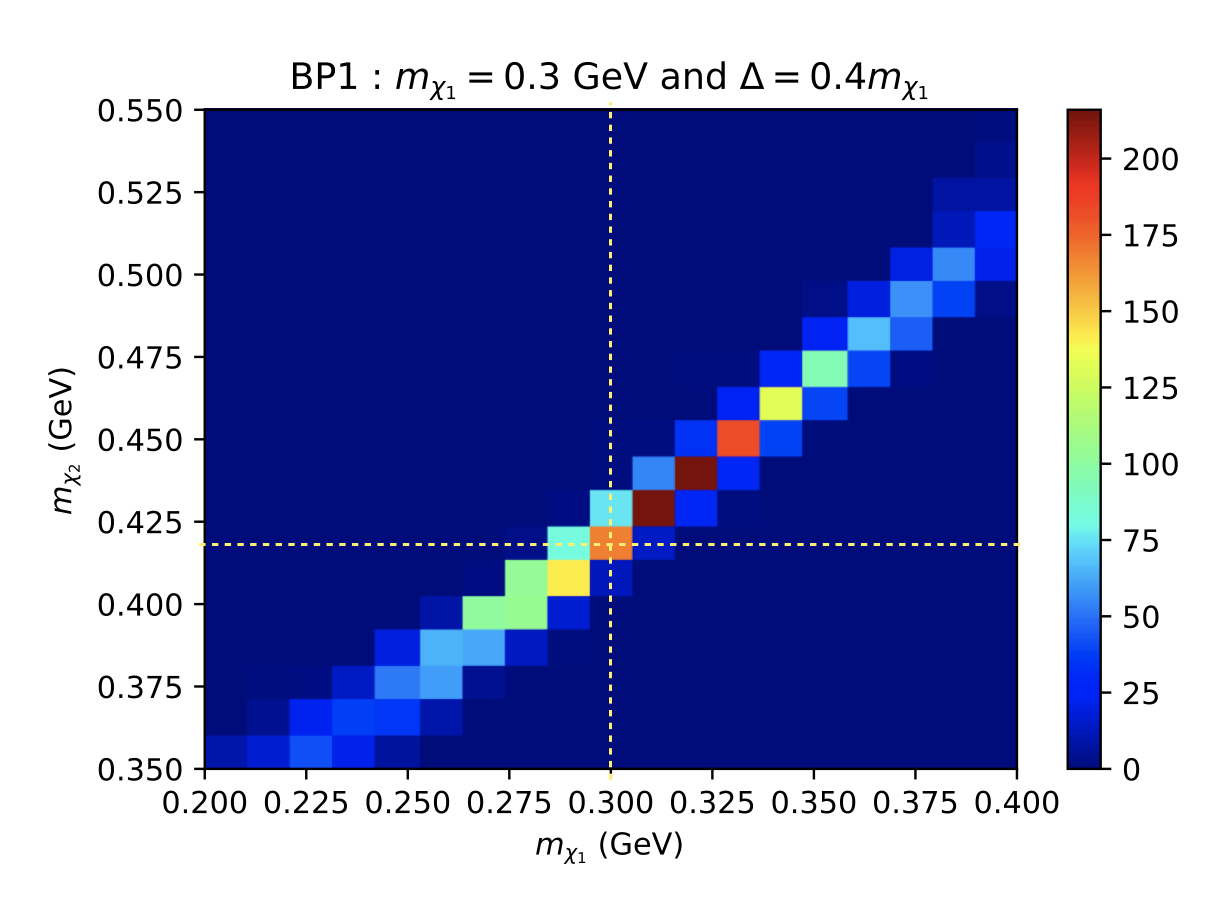}
\includegraphics[width=3.0in]{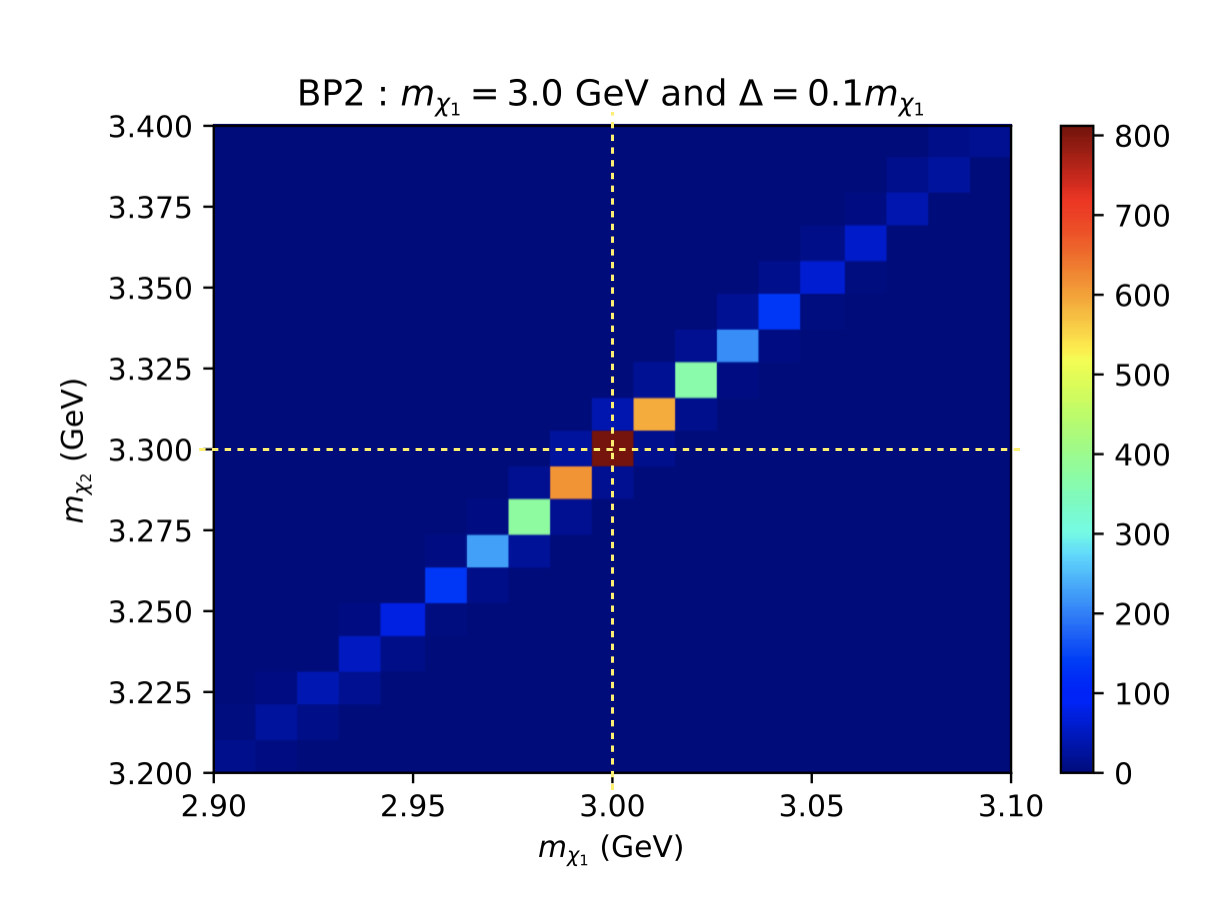}
\includegraphics[width=3.0in]{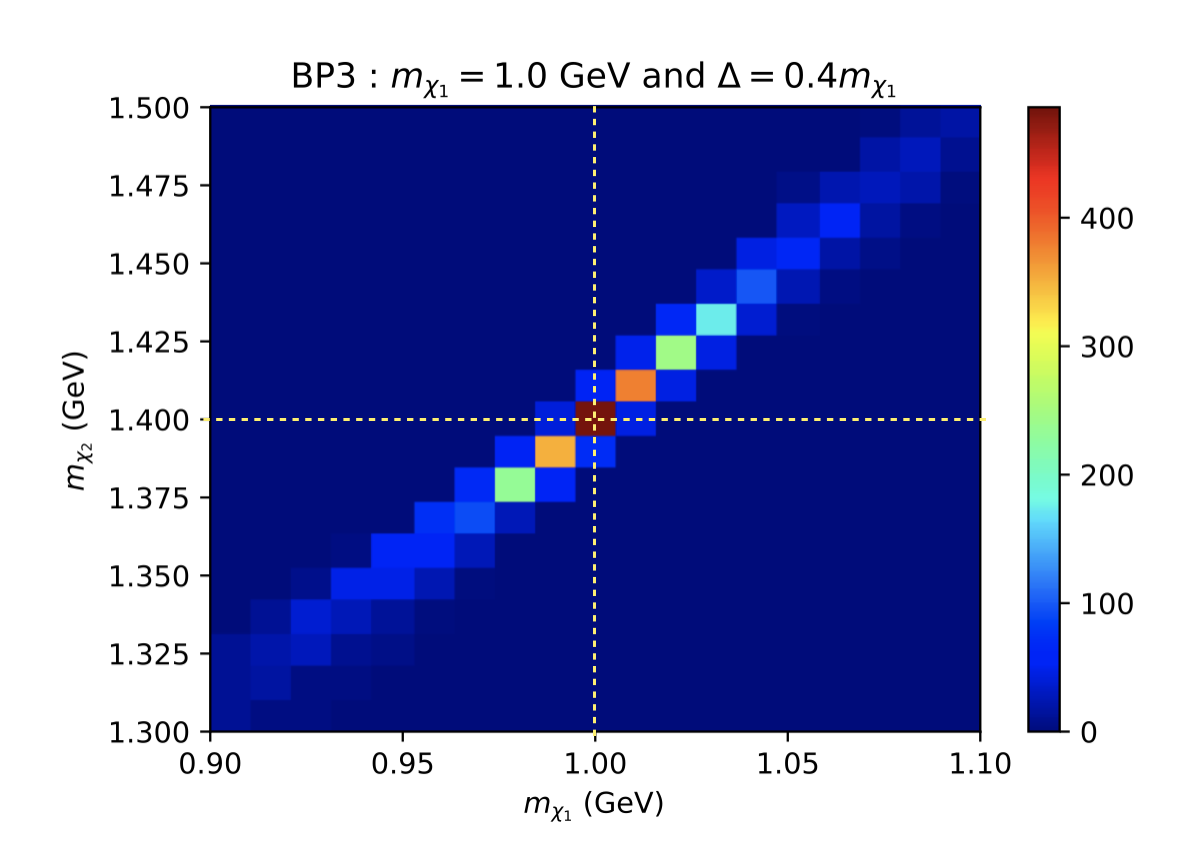}
\includegraphics[width=3.0in]{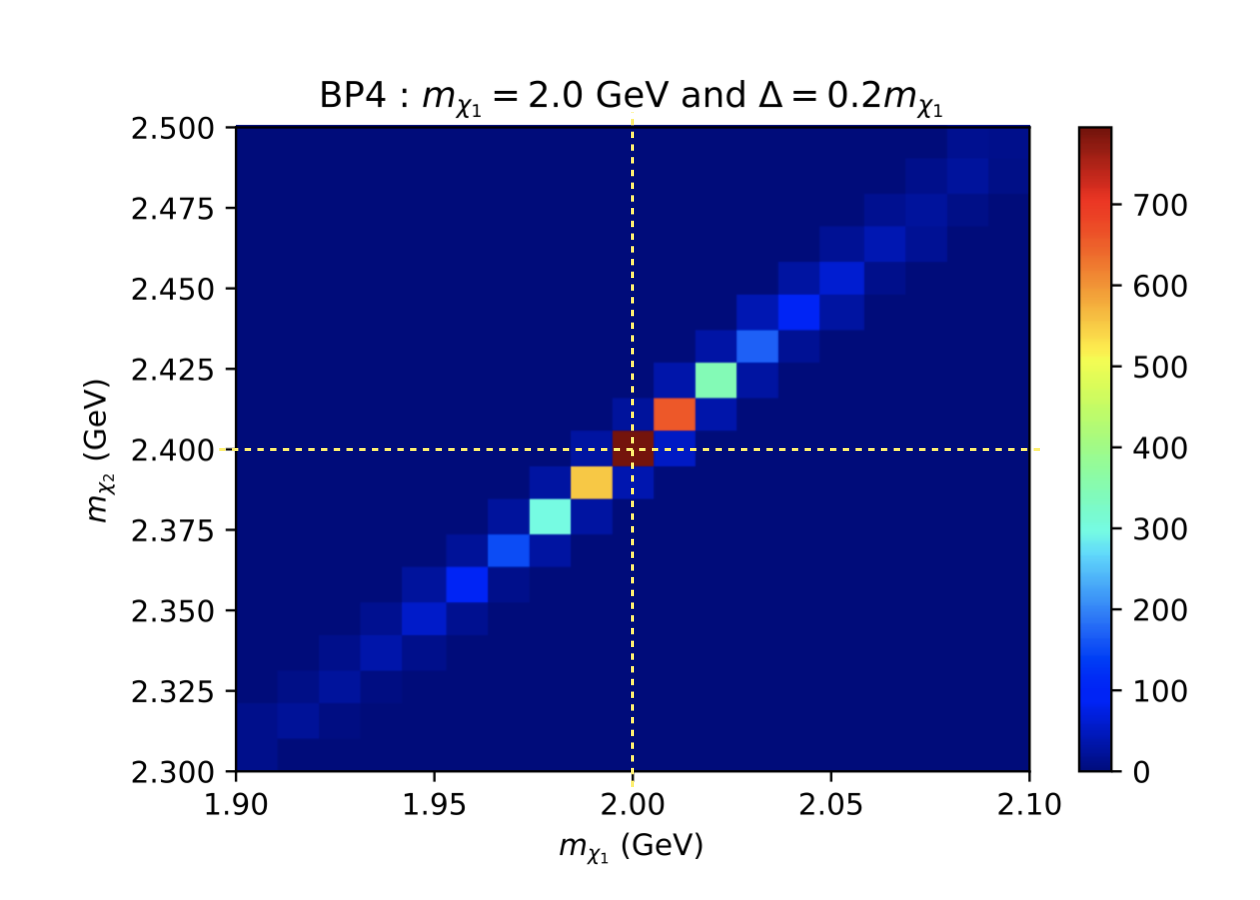}
\caption{
The solutions for Eq.(\ref{eq:Kinematic2}) with $ E_{\chi_2} $ in Eq.(\ref{eq:Echi2}) of $ e^+ e^-\rightarrow\chi_1\chi_2\rightarrow\chi_1\chi_1 l^{+}l^{-} $ process on the $ (M_{\chi_1}, M_{\chi_2}) $ plane for four BPs after involving detector resolution effects and event selections. Here 100 signal events are taken into account for each BP and the bin size for both $x$ and $y$ axes are set to be $ 10 $ MeV. 
}\label{fig:KFP3}
\end{figure}

\begin{figure}
\centering
\includegraphics[width=3.0in]{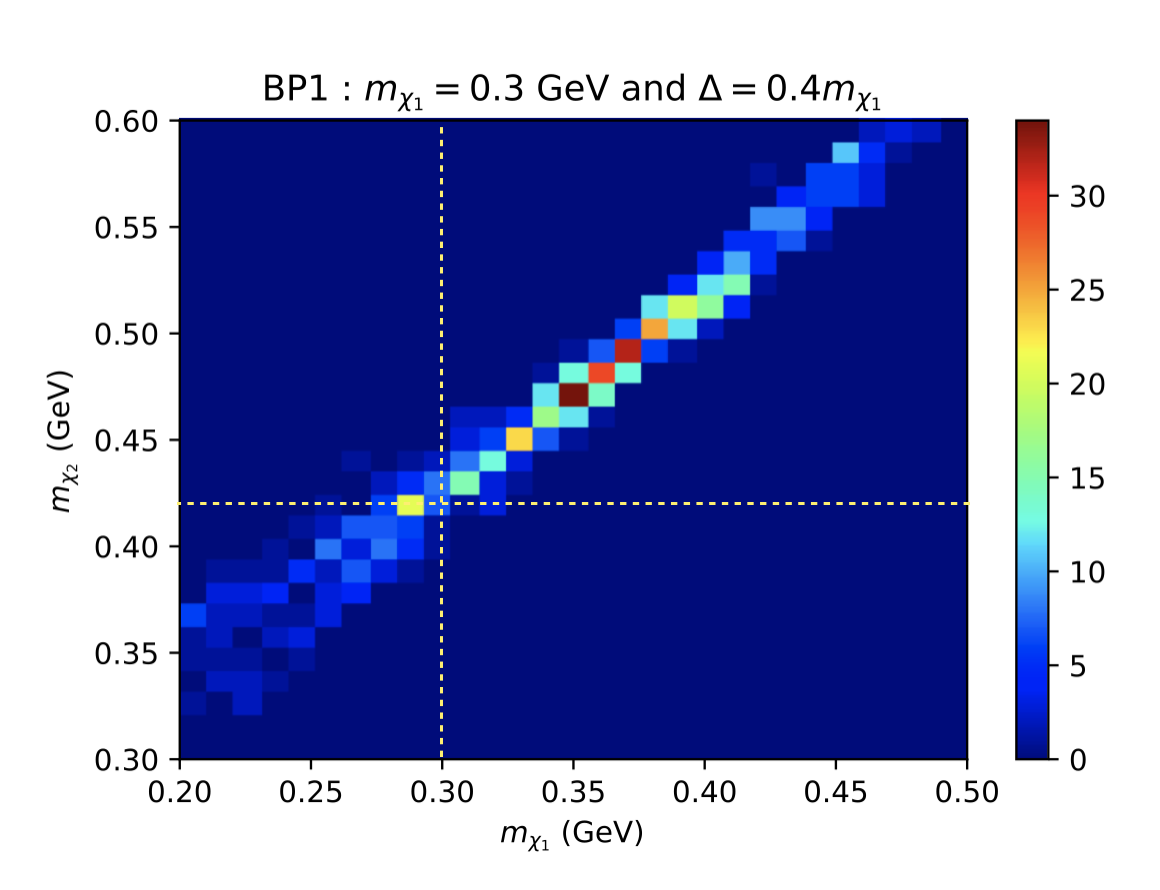}
\includegraphics[width=3.0in]{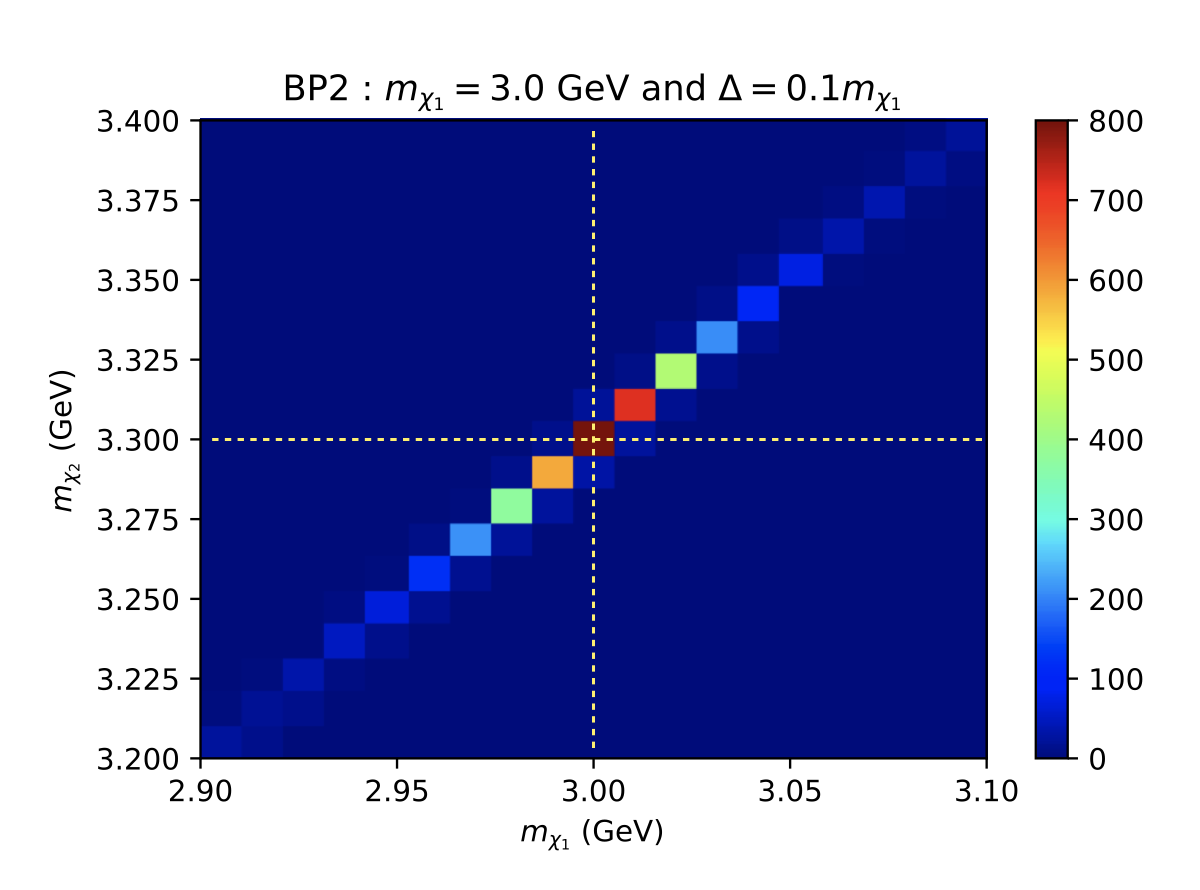}
\includegraphics[width=3.0in]{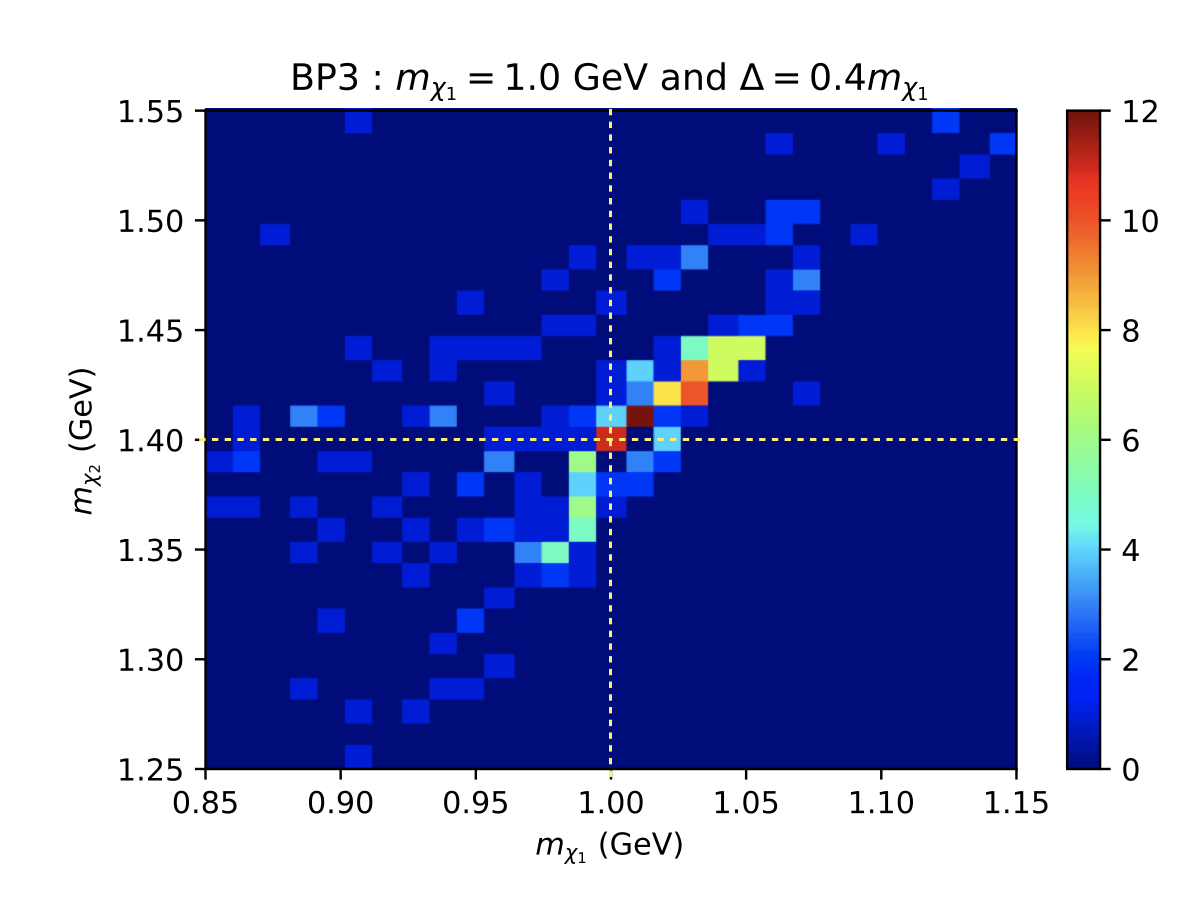}
\includegraphics[width=3.0in]{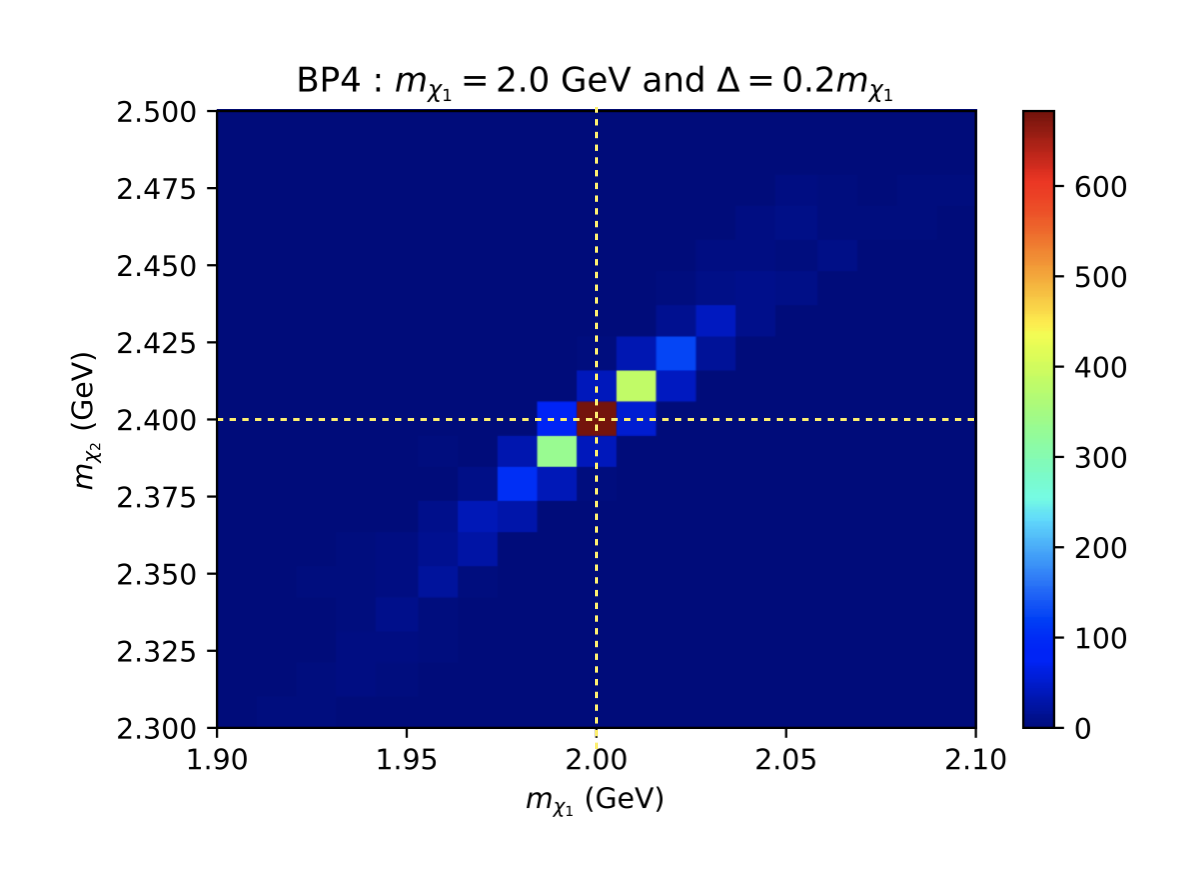}
\caption{
The same as Fig.~\ref{fig:KFP3}, but the solutions for Eq.(\ref{eq:Kinematic2}) with $ E_{\chi_2} $ in Eq.(\ref{eq:Echi2_ISR}) of $ e^+ e^-\rightarrow Z^{\prime}\gamma\rightarrow\chi_1\chi_2\gamma\rightarrow\chi_1\chi_1 l^{+}l^{-}\gamma $ process.  
}\label{fig:KFP4}
\end{figure}

\begin{table}[t!]
\begin{center}\begin{tabular}{|c|c|c|c|}\hline 
\multirow{2}{*}{BP} & \multirow{2}{*}{$N_{phys}$} & $(M_{\chi_2},\, M_{\chi_1})^\textrm{true}$ & \multirow{2}{*}{rms} \\
\cline{3-2}
& & $(M_{\chi_2},\, M_{\chi_1})^\textrm{peak}$  &  \\ 
\hline
\multirow{2}{*}{BP1} & \multirow{2}{*}{$4473$} & $(0.42, \, 0.30)$ &  \multirow{2}{*}{ $(0.168,\,0.175)$ } \\
\cline{3-2}
& & $(0.43,\,0.32)$ &  \\ 
\hline
\multirow{2}{*}{BP2} & \multirow{2}{*}{$4915$} & $(3.30, \, 3.00)$ &  \multirow{2}{*}{ $(0.175,\, 0.190)$ } \\
\cline{3-2}
& & $(3.30,\,3.00)$ &  \\
\hline
\multirow{2}{*}{BP3} & \multirow{2}{*}{$4856$} & $(1.40, \, 1.00)$ &  \multirow{2}{*}{ $(0.172,\, 0.192)$ } \\
\cline{3-2}
& & $(1.40,\,1.00)$ &  \\
\hline
\multirow{2}{*}{BP4} & \multirow{2}{*}{$4918$} & $(2.40, \, 2.00)$ &  \multirow{2}{*}{ $(0.155,\, 0.170)$ } \\
\cline{3-2}
& & $(2.40,\,2.00)$ &  \\
\hline \end{tabular} \caption{The peak measured values and root mean square (rms) for $ (M_{\chi_2},M_{\chi_1}) $ of four benchmark points in $ e^+ e^-\rightarrow\chi_1\chi_2\rightarrow\chi_1\chi_1 l^{+}l^{-} $ process. $ N_{phys} $ is the number of physical solutions from 100 signal events. All numbers are in GeV unit. }
\label{Tab:result1}
\end{center}
\end{table}
\begin{table}[t!]
\begin{center}\begin{tabular}{|c|c|c|c|}\hline 
\multirow{2}{*}{BP} & \multirow{2}{*}{$N_{phys}$} & $(M_{\chi_2},\, M_{\chi_1})^\textrm{true}$ & \multirow{2}{*}{rms} \\
\cline{3-2}
& & $(M_{\chi_2},\, M_{\chi_1})^\textrm{peak}$  &  \\ 
\hline
\multirow{2}{*}{BP1} & \multirow{2}{*}{$901$} & $(0.42, \, 0.30)$ &  \multirow{2}{*}{ $(0.114,\,0.138)$ } \\
\cline{3-2}
& & $(0.47,\,0.35)$ &  \\ 
\hline
\multirow{2}{*}{BP2} & \multirow{2}{*}{$4914$} & $(3.30, \, 3.00)$ &  \multirow{2}{*}{ $(0.121,\, 0.128)$ } \\
\cline{3-2}
& & $(3.30,\,3.00)$ &  \\
\hline
\multirow{2}{*}{BP3} & \multirow{2}{*}{$377$} & $(1.40, \, 1.00)$ &  \multirow{2}{*}{ $(0.216,\, 0.402)$ } \\
\cline{3-2}
& & $(1.41,\,1.01)$ &  \\
\hline
\multirow{2}{*}{BP4} & \multirow{2}{*}{$2824$} & $(2.40, \, 2.00)$ &  \multirow{2}{*}{ $(0.126,\, 0.173)$ } \\
\cline{3-2}
& & $(2.40,\,2.00)$ &  \\
\hline \end{tabular} \caption{The same as Table.~\ref{Tab:result1}, but in $ e^+ e^-\rightarrow Z^{\prime}\gamma\rightarrow\chi_1\chi_2\gamma\rightarrow\chi_1\chi_1 l^{+}l^{-}\gamma $ process.}
\label{Tab:result2}
\end{center}
\end{table}

In order to make the Kinematic Focus Point Method fit to reality, we involve the detector resolution effects from Sec.~\ref{Sec:detector} and event selections in Table.~\ref{Tab:selection}. Here we conservatively assume there are 100 signal events for each BP can be recorded. Since we can solve the above kinematic equations for any two signal events, we will get $ C^{100}_2 = 4950 $ solutions from 100 signal events. After removing the unphysical solutions, we show the distributions of these solutions on the $ (M_{\chi_1}, M_{\chi_2}) $ plane for four BPs of $ e^+ e^-\rightarrow\chi_1\chi_2\rightarrow\chi_1\chi_1 l^{+}l^{-} $ and $ e^+ e^-\rightarrow Z^{\prime}\gamma\rightarrow\chi_1\chi_2\gamma\rightarrow\chi_1\chi_1 l^{+}l^{-}\gamma $ processes in Figs.~\ref{fig:KFP3} and~\ref{fig:KFP4}, respectively. 
Here the bin size for both $x$ and $y$ axes are set to be $ 10 $ MeV. 
Moreover, the most probable mass values and their statistical errors for four BPs in our simulation are summarized in Table.~\ref{Tab:result1} and~\ref{Tab:result2}. The statistical errors are estimated as the root mean square (rms) value with respective to the most probable values $ rms = \sqrt{\sum^{N_{phys}}_{i=1}(M_i -M^{peak})^2 /N_{phys}} $, where $ N_{phys} $ is the number of physical solutions from 100 signal events and $ M^{peak} $ is the most probable value.

In Fig.~\ref{fig:KFP3}, except for BP1 on top left panel, $ M_{\chi_1} $ and $ M_{\chi_2} $ can be determined within the percentage of deviation for other three BPs. As shown in Fig.~\ref{fig:kinematic1}, both the decay length of $ \chi_2 $ is relative short and $ E_{ll} $ is 
relative large for BP1 compared with other three BPs such that events in BP1 cannot be avoided to 
severely suffer from detector resolution effects. This explains why the physical solutions are reduced 
and the reconstructed $ M_{\chi_1} $ and $ M_{\chi_2} $ are shifted to larger values with higher 
smearing for BP1. 
In Fig.~\ref{fig:KFP4}, $ M_{\chi_1} $ and $ M_{\chi_2} $ can still be determined within the percentage 
of deviation for BP2 and BP4. However, apart from detector resolution effects from charged leptons 
and displaced vertex, the ability to precisely pin down $ M_{\chi_1} $ and $ M_{\chi_2} $ for the process 
in Eq.(\ref{eq:wISR}) also relies on how well the on-shell $ Z^{\prime} $ can be reconstructed. As shown in Fig.~\ref{fig:kinematic2}, the peak of $ m_{Z^{\prime}} $ is rather spread out (or even shifted) 
for BP1 and BP3 than for BP2 and BP4. 
According to Eq.(\ref{eq:Zp4momenta}), the reconstructed 
$ Z^{\prime} $ four-momentum is highly dependent on the detector resolution effects of ISR photon. 
Once the $ E(\gamma) $ increases, its smearing is also enhanced such that the reconstruction of 
the on-shell $ Z^{\prime} $ becomes poor. In the end, the physical solutions of BP1 and BP3 are 
largely reduced and shifting behaviors of $ M_{\chi_1} $ and $ M_{\chi_2} $ become severe in 
Fig.~\ref{fig:KFP4} compared with the ones in Fig.~\ref{fig:KFP3}.
This is another reason to encourage our experimental colleagues to search for not only the usual process in Eq.(\ref{eq:wISR}) at B-factories, but also the process in Eq.(\ref{eq:woISR}) with the displaced vertex trigger which is more sensitive to determine the mass and mass splitting of DM sector for some parameter space in the inelastic DM models.

\section{Conclusion}\label{Sec:Conclusion}

The dark sectors with light DM candidate have received substantial attention as the null signal result from DM direct detetion has been reported until now. On the other hand, the null results from beyond the SM searches 
at the LHC also hint new physics signatures may hide in the elusive corner. New kind of signatures 
such as long-lived particles at colliders become more and more popular and may guide a royal road 
for new physics evidences.
Among these dark sector models, the inelastic DM model is an appealing example which can both 
allow light DM candidate and predict long-lived particle which can be searched for at colliders. 
Therefore, we focus on the displaced vertex signatures in inelastic DM models at Belle II in this work.

In the inelastic DM models, if the mass splitting between DM excited and ground states is small enough, the co-annihilation becomes the dominant channel for thermal relic density and the DM excited state can be long-lived at the collider scale. 
We first review scalar and fermion inelastic DM models with $ U(1)_D $ gauge symmetry in Sec.~\ref{Sec:Model} and point out the dark Higgs sector caused the mass splitting between DM 
excited and ground states and provided a mechanism to the $ Z^{\prime} $ mass. 
Besides, the off-diagonal interaction between $ Z^{\prime} $ and DM sector is derived from the 
covariant derivative term of DM sector.  

The analytical representations and numerical results for cross sections of $ e^{+}e^{-}\rightarrow\phi_1\phi_2 (\chi_1\chi_2) $ and $ e^{+}e^{-}\rightarrow\phi_1\phi_2 (\chi_1\chi_2)\gamma $ processes are studied in Sec.~\ref{Sec:Xsec}. In the first process, there is an extra $ \beta $ factor suppression for boson pair cross sections compared with fermion pair ones. In the second process, 
the dominant channel comes from $ e^{+}e^{-}\rightarrow Z^{\prime}\gamma\rightarrow\phi_1\phi_2 (\chi_1\chi_2)\gamma $. Therefore, cross sections in this process for scalar and fermion inelatic DM models are very close to each other. In addition, the polar angle distributions of $ \phi_2 (\chi_2) $ 
at the Belle II LAB frame as shown in Fig.~\ref{fig:theta} are helpful to distinguish these two models. 

The novel dilepton displaced vertex signatures at Belle II are the main targets of this study. 
We include detector resolution effects in Sec.~\ref{Sec:detector} and event selections in Table.~\ref{Tab:selection} for signal processes in Eq.(\ref{eq:woISR}) and~(\ref{eq:wISR}). 
Our analysis results are summarized in Figs.~\ref{fig:summary1} and ~\ref{fig:summary2},  
which indicate that the future bounds from dilepton displaced vertex searches in inelastic 
DM models at Belle II are stronger than previous constraints for two benchmark mass windows 
$ 0.4\lesssim M_{\phi_1,\chi_1}\lesssim 4.9 $ GeV and 
$ 0.1\lesssim M_{\phi_1,\chi_1}\lesssim 1.1 $ GeV, respectively. 
For the latter case, our results can cover the allowed parameter space which can explain the muon 
$ (g-2)_\mu $. Therefore the early stage of the Belle II experiment can explicitly close off this area
in the parameter space. We further apply the Kinematic Focus Point method for $ e^+ e^-\rightarrow\chi_1\chi_2\rightarrow\chi_1\chi_1 l^{+}l^{-} $ and $ e^+ e^-\rightarrow Z^{\prime}\gamma\rightarrow\chi_1\chi_2\gamma\rightarrow\chi_1\chi_1 l^{+}l^{-}\gamma $ processes to determine both $ M_{\chi_1} $ and $ M_{\chi_2} $ masses. Especially, we find the mass and mass splitting of DM sector can be determined within the percentage of deviation as shown in Table.~\ref{Tab:result1} and~\ref{Tab:result2}. 
Before closing, we would like to mention that our analysis in this work is quite general and can be applied to other models such as~\cite{Berlin:2018tvf,Ballett:2019pyw,Abdullahi:2020nyr,Wang:2019orr,Dey:2020juy}.

\section*{Acknowledgment} 
We thank Myeonghun Park and Youngjoon Kwon for very helpful discussions and useful information.
The work is supported in part by KIAS Individual Grants, Grant No. PG076201 (DWK), PG021403 (PK) and PG075301 (CTL) at 
Korea Institute for Advanced Study, and by National Research Foundation of Korea (NRF) Grant No. 
NRF-2019R1A2C3005009 (PK), funded by the Korean government (MSIT).

\newpage

\appendix

\section{The analytical representations for cross sections of $ e^+ e^-\rightarrow \phi_1\phi_2 (\chi_1\chi_2) $ in the LAB frame}\label{Sec:LAB_rep}

In this appendix, we show the analytical representations for cross sections of $ e^+ e^-\rightarrow \phi_1\phi_2 (\chi_1\chi_2) $ in the LAB frame which have been mentioned in Sec.~\ref{Sec:Xsec}. 
We first set the four momentum of the initial $ e^{+} $, $ e^{-} $ and $ \phi_2 (\chi_2) $, 
$ \phi_1 (\chi_1) $ in the LAB frame as
\begin{align}
& p_{e^{+}} = (E_{+},0,0,E_{+}),
\nonumber  \\ &
p_{e^{-}} = (E_{-},0,0,-E_{-}),
\nonumber  \\ &
p_{\phi_2 (\chi_2)} = (E_{\phi_2 (\chi_2)},p_{x,\phi_2 (\chi_2)},p_{y,\phi_2 (\chi_2)},p_{z,\phi_2 (\chi_2)}),
\nonumber  \\ &
p_{\phi_1 (\chi_1)} = (E_{\phi_1 (\chi_1)},p_{x,\phi_1 (\chi_1)},p_{y,\phi_1 (\chi_1)},p_{z,\phi_1 (\chi_1)}).
\end{align}
According to the four momentum conservation for  this scattering process, $ p_{\phi_1 (\chi_1)} $ can be written as
\begin{equation}
p_{\phi_1 (\chi_1)} = (E_{+}+E_{-}-E_{\phi_2 (\chi_2)},-p_{x,\phi_2 (\chi_2)},-p_{y,\phi_2 (\chi_2)},E_{+}-E_{-}-p_{z,\phi_2 (\chi_2)}).
\end{equation}
The $ d\sigma /d\cos\theta $ for $ e^+ e^-\rightarrow \phi_1\phi_2 $ via s-channel $ Z^{\prime} $ in the LAB frame can be written as
\begin{align}
\left.\frac{d\sigma (e^+ e^-\rightarrow \phi_1\phi_2)}{d\cos\theta}\right|_{LAB}= & \frac{\epsilon^2 e^2 g^2_D}{256\pi E_{+}E^2_{-} [(4E_{+}E_{-}-m^2_{Z'})^2 +m^2_{Z'}\Gamma^2_{Z'}]}\times 
\nonumber  \\ &
[8E_{+}E_{-}\left( E_{\phi_2}(E_{+} +E_{-})+(E_{-} -E_{+})|\overrightarrow{p_{\phi_2}}| \cos\theta \right)
\nonumber  \\ &
-4E_{+}E_{-}(M^2_{\phi_1}+3M^2_{\phi_2})+(M^2_{\phi_2}-M^2_{\phi_1})^2 
\nonumber  \\ &
+\left(4E_{+}(E_{\phi_2}-E_{-}-|\overrightarrow{p_{\phi_2}}| \cos\theta)+M^2_{\phi_1}-M^2_{\phi_2}\right)\times
\nonumber  \\ &
\left(4E_{-}(E_{\phi_2}-E_{+}+|\overrightarrow{p_{\phi_2}}| \cos\theta)+M^2_{\phi_1}-M^2_{\phi_2}\right)]\times
\nonumber  \\ &
\left(-\frac{dE_{\phi_2}}{d\cos\theta} +\cos\theta\frac{d |\overrightarrow{p_{\phi_2}}|}{d\cos\theta}-\sin\theta |\overrightarrow{p_{\phi_2}}|\right),
\label{eq:22scalarLAB}
\end{align}
where $ |\overrightarrow{p_{\phi_2}}| = \sqrt{E^2_{\phi_2} -M^2_{\phi_2}} $ and $ E_{\phi_2} $ is the same as Eq.(\ref{eq:Echi2}) with $ \chi_{1,2}\rightarrow\phi_{1,2} $.

On the other hand, the $ d\sigma /d\cos\theta $ for $ e^+ e^-\rightarrow \chi_1\chi_2 $ via s-channel $ Z^{\prime} $ in the LAB frame can be written as
\begin{align}
\left.\frac{d\sigma (e^+ e^-\rightarrow \chi_1\chi_2)}{d\cos\theta}\right|_{LAB}= & \frac{\epsilon^2 e^2 g^2_D}{32\pi E_{+}E^2_{-} [(4E_{+}E_{-}-m^2_{Z'})^2 +m^2_{Z'}\Gamma^2_{Z'}]}\times 
\nonumber  \\ &
[2(E^2_{+}+E^2_{-})E^2_{\chi_2}(1+\cos^2\theta)-(E_{-}-E_{+})(M^2_{\chi_2}-M^2_{\chi_1}) |\overrightarrow{p_{\chi_2}}| \cos\theta
\nonumber  \\ &
+(E_{+}+E_{-})\left( 4(E_{-}-E_{+}) |\overrightarrow{p_{\chi_2}}| \cos\theta +M^2_{\chi_1}-M^2_{\chi_2}\right) E_{\chi_2}
\nonumber  \\ &
-2(E^2_{+}+E^2_{-})M^2_{\chi_2}\cos^2\theta +4E_{+}E_{-}M_{\chi_1}M_{\chi_2}]\times
\nonumber  \\ &
\left(-\frac{dE_{\phi_2}}{d\cos\theta} +\cos\theta\frac{d |\overrightarrow{p_{\phi_2}}|}{d\cos\theta}-\sin\theta |\overrightarrow{p_{\phi_2}}|\right),
\label{eq:22fermionLAB}
\end{align}  
where $ |\overrightarrow{p_{\chi_2}}| = \sqrt{E^2_{\chi_2} -M^2_{\chi_2}} $ and $ E_{\chi_2} $ can be found in Eq.~(\ref{eq:Echi2}).

\section{The full analytical representations of $ e^+ e^-\rightarrow \phi_1\phi_2 (\chi_1\chi_2)\rightarrow \phi_1\phi_1 (\chi_1\chi_1) f\overline{f} $}\label{Sec:2to4_rep}  

In this appendix, we display the full analytical representations of $ e^+ e^-\rightarrow \phi_1\phi_2 (\chi_1\chi_2)\rightarrow \phi_1\phi_1 (\chi_1\chi_1) f\overline{f} $ as mentioned in Sec.~\ref{Sec:Xsec}.   
The differential cross section for $ e^+(p_1) e^-(p_2)\rightarrow\phi_1(p_3)\phi_1(p_4)f(p_5)\overline{f}(p_6)$ can be represented as
\begin{equation}
d\sigma (e^+ e^-\rightarrow\phi_1\phi_1 f\overline{f})=\frac{(2\pi)^4 |{\cal M}_{2\rightarrow 4}|^2}{2s}\times d\Phi_4 (p_1+p_2 ;p_3,p_4,p_5,p_6),
\label{eq:s2to4_1}
\end{equation} 
where $ {\cal M}_{2\rightarrow 4} $ is the scattering amplitude for $ e^+ e^-\rightarrow\phi_1\phi_1 f\overline{f} $, $ s $ is the square of centre-of-mass energy and $ \Phi_4 (p_1+p_2 ;p_3,p_4,p_5,p_6) $ is the four-body phase space. 
Since $\phi_2$ is on-shell produced in this process, we apply the narrow width approximation for on-shell $\phi_2$ to separate $ e^+ e^-\rightarrow\phi_1\phi_1 f\overline{f} $ to $ e^+ e^-\rightarrow\phi_1\phi_2 $ with $\phi_2\rightarrow\phi_1 f\overline{f}$. The differential phase space $ d\Phi_4 (p_1+p_2 ;p_3,p_4,p_5,p_6) $ can be expanded as
\begin{equation}
d\Phi_4 (p_1+p_2 ;p_3,p_4,p_5,p_6)=d\Phi_2 (p_1+p_2 ;p_3,l)\times d\Phi_3 (l;p_4,p_5,p_6)(2\pi)^3 dl^2,
\label{eq:s_phase}
\end{equation}
where $l$ is the four momentum of on-shell $\phi_2$.
Therefore, we can disassemble the differential cross section in the following form,
\begin{equation}
d\sigma (e^+ e^-\rightarrow\phi_1\phi_1 f\overline{f})=d\sigma (e^+ e^-\rightarrow\phi_1\phi_2)dB(\phi_2\rightarrow\phi_1 f\overline{f}),
\label{eq:s2to4_2}
\end{equation}
where 
\begin{equation}
d\sigma (e^+ e^-\rightarrow\phi_1\phi_2)=\frac{(2\pi)^4 |{\cal M}_{2\rightarrow 2}|^2}{2s}\times d\Phi_2 (p_1+p_2 ;p_3,l),
\label{eq:s2to4_3}
\end{equation}
and
\begin{align}
dB(\phi_2\rightarrow\phi_1 f\overline{f}) &
=\frac{2\pi}{\Gamma_{\phi_2}}\delta(l^2-M^2_{\phi_2})\frac{1}{2M_{\phi_2}}|{\cal M}_{1\rightarrow 3}|^2 d\Phi_3 (l;p_4,p_5,p_6)(2\pi)^3 dl^2 \nonumber \\ &
=\frac{d\Gamma_{\phi_2\rightarrow\phi_1 f\overline{f}}}{\Gamma_{\phi_2}}\delta(l^2-M^2_{\phi_2})(2\pi)^4 dl^2,
\label{eq:s2to4_4}
\end{align} 
where $ \Gamma_{\phi_2} $ is the total width of $\phi_2$, $ {\cal M}_{2\rightarrow 2} $ is the scattering amplitude for $ e^{+} e^{-}\rightarrow\phi_1\phi_2 $ and $ {\cal M}_{1\rightarrow 3} $ is the decay amplitude for $\phi_2\rightarrow\phi_1 f\overline{f}  $.
After applying the narrow width approximation for on-shell $\phi_2$,  $ |{\cal M}_{2\rightarrow 4}|^2 $ can be transferred to 
\begin{equation}
|{\cal M}_{2\rightarrow 4}|^2 =
|{\cal M}_{2\rightarrow 2}|^2\times\frac{\pi\delta(l^2-M^2_{\phi_2})}{M_{\phi_2}\Gamma_{\phi_2}}|{\cal M}_{1\rightarrow 3}|^2.
\label{eq:s2to4_5}
\end{equation}

Furthermore, we integrate out some phase space variables and leave the following kinematic variables :
\begin{itemize}
\item $ \cos\theta $ : $\theta$ is the direction of $ \phi_2 $ relative to the positron beam direction.
\item $ x_{-} $ : $ x_{-}=2E_f /M_{\phi_2} $ in the $ \phi_2 $ rest frame.
\item $ x_{+} $ : $ x_{+}=2E_{\overline{f}}/M_{\phi_2} $ in the $ \phi_2 $ rest frame.
\end{itemize}  
for the representation of differential cross section,
\begin{equation}
\frac{d\sigma (e^+ e^-\rightarrow\phi_1\phi_1 f\overline{f})}{d\cos\theta dx_{-}dx_{+}}=\frac{d\sigma (e^+ e^-\rightarrow\phi_1\phi_2)}{d\cos\theta}\frac{dB(\phi_2\rightarrow\phi_1 f\overline{f})}{dx_{-}dx_{+}}.
\label{eq:s2to4_6}
\end{equation}
Notice the opening angle between the fermion pair can be written as 
\begin{equation}
\cos\theta_{f\overline{f}} = 1-2(x_{-}+x_{+}-1+M^2_{\phi_1}/M^2_{\phi_2})/(x_{-}x_{+}).
\end{equation}
The $ d\sigma (e^+ e^-\rightarrow\phi_1\phi_2) / d\cos\theta $ can be found in Eq.(\ref{eq:22scalar1}) in the CM frame and Eq.(\ref{eq:22scalarLAB}) in the LAB frame and $ d\Gamma(\phi_2\rightarrow\phi_1 f\overline{f}) / dx_{-} / dx_{+} $ can be written as
\begin{equation}
\frac{d\Gamma(\phi_2\rightarrow\phi_1 f\overline{f})}{dx_{-}dx_{+}}=\frac{g^2_D \epsilon^2 \alpha M^3_{\phi_2}}{8\pi^2}\frac{(1-x_{+})(1-x_{-})M^2_{\phi_2}-(M_{\phi_2}-\Delta_{\phi})^2}{\left[M^2_{\phi_2}(x_{+}+x_{-}-1)-m^2_{Z'}\right]^2 +m^2_{Z'}\Gamma^2_{Z'}}.
\label{eq:s2to4_7}
\end{equation}
One can find $ x_{+} $ and $ x_{-} $ are symmetric in this formula.

The associated formulae for $ e^+(p_1) e^-(p_2)\rightarrow\chi_1(p_3)\chi_1(p_4)f(p_5)\overline{f}(p_6)$ can be obtained from Eq.(\ref{eq:s2to4_1}) to~(\ref{eq:s2to4_6}) by replacing $ \phi_{1,2} $ to $ \chi_{1,2} $. 
The $ d\sigma (e^+ e^-\rightarrow\chi_1\chi_2) / d\cos\theta $ can be found in Eq.(\ref{eq:22fermion1}) in the CM frame and Eq.(\ref{eq:22fermionLAB}) in the LAB frame and $ d\Gamma(\chi_2\rightarrow\chi_1 f\overline{f}) / dx_{-} / dx_{+} $ can be written as
\begin{align}
\frac{d\Gamma(\chi_2\rightarrow\chi_1 f\overline{f})}{dx_{-}dx_{+}}= & 
\frac{g^2_D \epsilon^2 \alpha M^3_{\chi_2}}{16\pi^2}
\frac{1}{\left[M^2_{\chi_2}(x_{+}+x_{-}-1)-m^2_{Z'}\right]^2 +m^2_{Z'}\Gamma^2_{Z'}} \nonumber \\ & 
\times\lbrace -M^4_{\chi_2}\left[x^2_{+}+x^2_{-}+2(x_{+}+x_{-})\right] +4M^3_{\chi_2}\Delta_{\chi} (1+x_{+}+x_{-}) \nonumber \\ &
-M^2_{\chi_2}\left[(6+x_{+}+x_{-})\Delta^2_{\chi} +2(6+x_{+}+x_{-})m^2_f \right] \nonumber \\ &
+2M_{\chi_2}(\Delta^3_{\chi} +4\Delta_{\chi} m^2_f)-m^2_f (\Delta^2_{\chi} +4m^2_f)\rbrace.
\label{eq:f2to4_1}
\end{align}

\section{The polar angular distributions for scalar and fermion inelastic DM models in Sec.~\ref{Sec:selections}}\label{Sec:pol_ang}

\begin{figure}
\centering
\includegraphics[width=3.0in]{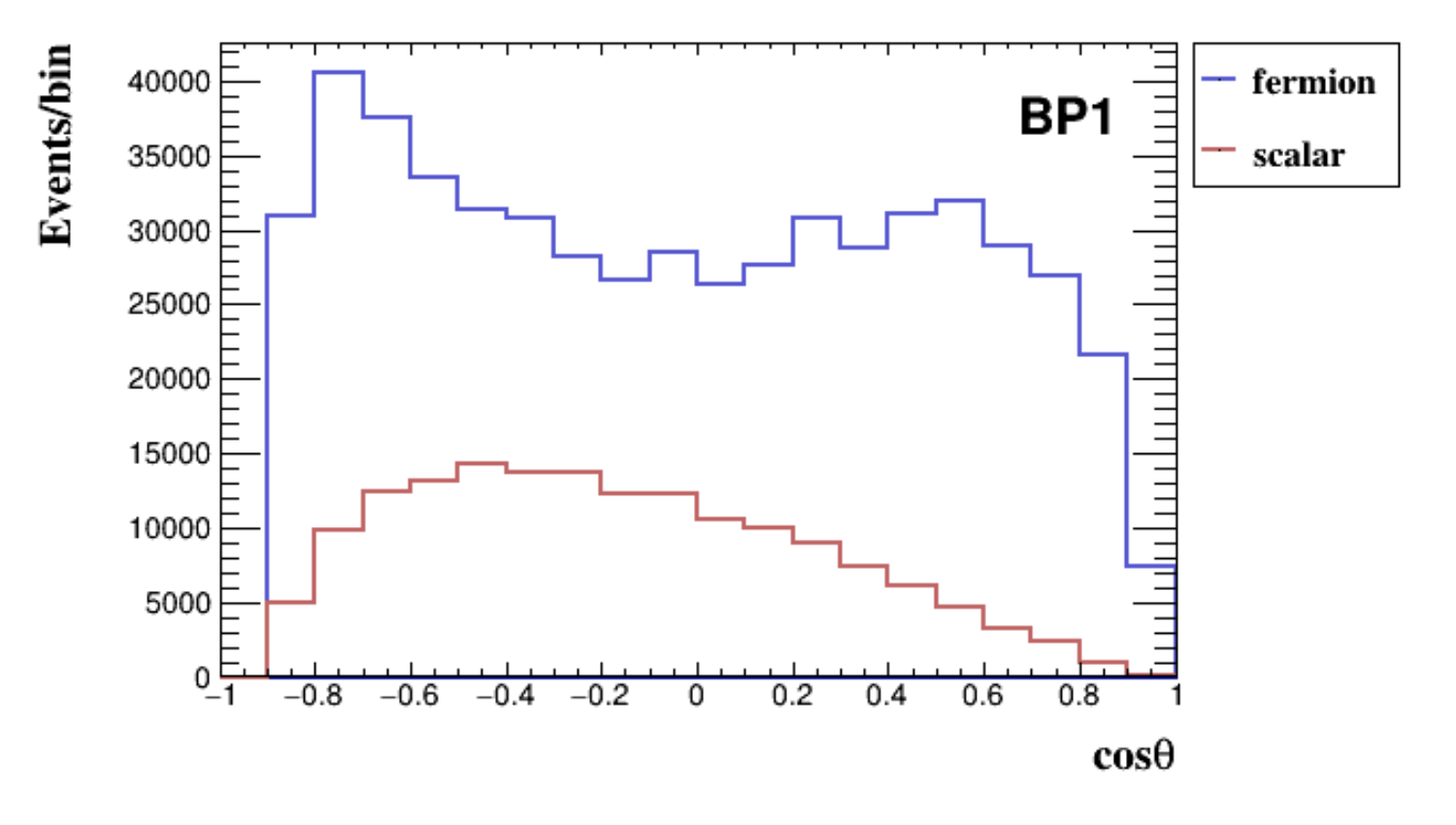}
\includegraphics[width=3.0in]{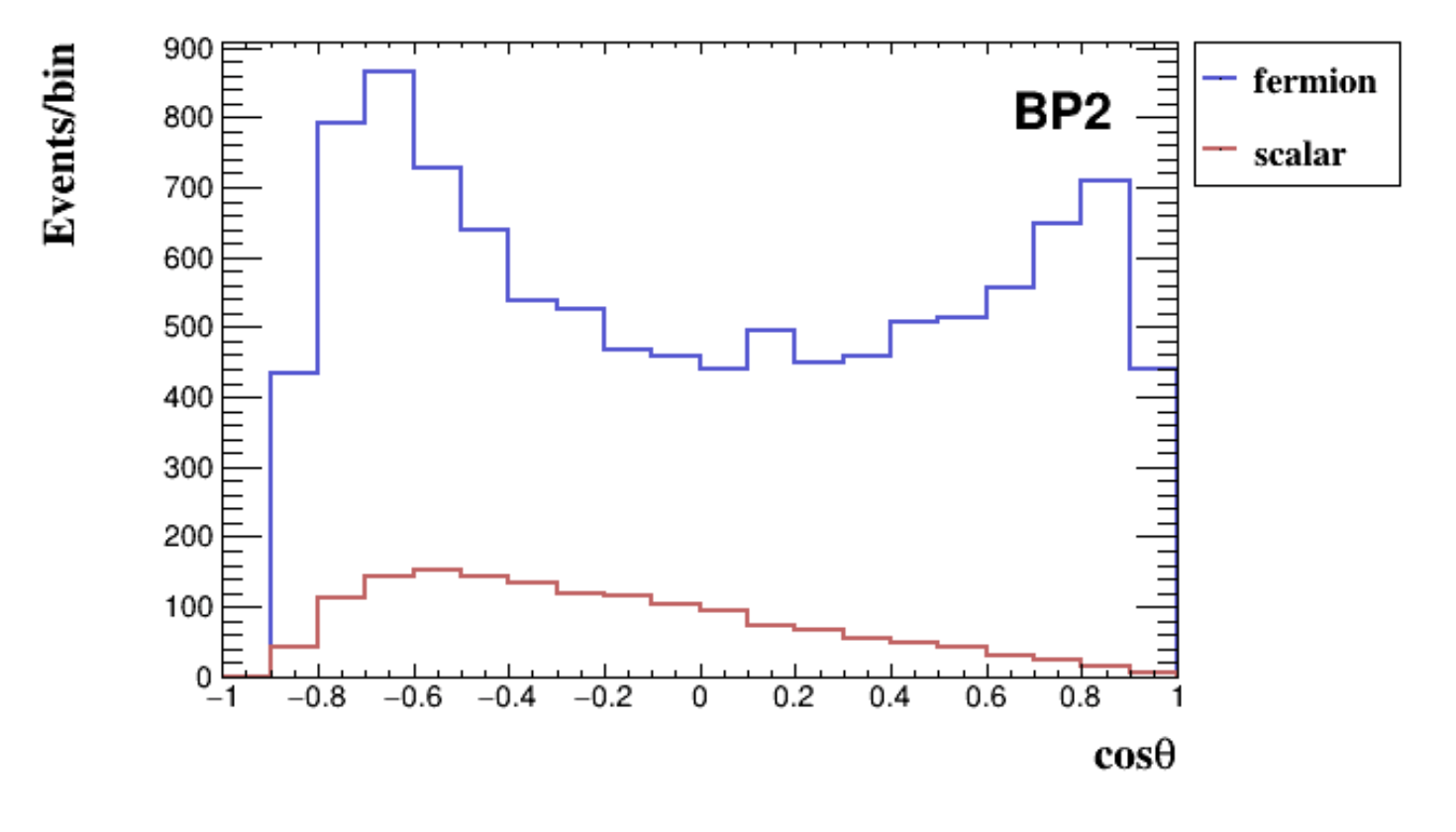}
\includegraphics[width=3.0in]{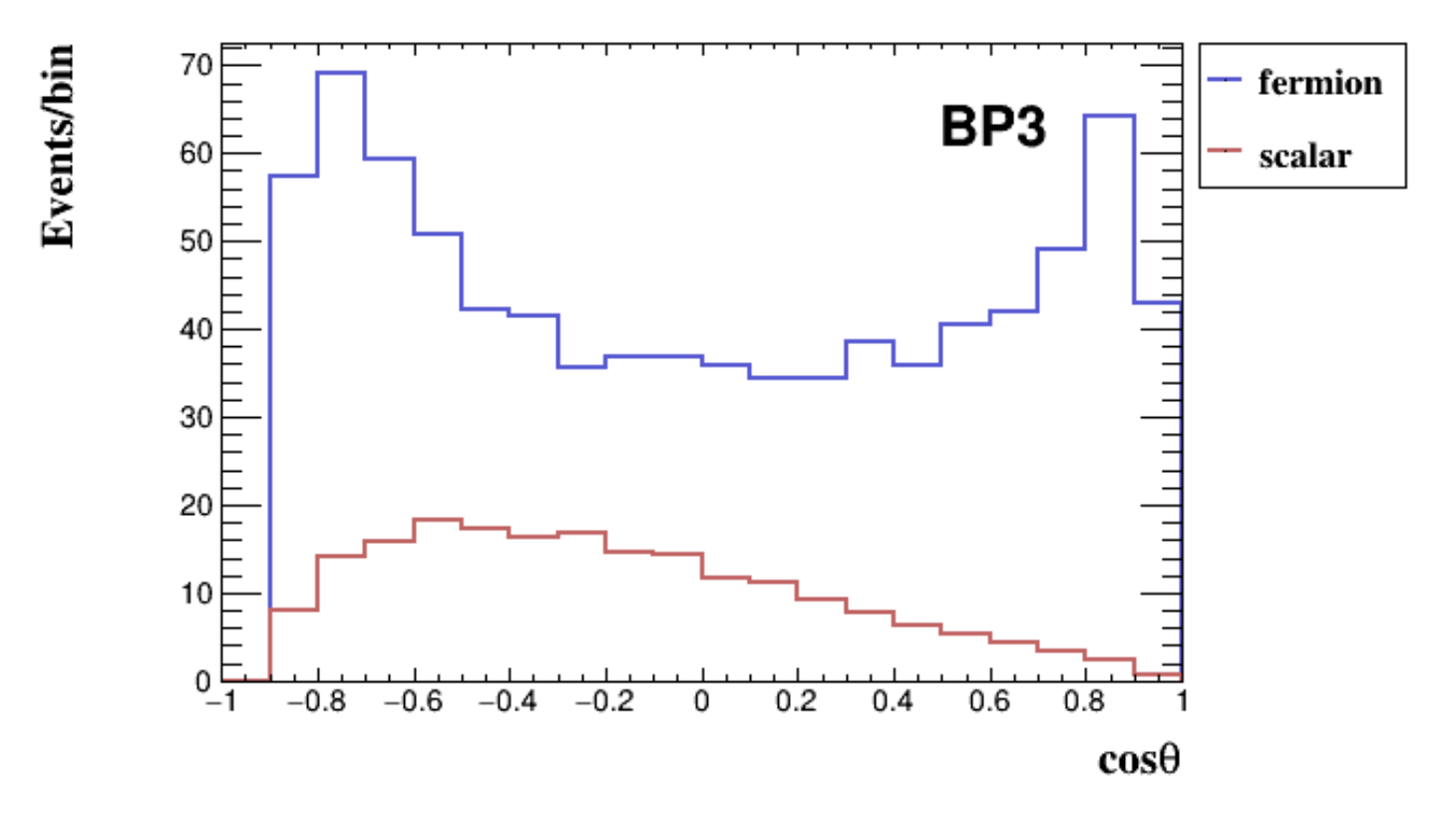}
\includegraphics[width=3.0in]{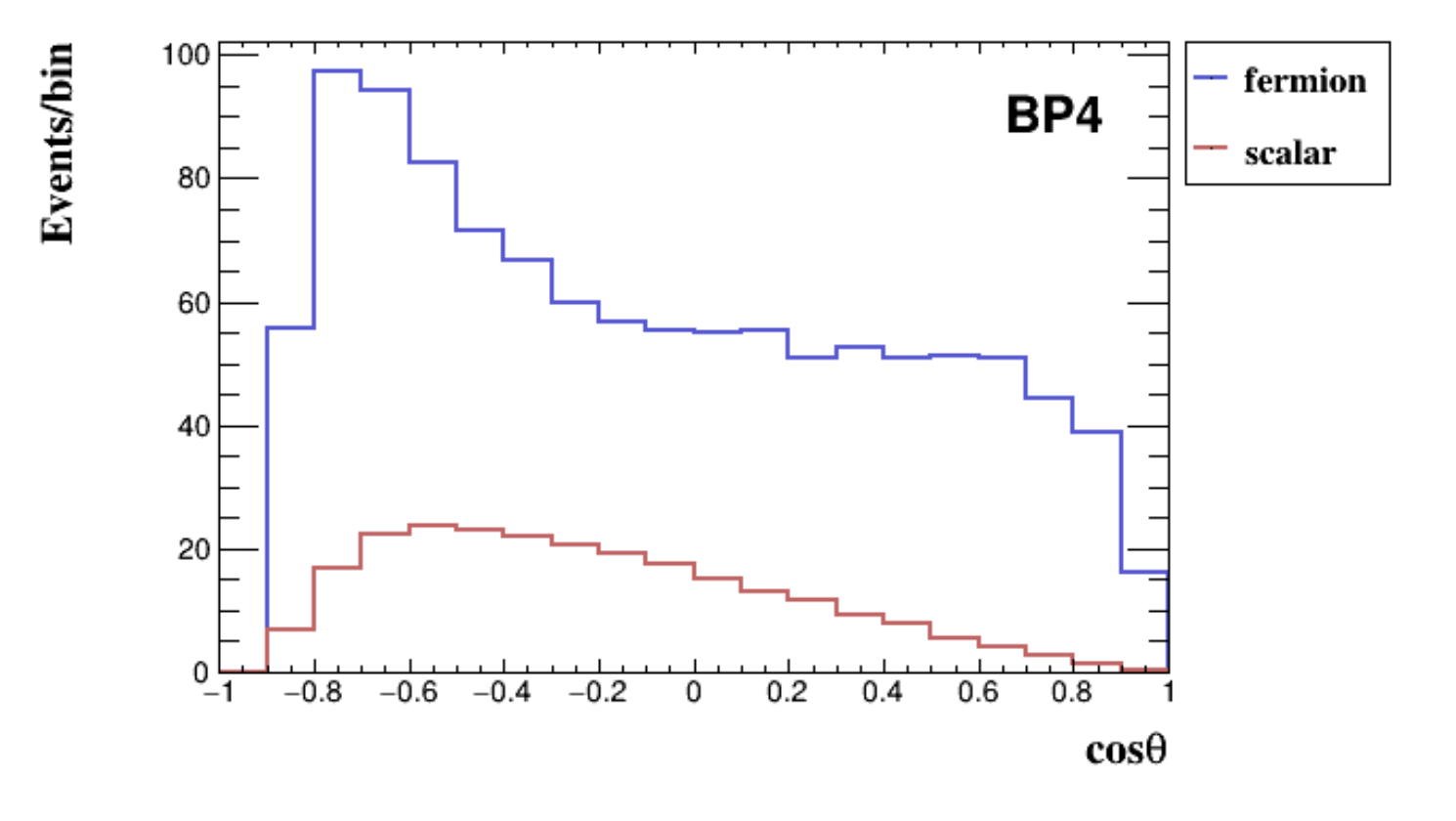}
\caption{
The polar angular distributions in the Belle II LAB frame after event selections in Table~\ref{Tab:selection} for process in Eq.(\ref{eq:woISR}) with electron pair in the final state and $N_{event}$ in Table~\ref{Tab:Eff1}. 
}\label{fig:polar}
\end{figure}

\begin{figure}
\centering
\includegraphics[width=3.0in]{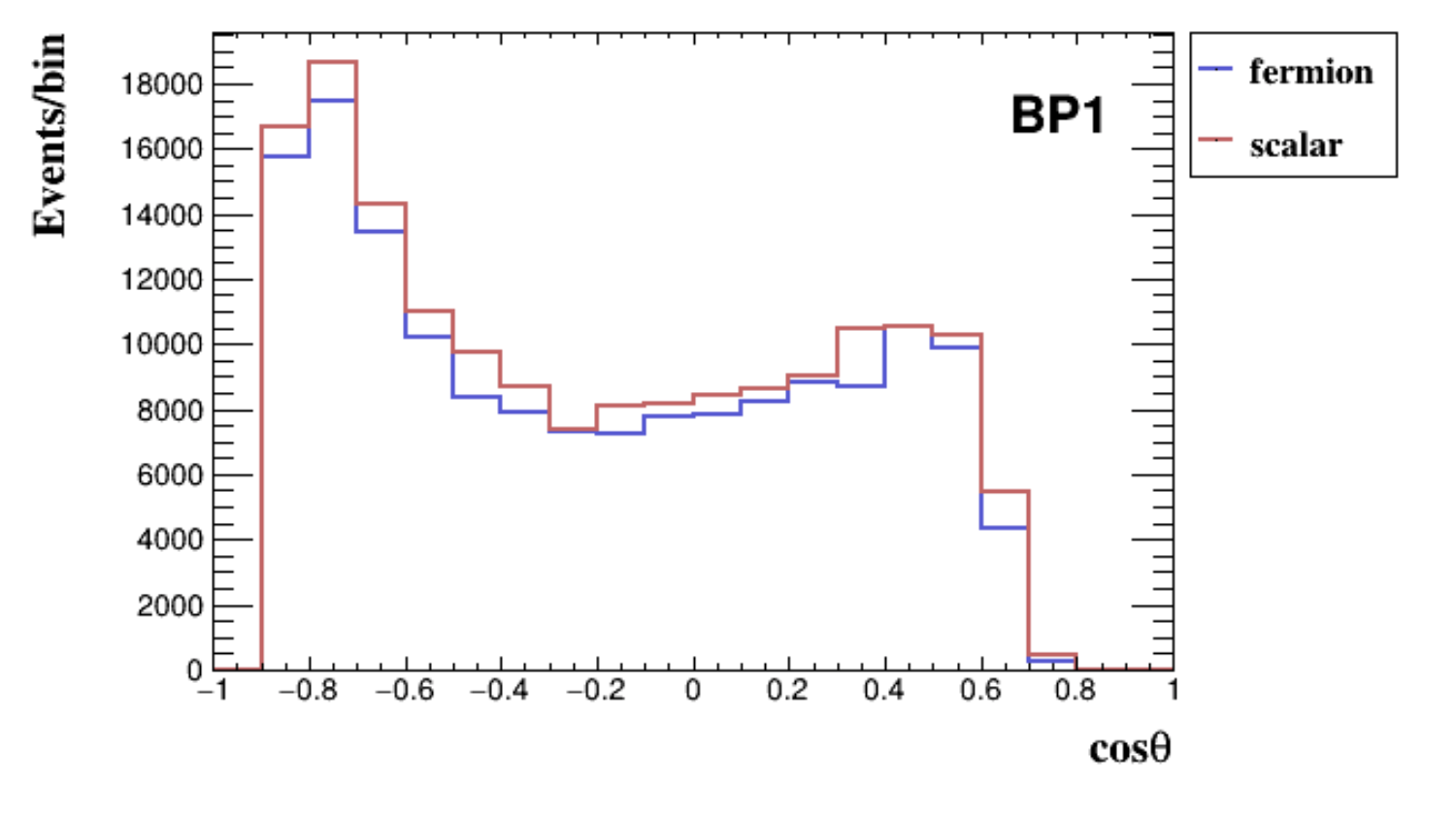}
\includegraphics[width=3.0in]{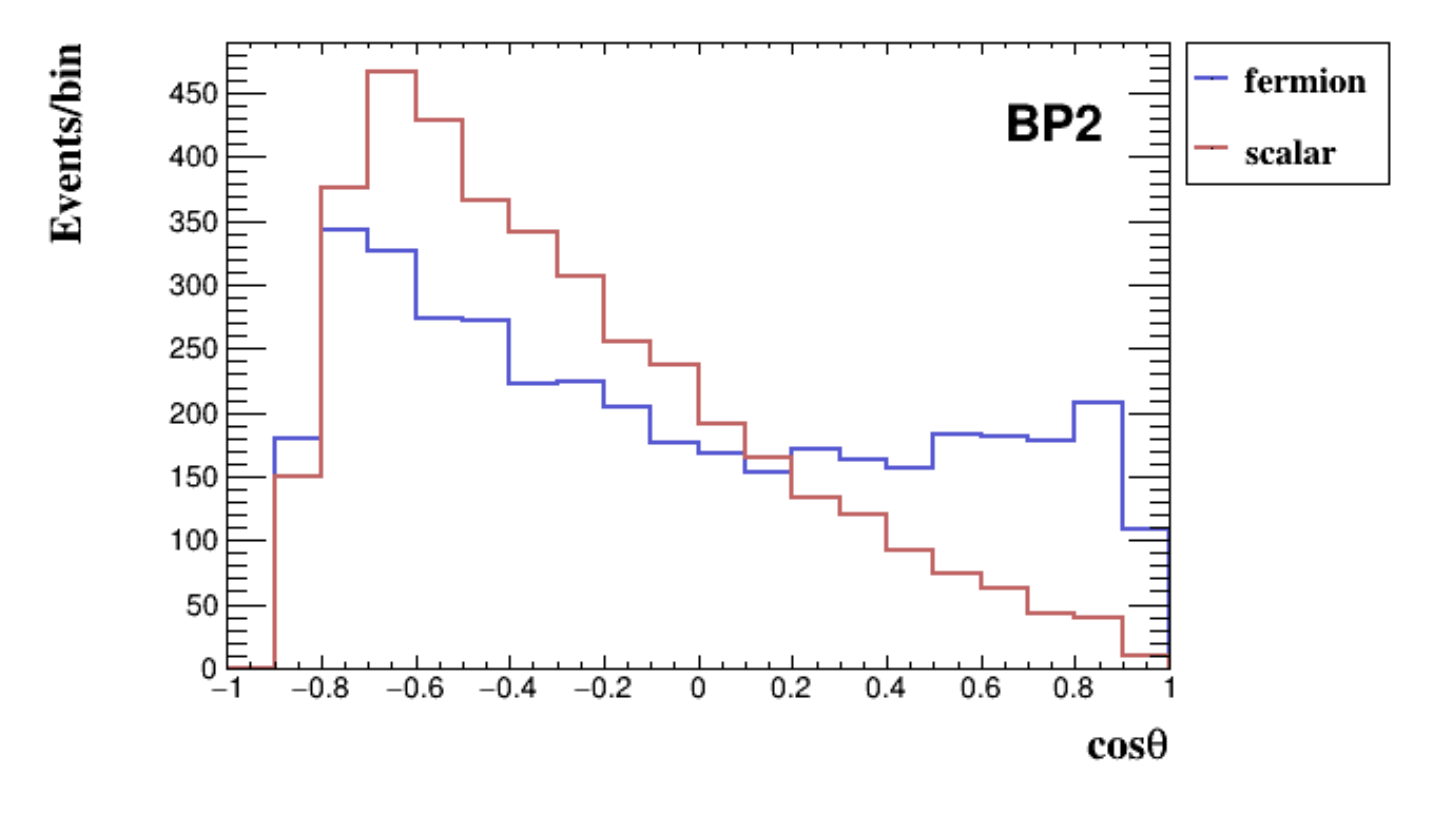}
\includegraphics[width=3.0in]{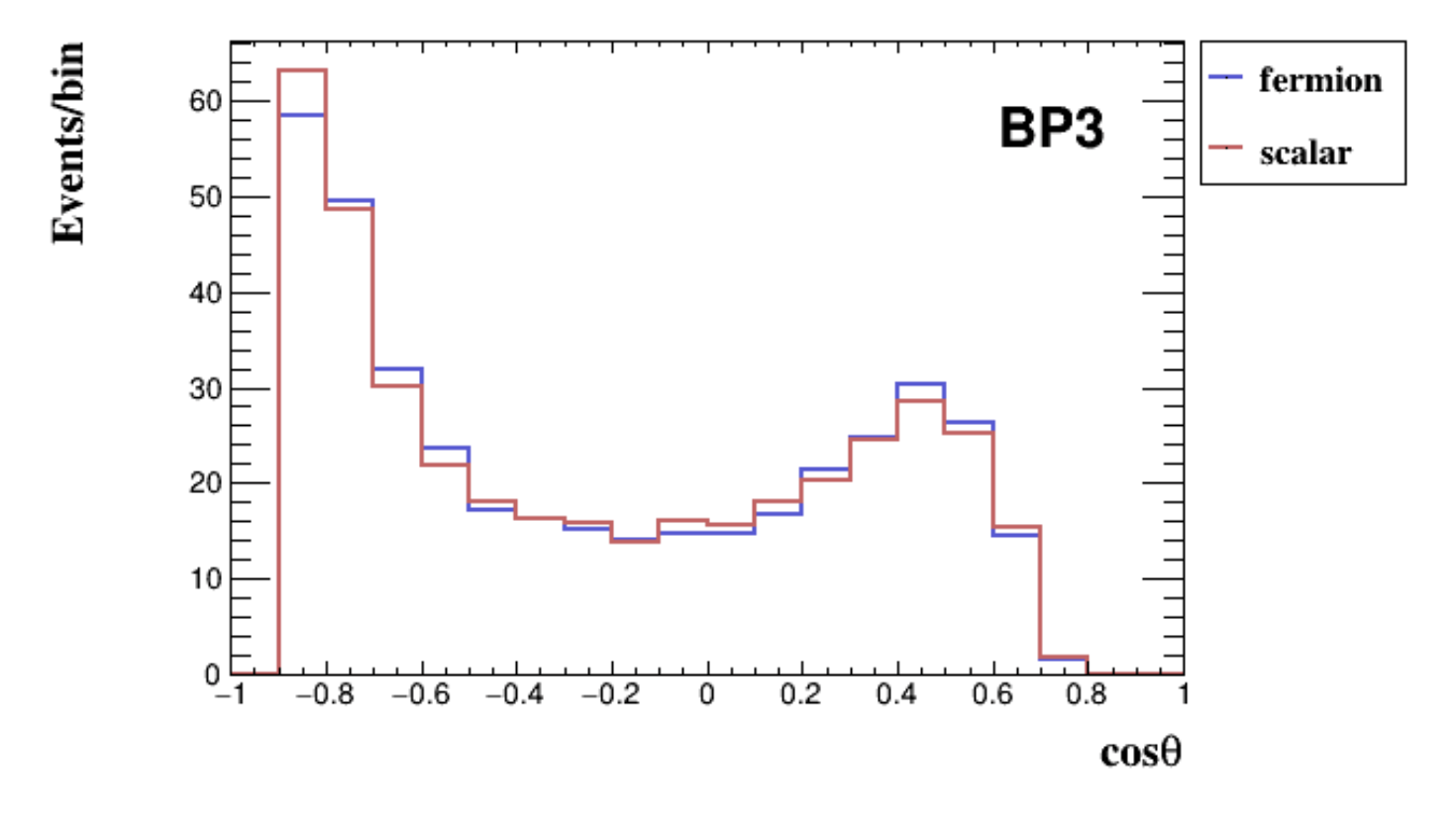}
\includegraphics[width=3.0in]{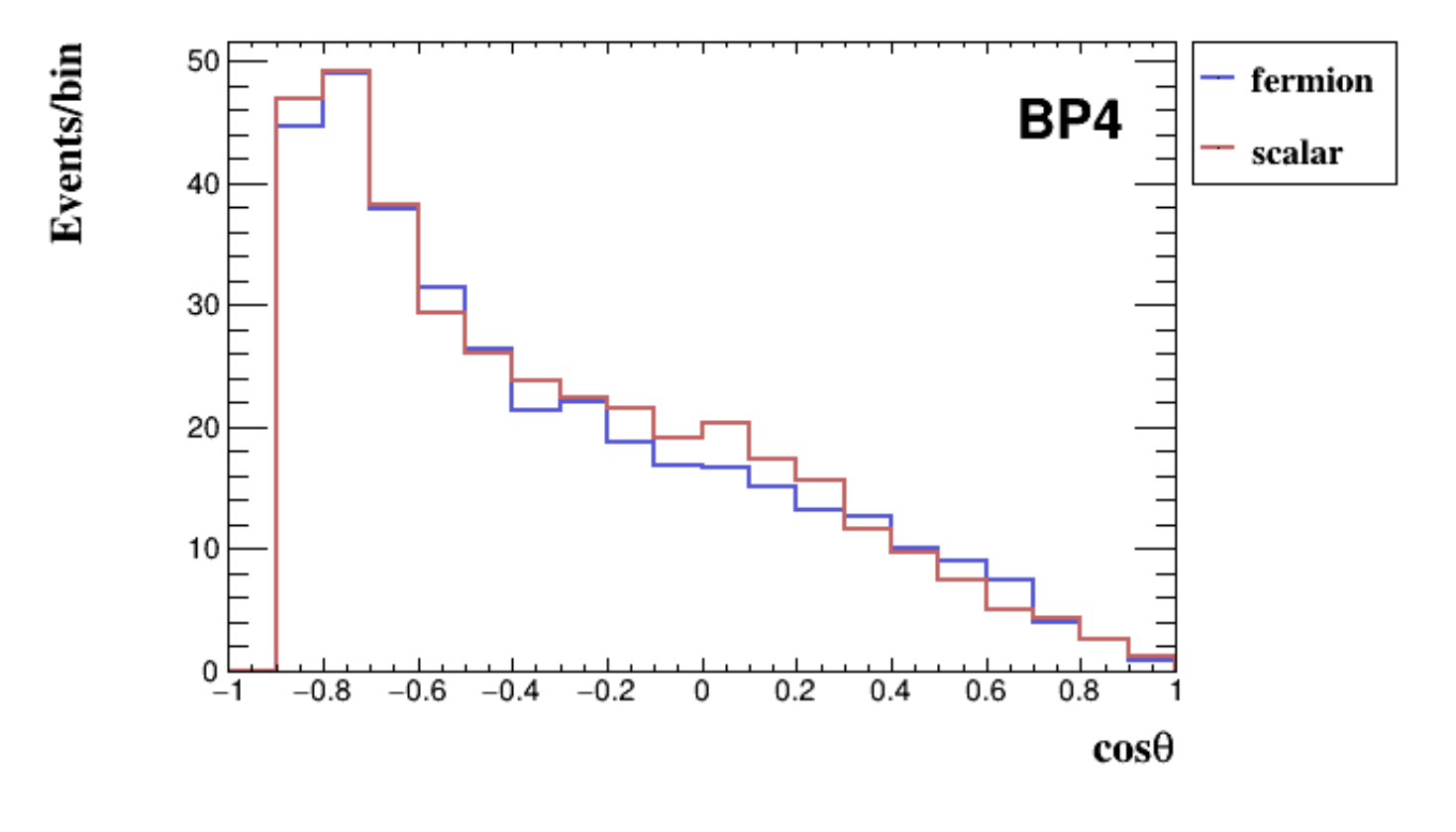}
\caption{
The same as Fig.~\ref{fig:polar}, but for process in Eq.(\ref{eq:wISR}) with electron pair in the final state
and $N_{event}$ in Table~\ref{Tab:Eff2}. 
}\label{fig:polar_ISR}
\end{figure}

In this appendix, we further compare polar angular distributions in the Belle II LAB frame after event selections in Table~\ref{Tab:selection} for processes in Eq.(\ref{eq:woISR}) and~(\ref{eq:wISR}) with electron pair 
in the final state in Fig.~\ref{fig:polar} and~\ref{fig:polar_ISR} with $N_{event}$ in Table~\ref{Tab:Eff1} and~\ref{Tab:Eff2}, separately. 
In Fig.~\ref{fig:polar}, since the selected regions can be approximately background free, we can directly count the events bin by bin to distinguish the signals are either from scalar or fermion inelastic DM models for these four BPs even only $ {\cal L} = 1~{\rm ab}^{-1} $ is used. 
However, in Fig.~\ref{fig:polar_ISR}, it is hard to distinguish the signals are either from scalar or fermion inelastic DM models except for the BP2. We have discussed this behavior in Sec.~\ref{Sec:Xsec} 
that we only have the chance to separate these two models if the ISR photon is soft enough. 
Therefore, in addition to the usual search process in Eq.(\ref{eq:wISR}) at B-factories, we propose to consider the process in Eq.(\ref{eq:woISR}) with the displaced vertex trigger suggested in Ref.~\cite{Duerr:2019dmv} which is more sensitive to distinguish the spin of DM in inelastic DM models.

\section{The kinematic equations for processes in Eq.(\ref{eq:woISR}) and~(\ref{eq:wISR})}\label{Sec:kin_eq}

In this appendix, we display detailed derivations of Eq.(\ref{eq:Echi2}), (\ref{eq:Kinematic2}), and (\ref{eq:Echi2_ISR}) in Sec.~\ref{Sec:kin_reco}. 
For $ e^+ e^-\rightarrow\chi_1\chi_2\rightarrow\chi_1\chi_1 l^{+}l^{-} $, we first set the direction of DV and following four momentum of $e^{+}$, $e^{-}$, $\chi_2$ and $\chi_1$ as
\begin{equation}
\widehat{r}_{DV} = (\sin\theta \cos\phi, \sin\theta \sin\phi, \cos\theta),
\label{eq:rDV}
\end{equation}
and
\begin{align}
& p_{e^{+}}=(E_{+},0,0,E_{+}), \nonumber \\ &
p_{e^{-}}=(E_{-},0,0,-E_{-}), \nonumber \\ &
p_{\chi_2}=(E_{\chi_2},|\overrightarrow{p_{\chi_2}}| \widehat{r}_{DV}), \nonumber \\ &
p_{\chi_1}=(E_{\chi_1},\overrightarrow{p_{\chi_1}}),
\label{eq:4momenta}
\end{align}
with $ E_{-}\neq E_{+} $ in the Belle II LAB frame, then according to energy and momentum conservation, we can write down 
\begin{align}
& E_{\chi_1}=E_{+}+E_{-}-E_{\chi_2}, \nonumber \\ &
\overrightarrow{p_{\chi_1}}=(-|\overrightarrow{p_{\chi_2}}|\sin\theta \cos\phi,-|\overrightarrow{p_{\chi_2}}|\sin\theta \sin\phi, E_{+}-E_{-}-|\overrightarrow{p_{\chi_2}}|\cos\theta).
\end{align}
After applying the above two equations to $ E^2_{\chi_i}=|\overrightarrow{p_{\chi_i}}|^2 +M^2_{\chi_i} $ for $ i=1,2 $, the $ E_{\chi_2} $ can be written as Eq.(\ref{eq:Echi2}). 
We then consider the energy and momentum conservation for $ \chi_2\rightarrow\chi_1 l^{+}l^{-} $, 
\begin{align}
& E_{\chi_2} = E_{\chi^{\prime}_1} + E_{l^{+}} + E_{l^{-}}\equiv E_{\chi^{\prime}_1} + E_{V^{\prime}},
\nonumber \\ &
|\overrightarrow{p_{\chi_2}}| \widehat{r}_{DV} =
 \overrightarrow{p_{\chi^{\prime}_1}} + \overrightarrow{p_{l^{+}}} + \overrightarrow{p_{l^{-}}}\equiv
\overrightarrow{p_{\chi^{\prime}_1}} + \overrightarrow{p_{V^{\prime}}}.
\end{align} 
Finally, we can receive Eq.(\ref{eq:Kinematic2}) for inputs of $ E_{V^{\prime}} $, $ \overrightarrow{p_{V^{\prime}}} $ and $ \widehat{r}_{DV} $ with $ E_{\chi_2} $ shown in Eq.(\ref{eq:Echi2}).

Similarly, for $ e^+ e^-\rightarrow Z^{\prime}\gamma\rightarrow\chi_1\chi_2\gamma\rightarrow\chi_1\chi_1 l^{+}l^{-}\gamma $, we use the same settings in Eq.(\ref{eq:rDV}),(\ref{eq:4momenta}) and set the four momentum of ISR photon in the Belle II LAB frame as 
\begin{equation}
p_{\gamma}=(E_{\gamma},\overrightarrow{p_{\gamma}}).
\end{equation}
 According to energy and momentum conservation, the four momentum of on-shell $ Z^{\prime} $ can be written as 
\begin{equation}
p_{Z^{\prime}} = (E_{Z^{\prime}},\overrightarrow{p_{Z^{\prime}}}) = (E_{+}+E_{-}-E_{\gamma},-p_{x,\gamma},-p_{y,\gamma},E_{+}-E_{-}-p_{z,\gamma}),
\label{eq:Zp4momenta}
\end{equation}
where $ \overrightarrow{p_{\gamma}} = (p_{x,\gamma},p_{y,\gamma},p_{z,\gamma}) $ and $ p^2_{Z^{\prime}} = m^2_{Z^{\prime}} $. Furthermore, we represent the four momentum of $ \chi_1 $ as 
\begin{equation}
p_{\chi_1}=(E_{Z^{\prime}}-E_{\chi_2},\overrightarrow{p_{Z^{\prime}}}-|\overrightarrow{p_{\chi_2}}| \widehat{r}_{DV}),
\end{equation} 
and write down the $ E_{\chi_2} $ as Eq.(\ref{eq:Echi2_ISR}). 
Finally, the kinematic equation for $ \chi_2\rightarrow\chi_1 f\overline{f} $ is the same as Eq.(\ref{eq:Kinematic2}) with $ E_{\chi_2} $ shown in Eq.(\ref{eq:Echi2_ISR}).


\end{document}